\documentclass{aa} 

\usepackage{graphicx}
\usepackage{txfonts}
\usepackage{xcolor}
\usepackage{hyperref}
\usepackage{epstopdf}

\newcommand{\sext}{\texttt{SExtractor}}

\newcommand{\re}{$\rm{R_e}$}
\newcommand{\logre}{$\log(\rm{R_e})$}   %%%%%% added by DFB
\newcommand{\muem}{$\rm{\langle\mu\rangle_e}$}
\newcommand{\BmV}{B--V}
\newcommand{\mV}{{$M_\mathrm{V}$}}
\newcommand{\DQ}{{$D_\mathrm{4000}$}}
\newcommand{\lM}{$\rm{\log(M^*)}$}

\newcommand{\bfilt}{$\rm{B}$}
\newcommand{\vf}{$\rm{V}$}

\newcommand{\Hbeta}{$\rm{H_{\beta}}$}

\newcommand{\nV}{{$n_\mathrm{V}$}}

\newcommand{\DCC}{$\Delta_{\rm CC}$}

\makeatother
\begin{document}
%\onecolumn
%\firstpage{1}

%\title{Cladistics on WINGS Galaxies}
\title{A Maximum Parsimony analysis of the effect of the environment on the evolution of galaxies}
%\author[\firstAuthorLast ]{\Authors}
%\address{}
%\correspondance{}
%\extraAuth{}% If there are more than 1 corresponding author, comment this line and uncomment the next one.
%\extraAuth{corresponding Author2 \\ Laboratory X2, Institute X2, Department X2, Organization X2, Street X2, City X2 , State XX2 (only USA, Canada and Australia), Zip Code2, X2 Country X2, email2@uni2.edu}
%\topic{}% If your article is part of a Research Topic, please indicate here which.

\author{D. Fraix-Burnet\inst{1}
\and 
M. D'Onofrio\inst{2}
\and
Paola Marziani\inst{3}
}

\institute{
Univ. Grenoble Alpes, CNRS, IPAG, Grenoble, France \\ \email{didier.fraix-burnet@univ-grenoble-alpes.fr}
\and 
Dipartimento di Fisica \& Astronomia, Universit\`a di Padova, Italy \\ \email{mauro.donofrio@unipd.it}
\and 
INAF, Osservatorio Astronomico di Padova, Italy \\ \email{paola.marziani@inaf.it}
}

\date{Received April 2019; accepted July 2019}

\abstract{Galaxy evolution and the effect of environment are most often studied using scaling relations or some regression analyses around some given property. These approaches however do not take into account the complexity of the physics of the galaxies and their diversification.}
{We here investigate the effect of cluster environment on the evolution of galaxies through multivariate unsupervised classification and phylogenetic analyses applied to two relatively large samples from the WINGS survey, one of cluster members and one of field galaxies (2624 and 1476 objects respectively).}
{These samples are the largest ones ever analysed with a phylogenetic approach in astrophysics. To be able to use the Maximum Parsimony (cladistics) method, we first performed a pre-clustering in 300 clusters with a hierarchical clustering technique, before applying it to these pre-clusters. All these computations used seven parameters: \BmV, \logre,  \nV, \muem, $H_{\beta}$, \DQ, \lM. }
{We have obtained a tree for the combined samples and do not find different evolutionary paths for cluster and field galaxies. However, the cluster galaxies seem to have accelerated evolution in the sense they are statistically more diversified from a primitive common ancestor. The separate analyses show a hint for a slightly more regular evolution of the variables for the cluster galaxies, which may indicate they are more homogeneous as compared to field galaxies in the sense that the groups of the latter appear to have more specific properties. On the tree for the cluster galaxies, there is a separate branch which gathers rejunevated or stripped-off groups of galaxies. This branch is clearly visible on the colour-magnitude diagram, going back from the red sequence towards the blue one. On this diagram, the distribution and the evolutionary paths of galaxies are strikingly different for the two samples. Globally, we do not find any dominant variable able to explain either the groups or the tree structures. Rather, co-evolution appears everywhere, and could depend itself on environment or mass.
	% We also do not find a correlation between our groups and the clusters.
}
{This study is another demonstration that unsupervised machine learning is able to go beyond the simple scaling relations by taking into account several properties together. The phylogenetic approach is invaluable to trace the evolutionary scenarii and project them onto any biavariate diagram without any a priori modelling. Our WINGS galaxies are all at low redshift, and we now need to go to higher redshfits to find more primitve galaxies and complete the map of the evolutionary paths of present day galaxies.} 


\maketitle

\section{Introduction}

Correlations between physical properties of galaxies   are helpful to extract key variables that may account for the variety in galaxy properties \citep{Lagos2016}.  The most fundamental of these correlations (called scaling relations) can effectively simplify the diversity of galaxy properties, however they do not enter in any model or numerical simulation of galaxy formation and evolution. They are used only a posteriori to check the goodness of our models in reproducing the observed correlations, and the galaxy formation models have still problems with many basic relations of galaxies, such as those including colours, metallicities, shapes, angular momentum, star formation rate and initial mass function \citep{DekelBirnboim2006,FallRomanowsky2013,Duttonetal2011,Cappellarietal2012}. 

The scaling relations are always in 2D or 3D because their identification is most often made from trials-and-errors, more rarely from statistical tools such as the principal component analysis (PCA) or the multivariate analyses that are explicitly aimed at minimising the number of variables needed to explain a large fraction of the variance in the $p$-dimensional (with $p \gg3$) space of variables. Although the scaling relations provide valuable insight, ultimately they cannot by themselves distinguish between cause and effect, past and future \citep[see e.g.,][]{Lagos2016}. This may be due to several reasons, one being that the scaling relations often show variability within or across the line/plane, this scatter being often correlated to another variable, showing the complexity of galaxy physics and evolution. Another reason can be that the scaling relations reveal some hidden (latent) variable which has not been or cannot be identified yet \citep{DFB2011}. 

Despite these problems, scaling relations provide very useful constraints for models; for example, the mass-metallicity relation
\citep[see e.g.][]{Faber1973} suggests the paths of chemical evolution and indicate the amount of inflow and outflow processes across cosmic epochs, the black-hole - bulge mass relation \citep[see e.g.][]{Magorrian1998} suggests the co-evolution of these two structures and the luminosity-size relation \citep[see e.g][]{Kormendy1977} constrains the epoch and location of star formation in the galaxy halo. Other relations, e.g. the velocity--luminosity relations  (the Faber-Jackson \citealt{FaberJackson1976}, the  Tully-Fisher relation \citealt{TullyFisher1977}), or the fundamental plane \citep{DjorgovskiDavis1987,Dressleretal1987},
indicate the main physical laws    driving the formation of galaxies and what are the main causes of the properties observed in the correlations. These information are extremely important for tuning models and simulations that today reproduce quite well the dark matter properties, but fail in reproducing the whole range of observable properties of stellar structures, probably for the complexity of baryon physics at local scales.

In this work we want to explore an alternative approach that we hope can permit to achieve a further progress in this research area. This approach can be summarised with the following question: what can we understand about the evolutionary paths of galaxies if we
start from the observed scaling relations? In other words, what can be deduced when a statistical multivariate analysis is applied to a set of galaxy variables? It must be realised that the multivariate analyses, especially the phylogenetic approaches, do reveal the equivalent of "scaling relations", but in a space whose dimension is given by the number of variables. This is much harder to visualise with plots representing real data and this is why there exists many visualisation tools (the typical cladistic tree being one of them) but they are still difficult to confront to physical models directly.
The phylogenetic approach can give a kind of chronology provided that the trees can be rooted (oriented) thanks to some ancestrality hypothesis.  This means that once we will have at our disposal data for galaxies at different cosmic epochs we could use the phylogenetic approach to identify the main roots (primitive kinds of galaxies) of evolution and derive useful constraints for our models. Different formation histories might in fact result in different types of correlations between galaxy variables.

Unfortunately galaxy variables across cosmic epochs are often difficult to obtain and are subject to numerous sources of biases. Good and abundant measures are available today only for nearby galaxies, being the high redshift objects still sparsely observed and at the limit of present day telescopes. However, in the last years thanks to the development of galaxy surveys, such as WINGS \citep{Fasanoetal2006}, SDSS \citep{Abazajianetal2003}, SAURON \citep{Baconetal2001} and many others, it was possible to collect a lot of galaxy measures for hundreds and thousands of galaxies that can be used for our statistical investigation. 

This work, based on the robust statistical sample of nearby galaxies in  WINGS clusters, tries to use the multivariate statistical analysis for checking the possibility of understanding the preferential paths of galaxy evolution hidden within the complex interrelations among the structural variables. At this moment, the sample involves galaxies in the redshift range 0.04 -- 0.07, at a typical radial comoving distance of $\approx$ 250 Mpc.  Future works with data for high redshift galaxies might in a near future complete the construction of the main paths of galaxy evolution.

The paper is designed as follow. In Sec.~\ref{Data} we describe the WINGS data sample used in this work and we give a short review of the galaxy variables used in our analysis. Sec.~\ref{Analyses} introduce the multivariate analysis approach, in particular the Maximum Parsimony method (Sec.~\ref{CladMeth}). We discuss the selection of the variables used in the phylogenetic analysis in Sec. \ref{paramselect} and we present our results in Sec.~\ref{Results} discussing three particular cases: the full sample (Sec.~\ref{Result4czb}), the cluster sample (Sec.~\ref{Result4czbM}) and the field galaxy sample (Sec.~\ref{Result4czbN}).
The discussion follows in Sec.~\ref{Discussion} and our conclusions in Sec.~\ref{Conclusion}.

\section{The data}
\label{Data}

\subsection{ The WINGS sample}
\label{WINGSsample}

The data used in this work are those belonging to the \textit{Wide-field Imaging of Nearby Galaxy-clusters Survey} (hereafter WINGS) survey.\footnote{See the WINGS web-site for all the details of this project at http://web.oapd.inaf.it/wings.} WINGS is a long term research project especially designed to provide a robust characterisation of the photometric and spectroscopic properties of galaxies in nearby clusters ($0.04<z<0.07$) \citep{Fasanoetal2006}. 
The core of the survey WINGS-OPT \citep{Varelaetal2009} is the database of optical \bfilt\ and \vf\ images of 77 clusters, obtained during dark time with the Wide Field Camera (WFC, $34'\times 34'$) mounted at the corrected f/3.29 prime focus of the INT-2.5 m telescope in La Palma (Canary Islands, Spain) and with the Wide Field Imager (WFI, $34'\times 33'$) mounted at the f/8 Cassegrain focus of the MPG/ESO-2.2 m telescope in La Silla (Chile), that have been analysed with \sext\ \citep{bertinarnouts96}. The optical photometric catalogues  are 90\% complete at \vf $\sim 21.7$, which translates to $M^{*}_{V} + 6$ at the mean redshift of the survey \citep{Varelaetal2009}. The WINGS-OPT database includes respectively 393013 galaxies in the \vf\ band and
391983 in the \bfilt\ band. 

The WINGS database has a spectroscopic follow-up WINGS-SPE for a subsample of 48 clusters (26 in the north and 22 in the south hemisphere) with the spectrographs WYFFOS@WHT ($\lambda$range $= 3800\div7000$ \AA, resolution   FWHM $=3$ \AA) and 2dF@AAT ($\lambda$range $= 3600\div8000$ \AA, resolution  FWHM $=6$ \AA). From these spectra we derived the redshift of 6132 galaxies \citep{Cava09}.  A subsample of this (5299 galaxies) have been analysed with spectrophotometric techniques deriving the star formation rate $sfr*$ at different epochs, the stellar masses $M^*$ and ages, the internal extinction $A_V$ and the equivalent widths of the main absorption features \citep{Fritz2011}. 

A full re-analysis of the WINGS spectra, aimed at measuring the emission line properties, has been also performed by \cite{Marzianietal2017}. These authors considered 46 clusters of the WINGS-SPE survey, and performed population synthesis from spectra that were not considered in \citep{Fritz2011}: spectrophotometric data were available for 5859 sources in the fields of 46 clusters. Of these, 3514 spectra were of cluster galaxies. All cluster spectra were then joined to form a stacked sample with 3514 spectra, while 2344  spectra of non-cluster sources were considered as the field sample.

\begin{figure}[ht]
\centering
\includegraphics[width=\linewidth]{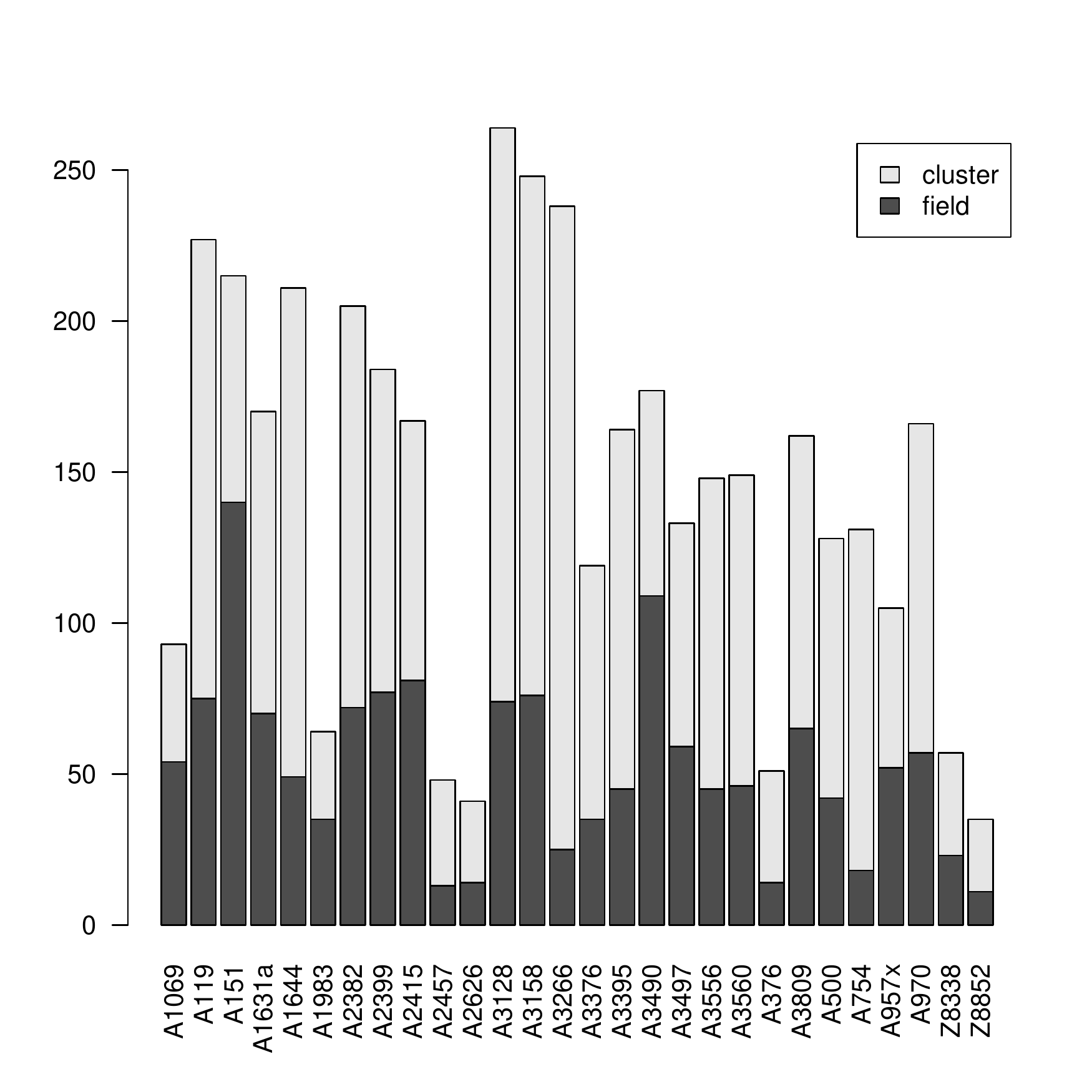}
\caption{Histograms of cluster and field sample galaxies into the clusters.}
\label{fig:histclusters}
\end{figure}

\begin{figure}[t]
\centering
\includegraphics[width=\linewidth]{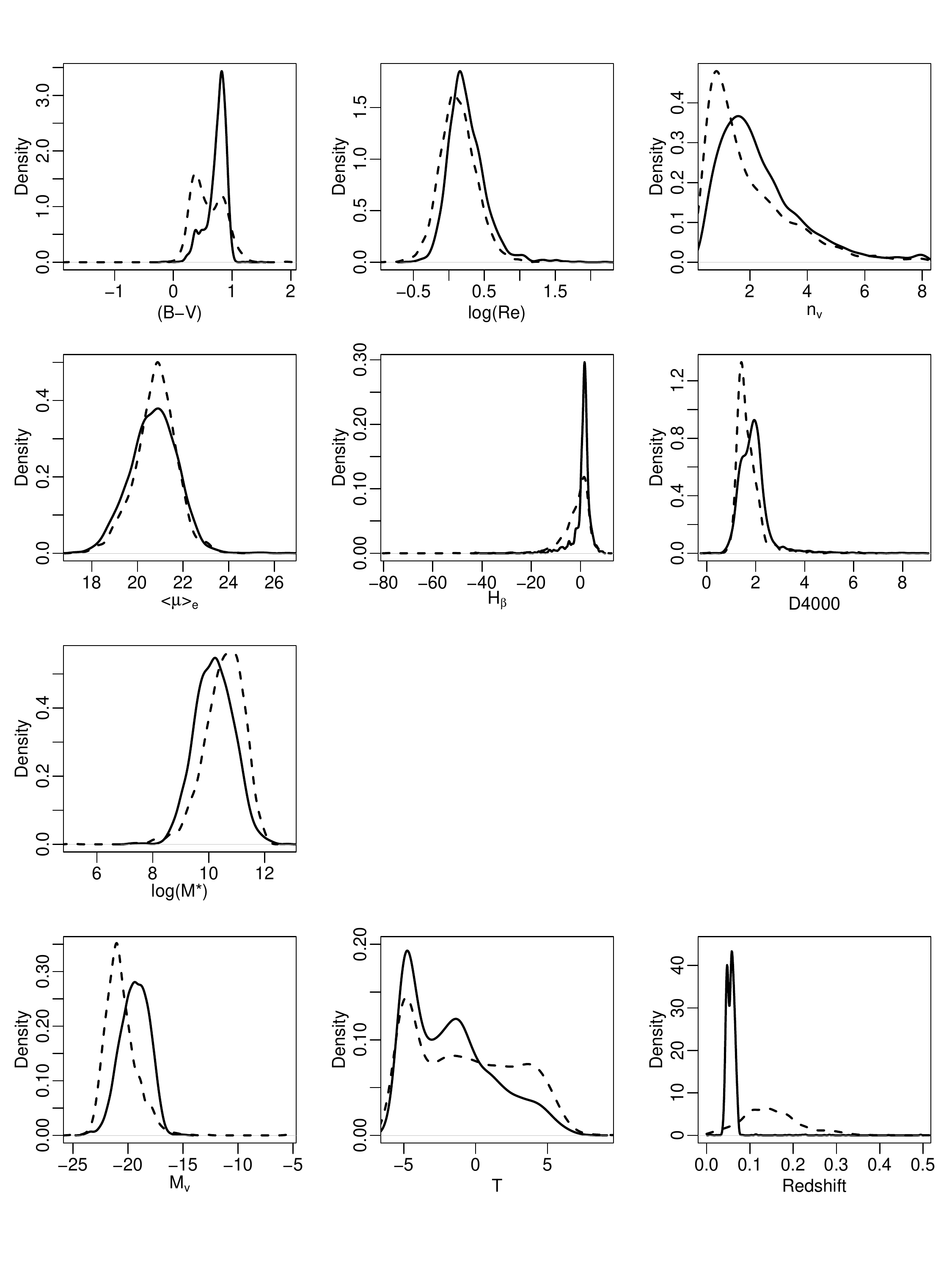}
\caption{Gaussian kernel density distribution of the seven variables used in the clustering and MP computation (Table~\ref{tab:param}), plus  the absolute V magnitude, the morphology $T$ and the redshift, for the cluster (solid line) and field (dotted line) samples. The scale on the y-axis is given by the integral of the kernel density estimation which is normalised to 1.}
\label{fig:redshifts}
\end{figure}

\begin{table}[t]
\caption{Variables available for the WINGS sample. Only the seventh first parameters (top panel) have been used for the clustering and MP analysis.}
\label{tab:param}
\begin{tabular}{l l l}
\hline
            \noalign{\smallskip}
\BmV\           &\BmV\ colour             \\
\logre\      & Logarithm of effective radius [kpc]   \\
\nV\          & Sersic index         \\
\muem\         & Mean effective surface brightness \\
$H_{\beta}$    & Equivalent Width of H$\beta$  \\
\DQ\  & Balmer Discontinuity index \\
\lM\     & Logarithm of stellar mass \\
            \noalign{\smallskip}
            \hline
            \noalign{\smallskip}
\mV         & Absolute V magnitude \\ 
$T$           &  Morphological type  \\ 
$\epsilon$     & Ellipticity             \\ 
$\Delta_{CC}$ & Projected distance to cluster centre \\
%$SFR$         & Star formation rate \\ 
$sfr1$         & Star formation rate in M$_{\sun} \mathrm{yrs}^{-1}$: stellar mass \\
               & produced within the age bin $0-2\cdot10^7$ yrs  \\ 
$sfr2$         & idem of $sfr1$ for the bin $2\cdot10^7-6\cdot10^8$ yrs \\ 
$sfr3$         & idem of $sfr1$ for the bin $6\cdot10^8-5.6\cdot10^9$ yrs \\ 
$sfr4$         & idem of $sfr1$ for the bin $5.6\cdot10^9-17.8\cdot10^9$ yrs \\ 
$M1/M4$     & Stellar production ratio: \\
            & $\left(sfr1\times 2\cdot10^7\right)/\left(sfr4 \times\left(17.8\cdot10^9-5.6\cdot10^9\right)\right)$ \\ 
W(H$\alpha$) & Rest-frame W(H$\alpha$) \\ 
$\left[\mathrm{NII}\right]$ /H$\alpha$& Intensity ratio [NII]${\lambda 6584}$/H$\alpha$ \\ 
            \noalign{\smallskip}
            \hline
\end{tabular}
\end{table}

\subsection{The adopted sample}
From the WINGS database, a sample of 4100 galaxies  has been extracted for this work. Distances were determined from their redshifts. 2624 galaxies were determined to be members of 28 galaxy clusters (Fig. ~\ref{fig:histclusters}) with redshifts around 0.05-0.06. The remaining 1476 galaxies are background galaxies with slightly larger redshifts around 0.15 (see Fig.~\ref{fig:redshifts}). They are considered as field galaxies.
This field sample is heterogeneous, since it is represented by galaxies in the field
of our clusters. Binary galaxies or small groups might be present. The possible
presence of subgroups have been analysed by \citet{Ramella2007}.
Cluster background galaxies have been eliminated on the basis of the colour-magnitude diagrams \citep[see][]{Valentinuzzi2011} and on the results obtained by \citet{Ramella2007}. In general, galaxies redder than 
\BmV=1.4, not belonging to background clusters have been kept. The reader should 
keep in mind this fact that could partly affect our conclusions, when the cluster and field samples are compared.

The GASPHOT tool used in WINGS to derive radius, surface brightness, mass, ellipticity etc, takes into account the convolution with the PSF. Hence, the model profiles includes the PSF measured on nearby stars. In addition the smallest galaxies cover always a surface of 200 pixels squares, so the minimum radius is about 8 pixels while the typical stellar PSF of the survey of around 3 pixels. The light profiles are well seen even if the pixel scale is different for field and cluster objects. All the parameters measured for Cluster and Field objects are thus equally well derived, alleviating any potential biases in the comparison of the properties between cluster and field objects.

In this paper, we consider 18 galaxy properties (see Table~\ref{tab:param}). More variables are available, but most of them have a large fraction of unavailable values.

The set of seven parameters used for the cladistic analysis itself (see Sect.~\ref{paramselect}) is reported on the top part of Tab. \ref{tab:param}. They are: the total \BmV\ galaxy colour corrected for galactic extinction and K-correction; the logarithm of the circularised effective radius \logre\ in kpc units; the Sersic index \nV\ in the $V$ band obtained by Gasphot \citep{Donofrioetal2014} from the fit of the seeing convolved major and minor axes surface brightness profiles; the mean effective surface brightness in the \vf\ band; the equivalent width  (EW) of the $H_{\beta}$ line visible in the spectrum,   following \citet{Fritz2011}; the Balmer discontinuity at 4000 \AA\ \DQ; the logarithm of the stellar galaxy mass \lM\ derived from the SED fitting \citep{Fritz2007}. The distribution of these seven parameters for the cluster and field samples are essentially the same except for \BmV, \nV\ and \lM, field galaxies being somewhat more massive, bluer and have a smaller Sersic index (Fig.~\ref{fig:redshifts}). This implies that any difference in clustering results between the two populations could not be explained simply some monovariate differences.

The other variables are used for describing the characteristics
of the groups and their mutual relationships. Among them, the total galaxy luminosity $M_V$, the morphological types derived with Morphot by \cite{Fasano2012}, the galaxy mean ellipticity derived from the Gasphot analysis \citep{Donofrioetal2014},  the projected distance from the cluster centre BCG, the star formation rates ($sfr1$, $sfr2$, $sfr3$ and $sfr4$)
computed in four age ranges ($ 0 - 2\cdot10^7, 2\cdot10^7-6\cdot10^8, 6\cdot10^8-5.6\cdot10^9$ and $5.6\cdot10^9-17.8\cdot10^9$ years), the ratio between the stellar mass formed at the latest epoch to the mass formed at the earliest epoch
(that we will call the stellar production ratio $M1/M4 = \left(sfr1\times 2\cdot10^7\right)/\left(sfr4 \times\left(17.8\cdot10^9-5.6\cdot10^9\right)\right)$),  
the equivalent width W(H$\alpha$) and the intensity ratio [NII]${\lambda 6584}$/H$\alpha$\footnote{The values for the emission lines H$\alpha$ and  [NII] are not available for all galaxies and have been simply disregarded in the corresponding plots.}. 

With the exception of the stellar mass $M^{*}$ all variables come from direct measurements made on images and spectra. They are not model dependent.

\section{Analyses}
\label{Analyses}

\subsection{The Maximum Parsimony analysis}
\label{CladMeth}
Phylogenetic approaches seek to establish relationships between objects or classes of objects. They are looking for the simplest evolutionary scenario to transform any object of the sample into any other one. The result is a graphical representation (most often a tree) on which objects are gathered according to their evolutionary closeness. The phylogenetic techniques can be distance-based if pairwise (dis)similarity matrices are computed from the parameters, or character-based if the parameters are used directly. These parameters are called characters if they can be given discrete values that represent evolutionary steps. In the ideal case, innovations occurring in the course of evolution are transmitted to all the descendants, allowing the reconstruction of the entire diversification history of the sample. The characters can be discrete, continuous, or categorical (nominal, qualitative) in the case of genetic data. In astrophysics, most variables are quantitative and continuous.

The Maximum Parsimony \citep[or cladistics, hereafter MP, ][]{hennig1965,Felsenstein1984} technique is character-based and the most general phylogenetic tool. Because it looks for the simplest path connecting the objects under study, it has some similarity with the Minimum Spanning Tree technique that has been heavily used to determine the galaxy clusters in order to map the spatial distribution of the baryonic matter \citep{Barrow1985,Bhavsar1996}. However, in the MP technique, internal nodes are introduced, creating a much larger variety of tree topologies at the expense of a much heavier computation.

Among all the possible arrangements of the objects on a tree, MP selects the one that has the lowest total number of step changes (the score). A step change is the absolute difference between the values of a parameter (character) between two nodes of a branch (edge) of the tree. By minimizing the score, the algorithm selects the simplest evolutionary path that links all the objects. Mathematically, the score of a tree corresponds, after labelling of the internal nodes, to the minimum number of edges $(u,v)$ with $c(u)\neq c(v)$, $c(u)$ being the character state at node $u$.  To each internal node is associated a real value $f(u)$. The score $s$ of a tree equals the sum over all edges of the absolute difference between those values:

\begin{equation}
\label{eq:MPs}
s = \sum_{e=(u,v) \epsilon E}  \lvert f(u) - f(v) \rvert
\end{equation}

This can be directly extended to continuous characters or values. 

As a consequence, two objects which are close on the tree share similar evolutionary histories. In this way, we can define evolutionary "families" that are called "clades" in biology but that we will simply call groups. This approach has been explained in detail in many papers for astrophysics \citep[e.g. ][]{jc1,jc2,Fraix-Burnet2015,Fraix-BurnetHouches2016,Holt2018}, and the simplest illustration of the power of the phylogenetic approach in astrophysics is the reconstruction of the stellar evolutionary paths and of the families of stars born together in open clusters, a problem known as "chemical tagging"  \citep{Blanco-Cuaresma2018}. 

In this paper, the parameter values have been discretised into 32 bins, and the MP computations were performed with PAUP 4.0 \citep{paup}. There are generally several to many equally most parsimonious trees, so that we make a consensus tree which keeps internal nodes appearing in at least 70\% of them. We aim at what is called a "resolved" tree (or binary tree) where each internal node has three branches and only three branches. In this case, the evolutionary scenario depicted by the tree is more informative (it is said that the phylogenetic signal is strong). This is the criterion we used to select the best subset of parameters (see Sect.~\ref{paramselect}).
The reliability of the result was then estimated through several methods:
\begin{enumerate}
   \item Other subsets of parameters: in theory the best indicator of robustness is a bootstrap approach. Since the computation time is too large, we simply ran several analyses with some parameters removed, and compared tree structures and the evolution of the parameters along the trees.
   \item We performed a Neighbour Joining Tree estimation \citep{NJ1987,NJ2006,Fraix-Burnet2015,Jofre2017} using both the Euclidean and the Manhattan (L1 norm) distances. This is a different but also quite popular phylogenetic technique, based on distances between objects. This is a bottom-up hierarchical clustering method that minimises a tree length, according to a criterion that can be viewed as a Balanced Minimum Evolution \citep{NJ2006}
   \item Using the median values of the parameters for each group defined from the MP analysis, we ran a Branch-and-Bound algorithm \citep{BandB1982} which thoroughly explores all the possible tree topologies and select the most parsimonious one. This algorithm is much more efficient in finding the optimal tree than the heuristic one used for the three samples, but it can only be used on a small number of objects.
\end{enumerate}
      The fact that all these analyses agree is a good indicator that a phylogenetic signal is present in the set of parameters used.

\subsection{Pre-clustering with the hierarchical clustering technique}
%\subsection{Maximum Parsimony with a large sample}
\label{MPlarge}

Since the computing time to explore the tree space is prohibitive beyond, say, 800 objects
or so, we decided to apply a pre-clustering technique. The idea is that we are not aiming at obtaining a genealogy of galaxies, which is a non-sense, but a phylogeny, in other word a diagram depicting the evolutionary relationships between classes of galaxies, these classes being unknown at this stage. Each galaxy of our sample is first considered as a potential exemplar representing its own class but similar objects are expected to be redundant. Hence, we can try to group them together into what we will call sub-clusters. A class is thus made of one or more sub-clusters. We can adapt the number of sub-clusters to the computational constraint of the MP analysis and to the expected or desired number of classes. The advantage of this approach is that we can use very efficient methods of multivariate clustering with the pairwise distances (between each pair of galaxies) using the $p$-parameters. The most logical technique for our particular goal seems to be the hierarchical clustering approach.

Each sub-cluster is then characterised by the median values of the parameters computed from all the galaxies belonging to this sub-cluster. The pre-clustering is done independently on the three samples: full, cluster and field, with the same number of pre-clusters each time.
	The MP analysis is performed on the sub-clusters, and the classes defined from the resulting tree gather one or several sub-clusters.

We are now faced with the choice of the number of sub-clusters. For the MP analysis, one thousand is a maximum otherwise the computing time becomes prohibitive and the search for the most parsimonious tree  may be less efficient.
 In any case, we do not expect that much classes of galaxies, at least for this still exploratory work, a few tens being already quite satisfactory. We considered that 300 pre-clusters is a good compromise. We also performed the same analyses with 100 and 200 pre-clusters to check the stability of the results on the full sample (see Appendix~\ref{App:hclustcompar}). We cannot expect a perfect match since the minimised paths between different sets of pre-clusters cannot be identical. However the global evolutionary scenario is preserved, while some groups might be displaced and/or split. 

The hierarchical clustering was performed with the function \textit{hclust} of the package \textit{stats} in \textsf{R}, with an euclidean distance and the method "ward.D2" as recommended in \citet{Murtagh2014}.

\begin{figure*}[ht]
\centering
%\renewcommand{\thefigure}{\arabic{figure}a}
%\begin{figure}[t]
\centering
\includegraphics[width=0.48\linewidth]{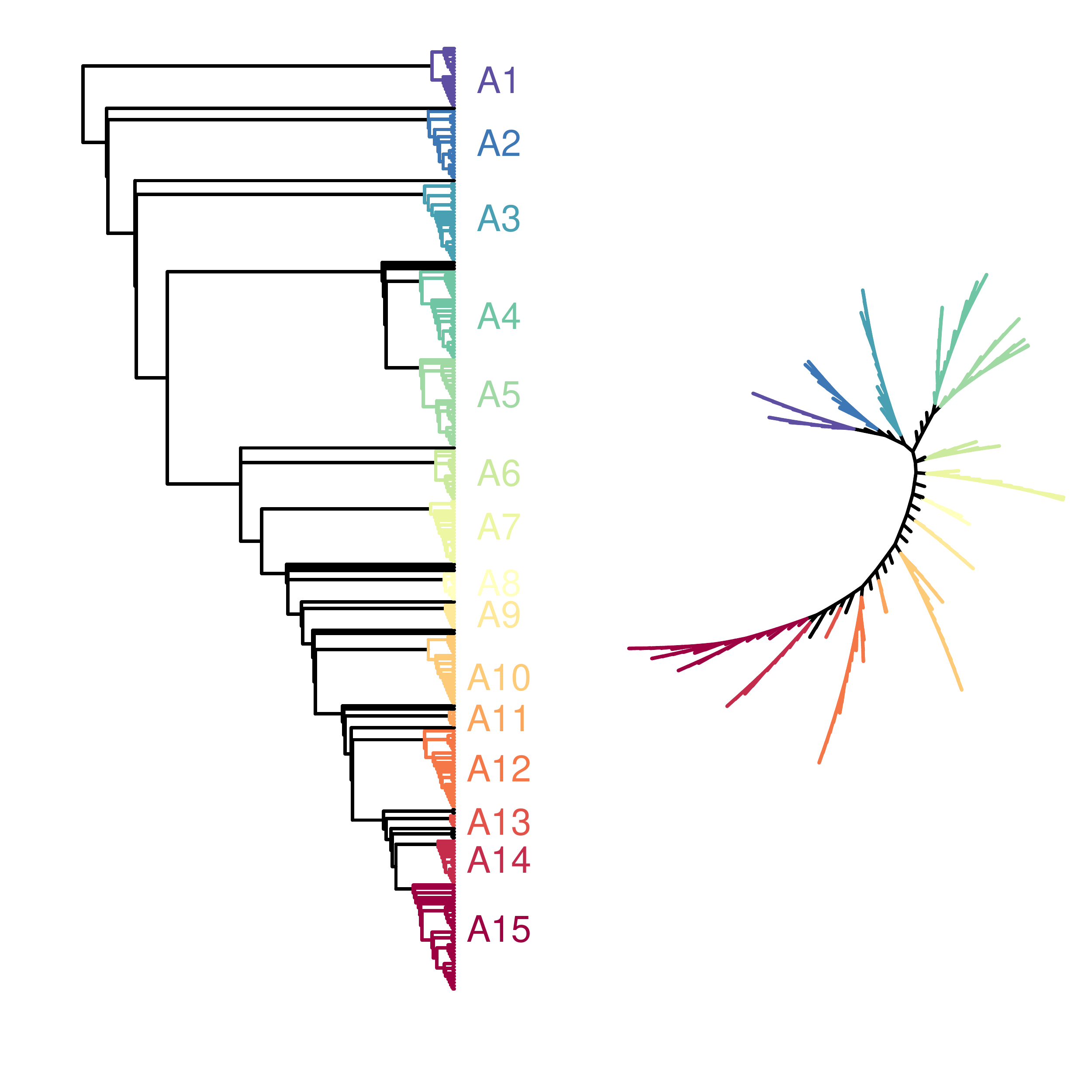}
%\caption{Tree for the full sample with group definition. Only groups with more than 50 galaxies are considered for this tree in this paper. The tree has been rooted with the group (at the top) having low values of \BmV, \nV\ and \lM.}
%\label{fig:TreeAll}
%\end{figure}
%
%\begin{figure}[t]
%\centering
%\renewcommand{\thefigure}{\arabic{figure}b}
%\addtocounter{figure}{-1}
\includegraphics[width=0.48\linewidth]{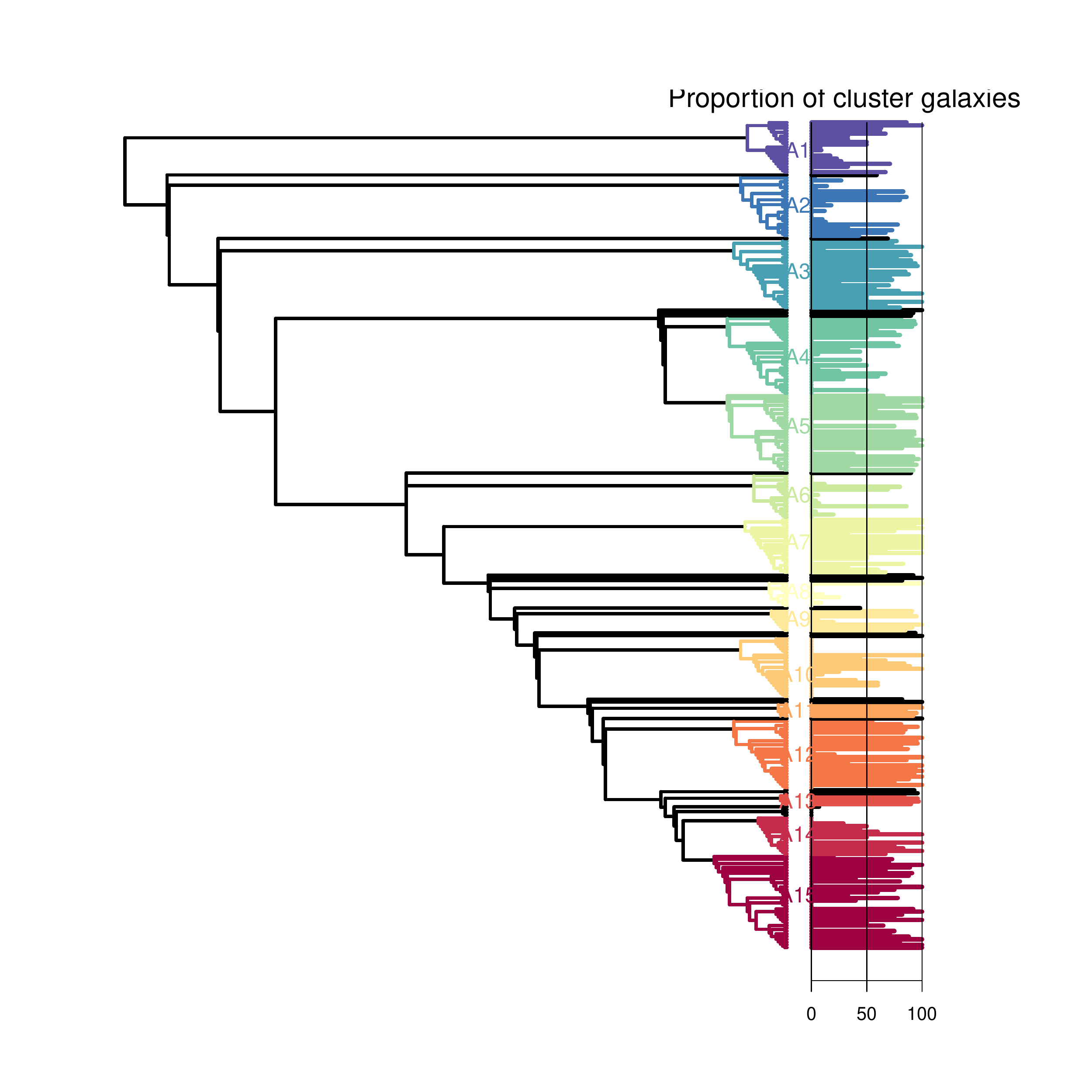}
\caption{\textit{(Left)} Tree for the full sample with group definition. Only groups with more than 50 galaxies are considered for this tree in this paper. The tree has been rooted with the group (at the top) having low values of \BmV, \nV\ and \lM. \textit{(Middle)} Same tree but unrooted. \textit{(Right)} Same tree with the proportion of cluster galaxies within each of the 300 pre-clusters indicated by the length of segments.}
\label{fig:TreeAll}
%\label{fig:AllPropMemb} 
%\end{figure}
\end{figure*}

\subsection{Defining groups from a tree}
\label{groupdef}

Unlike partitioning techniques, clustering methods based on graphs, such as hierarchical clustering and phylogenetic tools, require some choices in the definition of classes or groups since there are many possible subgraphs. Nevertheless this reflects the fact that there are several possible classifications for the same data set depending on the purpose. This is well known in biology with the hierarchical organisation in species, genus, family, etc. The choices depend on the granularity needed for the study and is necessarily a compromise between the two extremes: there can be as many classes as galaxies implying that each galaxy is considered to be unique, or there can be only one class with all galaxies. 

In our case, we have first reduced the granularity by performing a pre-clustering through the hierarchical clustering technique (Sect.~\ref{MPlarge}) to ease the visualisation of the relationships between different kinds of galaxies. The number of pre-clusters is constrained by the computation time of the subsequent MP analysis and a good representation of the diversity of the sample.

The definition of the groups from the MP tree is firstly guided by the structure of the tree: a group can be defined when a bunch of branches starting from a node is clearly apart from other bunches of branches (see Fig.~\ref{fig:TreeAll} for an illustration). Such groups may potentially define a kind of family or species sharing some properties resulting from the evolution. The goal here is not necessarily to establish a firm general classification but to find the relevant level of granularity that helps in understanding the diversification of galaxies. 

We must point out that classes of galaxies could be defined by a partitioning or a hierarchical clustering technique (that is our pre-clusters). However, the MP analysis provides a complementary information with the relationships between these classes. As a matter of fact, this is the first time that a hierarchical (clustering and phylogenetic) classification is attempted on such a large sample.

\subsection{Selection of the parameters}
\label{paramselect}

Ideally we would like to keep all available variables but there is no guarantee that a phylogenetic signal is present or can be found since they must bear some reliable information on the evolutionary history of the galaxies. As a consequence, we should disregard variables with convergent and regressive behaviours, as well as redundancies, because they bring confusion in ancestorship and can destroy the tree structure. Since it is not always easy to detect these "flaws" a priori, we are bound to try different subsets and keep the results that show the most parsimonious trees that are as much resolved as possible. We always try to choose the subset with the largest number of variables to be more informative, more objective, and in better accordance with the complexity of galaxies and their evolution.

For the analysis itself, we did not use the star formation rates $sfr*$ at different epochs, which are derived, model-dependent variables, and the total galaxy luminosity $M_V$ which is related with %$\log(\re)$  
\logre\ and \muem\ by a linear formula. We also do not consider the De Vaucouleurs'  morphological type index $T$ because it is a categorical value, and the distance to the cluster centre $\Delta_{CC}$ for the galaxies belonging to a cluster because it is not an intrinsic descriptor of a galaxy. A first analysis included the ellipticity $\epsilon$ which was found to be somewhat erratic on the tree as compared to the other parameters. We decided to disregard it since it is very uncertain because of the orientation effects. 

We end up with the seven parameters described on the top part of Table~\ref{tab:param}: \BmV, \logre,  \nV, \muem, $H_{\beta}$, \DQ, \lM. The disregarded variables are also reported on the bottom part of Table~\ref{tab:param} and are used for the interpretation.

It should be clear that the selection of the parameters is not as arbitrary as it might look. Firstly there are solid arguments to disregard some the observables as explained above. Secondly, the robustness of the tree is a strong statistical argument. Finally, any multivariate statistical analysis is an objective and powerful way to observe the data space, but only the physical interpretation of the results is able to tell its real interest and relevance. Any obvious contradiction should lead to reconsider the selection of the parameters as we did for the ellipticity.

The seven parameters that have been selected mainly from a statistical point of view constitute in fact a reasonable description of the physics related to the stellar content of galaxies: colour (metallicity and age of the stellar population), structure (radius and light distribution), star forming activity, relative importance of old and young stars, and total stellar mass. 

We here stress that the selection of the parameters is based on purely statistical arguments, that is the robustness of the tree we can obtain. There is no physical assumption here except for the fact that the "simplest" tree chosen in the Maximum Parsimony analysis corresponds to the smoothest evolution of the seven parameters. This means that, as said at the beginning of this section, we consider that these parameters are tracers of the evolution of galaxies. For instance, the intensity of $H_{\beta}$ in emission is a short-lived feature depending on stellar bursts, while its equivalent width (used in this paper) is a more stable feature of the dynamics of the galaxy. 

The tree obtained is an hypothesis for an evolutionary scheme and its interpretation provides physical justification or rebuttal \citep[see an example in][ for a physical inconsistency of the resulting tree due to burst indicators]{jc1}.

\section{Results}
\label{Results}

We have performed three identical analyses (same seven parameters (Sect.~\ref{paramselect}), pre-clustering with 300 pre-clusters (Sect.~\ref{MPlarge}) followed by the MP analysis (Sect.~\ref{CladMeth})) separately on the full sample (4100 galaxies, Sect.~\ref{Result4czb}), the cluster sample (2624, Sect.~\ref{Result4czbM}) and the field sample (1476, Sect.~\ref{Result4czbN}).

 Our goal with the full sample was to look for possible separate evolutionary paths between cluster objects and field ones. In other words, are they evolutionary disconnected, do they share some common ancestor from which their evolutions diverged, or did they evolve similarly but at a different pace due to their environment? Since we find that the latter is true, we decided to conduct analyses of the two sub-samples to look for more subtle differences in their evolution as separate populations and shed some light on the precise role of the environment on the statistical evolution of galaxies.

\subsection{Full sample}
\label{Result4czb}

\begin{table*}[ht]
	\caption{Distribution of galaxies in the different groups from the analysis with the full sample, with numbers and proportions of cluster and field galaxies for each group. The table also lists the numbers of galaxies in the 21 small groups not considered in this paper.}
	\label{tab:ADist} 
		\resizebox{\textwidth}{!}{%
		\begin{tabular}{lrrrrrrrrrrrrrrrr}%\toprule
			\hline
			\noalign{\smallskip}
		   & A1 & A2 & A3 & A4 & A5 & A6 & A7 & A8 & A9 & A10 & A11  & A12 & A13 & A14 & A15 & \textit{Total} \\
			\hline
			\noalign{\smallskip}
			 All galaxies & 208 & 434 & 344 & 297 & 477 & 231 & 162 & 128 & 182 & 172 & 139 & 362 &  94 &  64 & 326 & 3620 \\ 
			 Cluster galaxies   & 77 & 116 & 265 & 144 & 405 &  55 & 143 &  41 & 146 &  70 & 134 & 299 &  86 &  33 & 239 & 2253 \\  
			 Field galaxies   & 131 & 318 &  79 & 153 &  72 & 176 &  19 &  87 &  36 & 102 &   5 &  63 &   8 &  31 &  87 & 1367 \\ 
			\hline
			\noalign{\smallskip}
			 \% cluster galaxies& 37 & 27 & 77 & 48 & 85 & 24 & 88 & 32 & 80 & 41 & 96 & 83 & 91 & 52 & 73 & 62 \\
			\noalign{\smallskip}	
			\hline
			\noalign{\smallskip}
			\multicolumn{17}{l}{\textit{Composition of the remaining 21 small groups:}} \\
			\multicolumn{17}{l}{	6 13 14 15 15 16 17 18 19 20 22 22 23 25 25 27 32 34 37 40 40}\\
			\noalign{\smallskip}	
			\hline
\end{tabular}
}
\end{table*}

\begin{figure*}[t]
\centering
\includegraphics[width=0.8\linewidth]{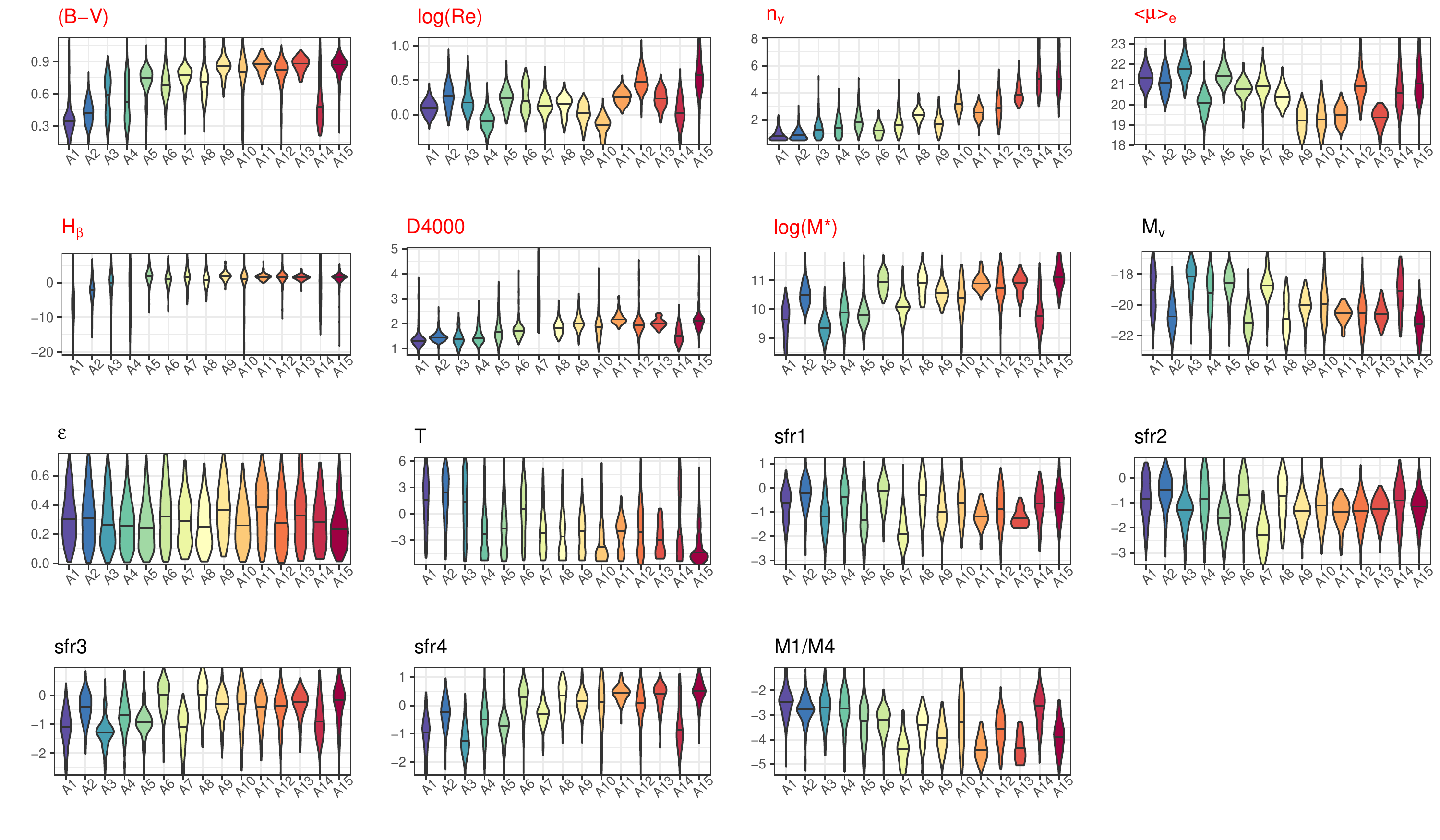}
\caption{Violinplots for fifteen variables and the groups A[1-15] of the full sample as defined on Fig.~\ref{fig:TreeAll}. The seven first parameters only (in red) were used for the analysis. The horizontal bar is the median value, and each "violin" symmetrically depicts the distribution of the variable.}
\label{fig:BoxplotAll} 
\end{figure*}

The tree with the 300 hierarchical clusters is nearly perfectly resolved and is well structured (Fig.~\ref{fig:TreeAll}). On this figure, the tree is shown in two graphical representations: the one to the left has been rooted with the group (at the top) having low values of \BmV, \nV\ and \lM, and the one to the right is unrooted and is the real outcome of the MP analysis. Rooting a tree is merely a help for an evolutionary interpretation. 

It is obvious that the evolution of the properties along the tree, hence the evolutionary scenario that is depicted by the tree, depends on the rooting. The analysis is independent on this choice. Rooting a tree consists in choosing the group that is considered as the most ancestral and draw the tree starting from this group. We have chosen blue colour, low Sersic index and low mass since their global evolution is known. We show in Figs~\ref{fig:treeotherroot} and \ref{fig:violinotherroot} a different rooting, on group A7, which could be justified as having a moderate mass and formed stars very early very low $sfr1$, $sfr2$ and $sfr3$. Clearly this new rooting does not yield a satisfactory evolutionary scenario.

 The groups are defined from the structure of the tree guided by bunches of branches which are clearly apart from other bunches of branches (Sect.~\ref{groupdef}). As we can see, more (smaller) groups (each corresponding to a branch in a bunch of brunches) could have been defined but this would complicate the discussion in the paper. We recall that each tip (leave) of the tree is a pre-cluster (see Sect.~\ref{MPlarge}) of one or several galaxies. For reason of simplicity in the discussion and interpretation in this paper, we only consider fifteen groups which have more than 50 galaxies (A1-A15, Fig.~\ref{fig:TreeAll}), thus leaving only 480 (11.7\%) galaxies aside.
 
The tree in Fig.~\ref{fig:TreeAll} is quite regular, with A4 and A5 together making a significant substructure. The distribution of the galaxies in the groups (Table~\ref{tab:ADist}) is roughly balanced, with the two smallest groups gathering 64 and 94 galaxies (A14 and A13 respectively) and the two largest ones gathering 477 and 434 objects (A5 and A2 resp.). We here note that parts of groups A4 and A11 are displaced towards the top on the tree with 200 pre-clusters (Appendix~\ref{App:hclustcompar}) suggesting some uncertainties on their precise placement.

The distribution of cluster and field galaxies is quite different in the groups (Table~\ref{tab:ADist}): eight groups (A3, A5, A7, A9, A11, A12, A13, A15) are composed mainly or nearly exclusively of cluster galaxies (from 73 to 96\%  to be compared to 62\% for the full sample), five groups (A1,A2, A6, A8 and A10) are dominated by field galaxies (since they have only 24 to 41\% of cluster galaxies), and two (A4 and A14) have an equal share (48 and 52\% of cluster galaxies). There seems to be an increase of the percentage of cluster galaxies from A1 to A15 with large fluctuations, which can be visualised from the proportion of cluster galaxies within each of the 300 pre-clusters (Fig.~\ref{fig:TreeAll}). These large fluctuations indicate that neither the hierarchical clustering nor the MP analysis are able to separate clearly the two galaxy populations despite some very few exceptions. 

The tree does not show separate evolutionary paths for cluster and field galaxies. However, assuming that the rooted tree reveals an evolutionary scenario, the increasing importance of cluster galaxies could be interpreted as them being somewhat more diversified on average than field galaxies. By diversification we mean that the objects become more different from a common ancestor supposedly situated at the top of the tree. \citet{Fraix2010} have shown that "diversification" is preferable to "evolution" since the latter is difficult to define and measure in a multi-parameter space for a whole population. In the present case, despite field galaxies are more massive on average (Fig.~\ref{fig:redshifts}), they are not the most "evolved" objects, probably because they are also still blue and concentrated. Member and field galaxies evolved but underwent different transforming process histories: since cluster galaxies appear more diversified, one obvious explanation is their environment that make them suffer from more frequent and/or violent events.

The properties of the groups are represented as violinplots in Fig.~\ref{fig:BoxplotAll}. This representation gives the median values and the distribution for each variable as a function of the group index in the order given by the tree on Fig.~\ref{fig:TreeAll}. This sequence provides the changes of the variables along the diversification process and should not be taken literally as the evolutionary steps followed by individual galaxies. 

The colour \BmV\ increases sharply up to group G5 and then becomes roughly constant except for a drop at Group A14. \logre\ wiggles with no clear evolutionary trend. The Sersic index \nV, also used to root the tree, increases regularly. \muem\ decreases (the effective surface brightness increases) globally despite some exceptions (A4 and A12 for instance). \Hbeta\ and \DQ\ have a very similar behaviour as \BmV, but the \DQ\ value of A7 is remarkably higher than for all the other groups. \lM\ seems to be lower from A1 to A5  and higher afterwards, but there are many exceptions, while \mV\ behaves more or less in an opposite manner.  $\epsilon$ does not show much significant difference between the groups since the intragroup variance is large. The groups A1, A2 and A3 are spiral dominated groups with the morphology variable T higher than for all the other ones.

The different star formation rates are shown in Fig.~\ref{fig:BoxplotAll} as well. $sfr1$ and $sfr2$ look quite diverse in groups from A1 to A8 and more similar in the subsequent groups. On the contrary, $sfr3$ and $sfr4$ seem globally lower in the first part (up to group A7) with few exceptions: A6 and A14. Finally, the stellar production ratio M1/M4 is lower from A7 to A15, except for A14 and possibly A10.

The properties as given by the violinplots explain the structure of the tree and allows a physical understanding of the groups. The gross picture of galaxy evolution is correctly recovered: galaxies become redder, with ageing stars, more massive, the most diversified galaxies formed stars earlier in the past, they tend to get early-type morphologies. We see an increase of the effective surface brightness, but no clear evolution of the effective radius. There are some peculiar groups that require further investigation (Sect.~\ref{Discussion}), such as  A7 with its high \DQ, or A14 with its astonishing low \BmV, low mass, low \DQ, and low star formation rates at earlier times ($sfr3$ and $sfr4$) suggesting they could be rejuvenated galaxies.

\subsection{Cluster galaxies}
\label{Result4czbM}

\begin{table*}[ht]
	\caption{Distribution of galaxies in the different groups from the analysis with the cluster sample.  The table also lists the numbers of galaxies in the 27 small groups not considered in this paper.}
	\label{tab:CDist} 
	\begin{tabular}{rrrrrrrrrrrrr}%\toprule
		\hline
		\noalign{\smallskip}
		C1 & C2 & C3 & C4 & C5 & C6 & C7 & C8 & C9 & C10 & C11 & C12 & \textit{Total}  \\
		\noalign{\smallskip}
		\hline
		\noalign{\smallskip}
		100 & 121 &  70 & 124 & 367 &  59 & 160 & 144 & 110 &  51 & 442 & 540 & 2288  \\ 
		\noalign{\smallskip}
		\hline
		\noalign{\smallskip}
		\multicolumn{13}{l}{\textit{Composition of the remaining 27 small groups:}} \\ 
		\multicolumn{13}{l}{1  1  3  4  7  7  8  8  9  9  9  9 10 10 11 11 12 14 15 16 16 18 19 21 23 24 41 }\\
		\noalign{\smallskip}	
		\hline
	\end{tabular}
\end{table*}

\begin{figure}[t]
\centering
\includegraphics[width=\linewidth]{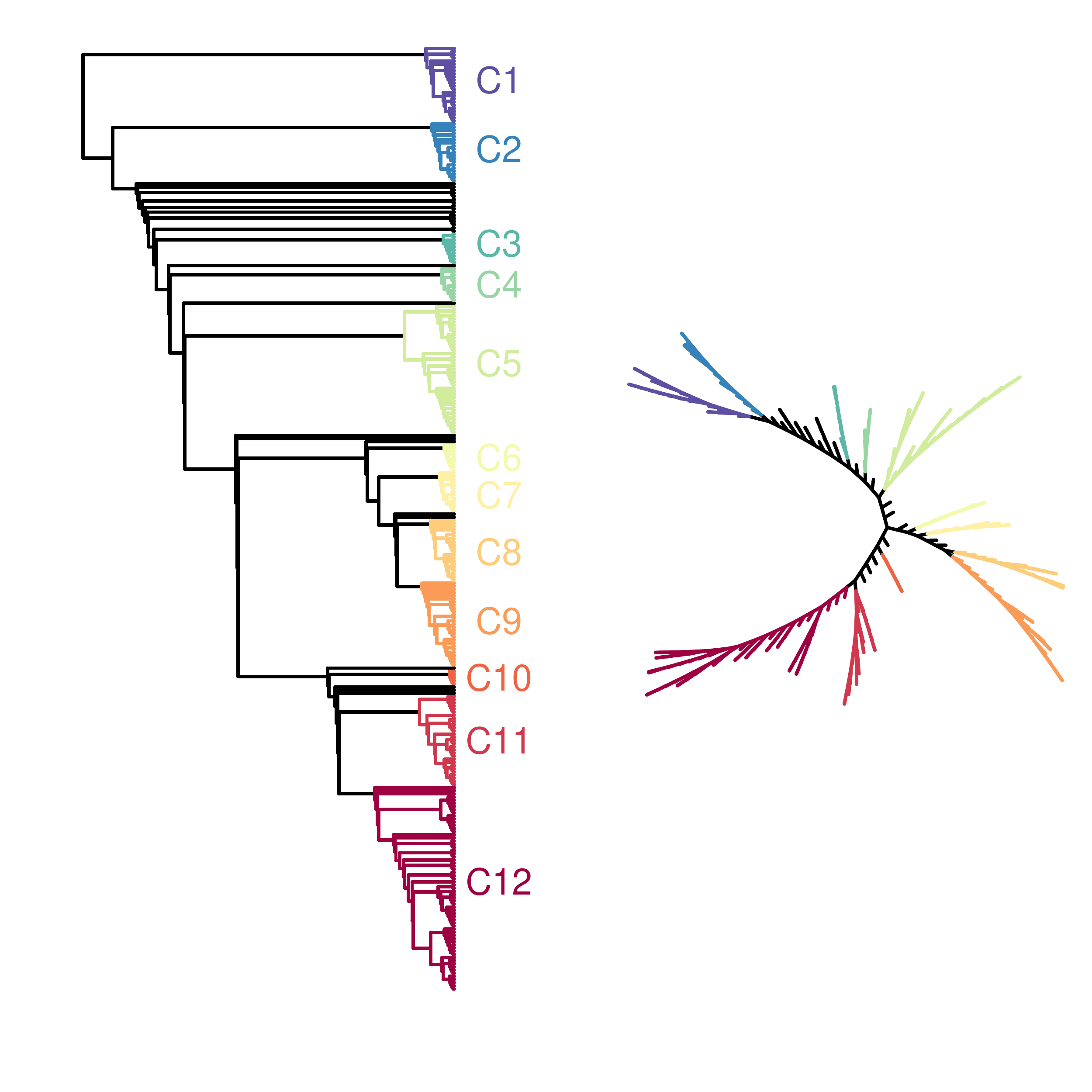}
\caption{Tree from the analysis with the cluster sample. Only groups with more than 50 galaxies are identified. The tree on the left has been rooted with the group (at the top) having low values of \BmV, \nV\ and \lM. To the right, same tree but unrooted.}
\label{fig:TreeMemb} 
\end{figure}

\begin{figure*}[ht]
\centering
\renewcommand{\thefigure}{\arabic{figure}a}
\includegraphics[width=0.7\linewidth]{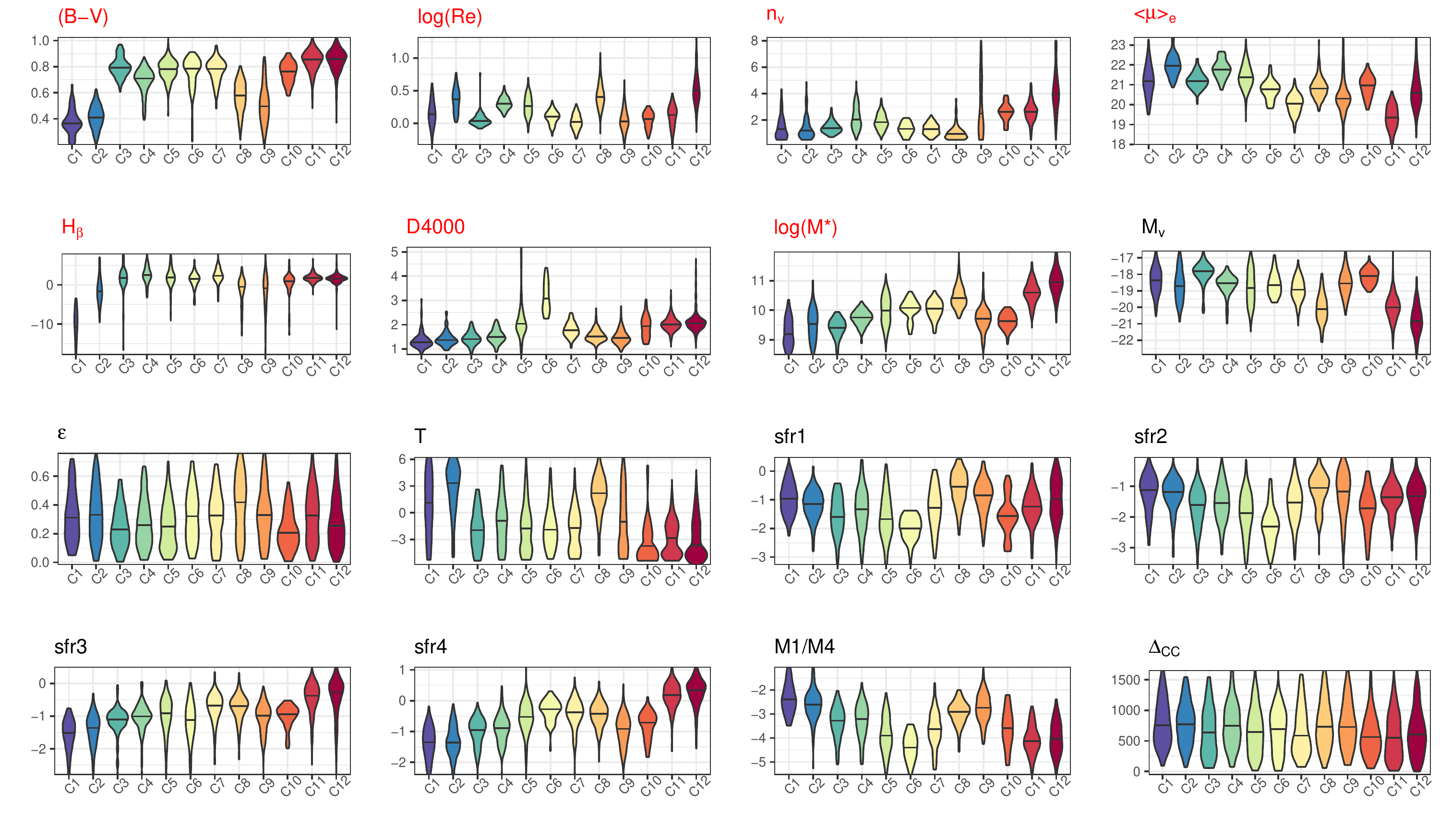}
\caption{Violinplots for sixteen variables and the groups C[1-12] for the cluster sample. \DCC\ is the projected distance of a galaxy to the centre of a cluster.%}
\label{fig:BoxplotMemb} }
%\begin{figure*}[t]
%\centering
\renewcommand{\thefigure}{\arabic{figure}b}
\addtocounter{figure}{-1}
\includegraphics[width=0.7\linewidth]{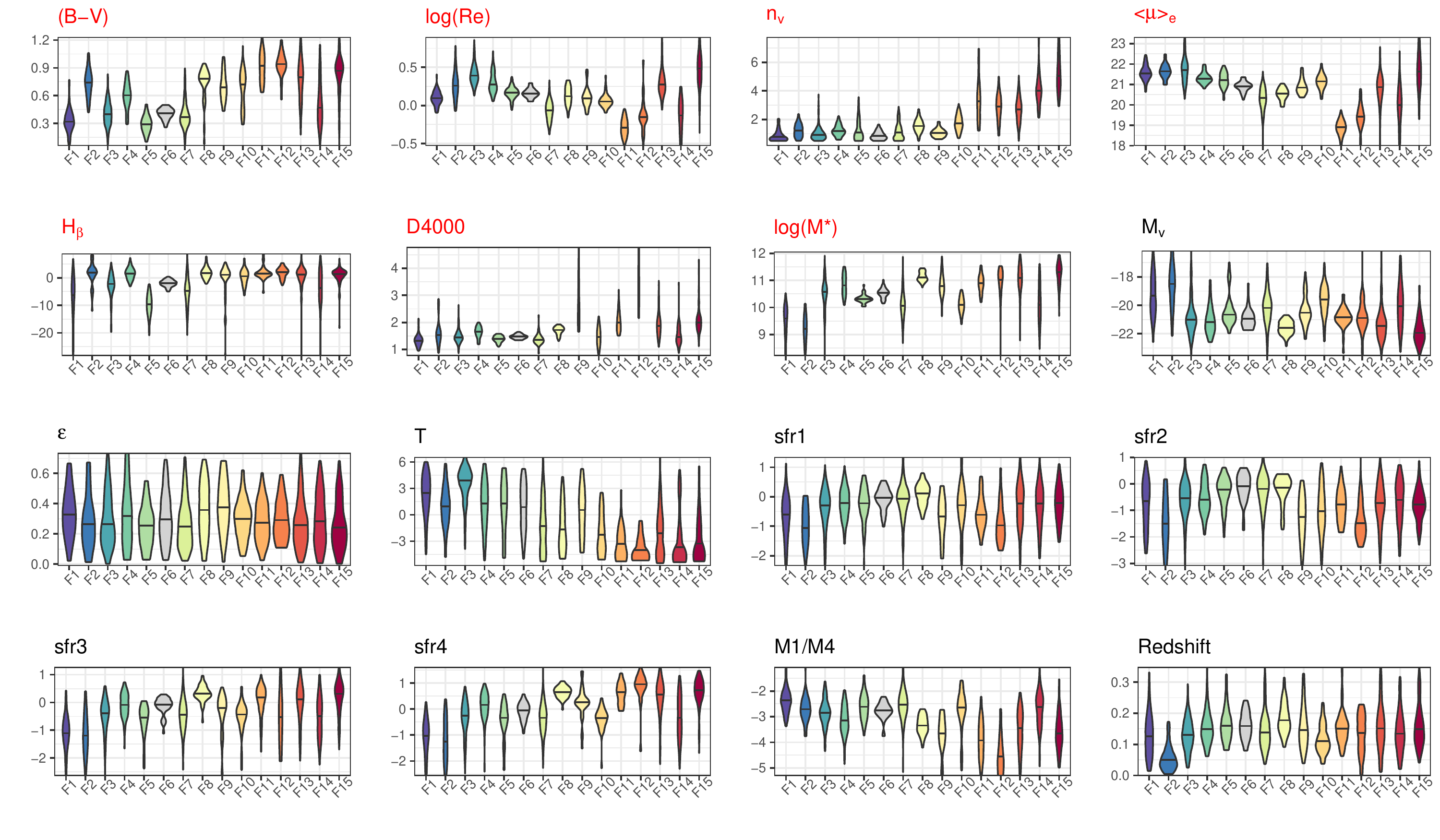}
\caption{Violinplots for sixteen variables and the groups F[1-15] for the field sample.  %}
\label{fig:BoxplotCont} }
%\end{figure*}
\end{figure*}
\renewcommand{\thefigure}{\arabic{figure}}

The tree resulting from the analyses of the 2624 cluster galaxies (Fig.~\ref{fig:TreeMemb}) and rooted as for the full sample (using low values for \BmV, \nV\ and \lM) is perfectly resolved and nicely structured.  We only consider the twelve groups which have more that 50 galaxies (C1-C12, Fig.~\ref{fig:TreeMemb}), leaving 336 (12.8\%) galaxies aside.

Three groups dominates (C5, C11 and C12, Table~\ref{tab:CDist}), and three are significantly smaller (C3, C6 and C10). On the tree, after C5, there is a split into two clear ensembles: C6 to C9 on one side, and C10 to C12 on the other. This is important to keep in mind when interpreting the evolution of the variables and not to be deceived by the labelling: after C5, the evolution of the variables splits into two branches, one towards C6, the other one towards C10.

The violin plots (Fig.~\ref{fig:BoxplotMemb}) show a similar evolution of all the variables along the tree as for the full sample tree, except that there is a clearer increase of mass \lM\ (and a decrease of \mV), as well of $sfr3$ and $sfr4$. There are also three clear regimes for the morphology T: spirals for C1 and C2, S0s for C3 to C7 and C9, and ellipticals for C10 to C12. Recall that the latter three form a specific ensemble on the tree. Departing from this scheme, C8 is composed by spirals.

It is tempting to associate C6 to A7 (being composed at 88\% of cluster galaxies) considering their peculiar high \DQ\ value. This is corroborated by low $sfr1$, $sfr2$ and M1/M4 in both cases. However C6 is made of 59 galaxies while A7 has 143 cluster galaxies.

The median distances from the centres of the clusters (Fig.~\ref{fig:BoxplotMemb}, last panel at the bottom right) show that the groups C3, C7, and the ensemble C10, C11 and C12, are the most central ones even though the differences are not highly significant. 

Despite the global similarities in the evolution of the variables which reflects the approximate picture of galaxy evolution, the tree for the cluster galaxies is characterised by two distinct evolutionary paths, one from C6 to C9, and one from C10 to C12 (Fig.~\ref{fig:TreeMemb}). The evolution of some variables are clearly distinct along these two branches. \BmV\ decreases in the first branch from C6 to C9 and increases in the second one from C10 to C12. \nV\ is very low in the first branch, but increases strongly in the second one. \DQ\ decreases in the first branch and increases in the second one. In the second branch, \lM\ increases strongly, $\epsilon$ is lower and $T$ is much lower. Interestingly, $sfr1$ and $sfr2$ increase in both branches, while $sfr3$ and $sfr4$ decrease in the first branch and increase in the second one. These different star formation history evolutions are clearly visible in the stellar production ratio M1/M4 that increases in the first branch and decrease in the second one. Finally, the second branch contain more central galaxies (lower $\Delta_{CC}$). 
All this might suggest that the first ensemble (C6 to C9) are galaxies that increase their stellar mass by forming stars, while the second ensemble (C10 to C12) are galaxies that increase their stellar mass by accreting old stars probably formed in other galaxies, that is by dry mergers.

\subsection{Field sample}
\label{Result4czbN}

\begin{table*}[ht]
	\caption{Distribution of galaxies in the different groups from the analysis with the field sample.  The table also lists the numbers of galaxies in the 51 small groups not considered in this paper.}
	\label{tab:FDist} 
	\begin{tabular}{rrrrrrrrrrrrrrrr}%\toprule 
		\hline
		\noalign{\smallskip}
		F1 & F2 & F3 & F4 & F5 & F6 & F7 & F8 & F9 & F10 & F11  & F12 & F13 & F14 & F15 & \textit{Total}  \\
		\noalign{\smallskip}\hline\noalign{\smallskip}
		75 &  34 & 176 &  53 &  35 & 31 & 219 &  40 &  33 &  40 &  36 &  24 & 182 &  79 & 101 & 1158 \\ 
		\noalign{\smallskip}
		\hline
		\noalign{\smallskip}
		\multicolumn{16}{l}{\textit{Composition of the remaining 50 small groups:}  }\\
		\multicolumn{16}{l}{1  1  1  2  3  3  3  3  3  3  4  4  4  5  5  5 5  5  5  5  5  5  6  6  6  6 7 7  7 7  7  7  7  8  8  8  8  8  9  9  9  9 10 10 10 10 11 11 11 16 }\\  
		\noalign{\smallskip}	
		\hline
	\end{tabular}
\end{table*}

\begin{figure}[t]
\centering
\includegraphics[width=\linewidth]{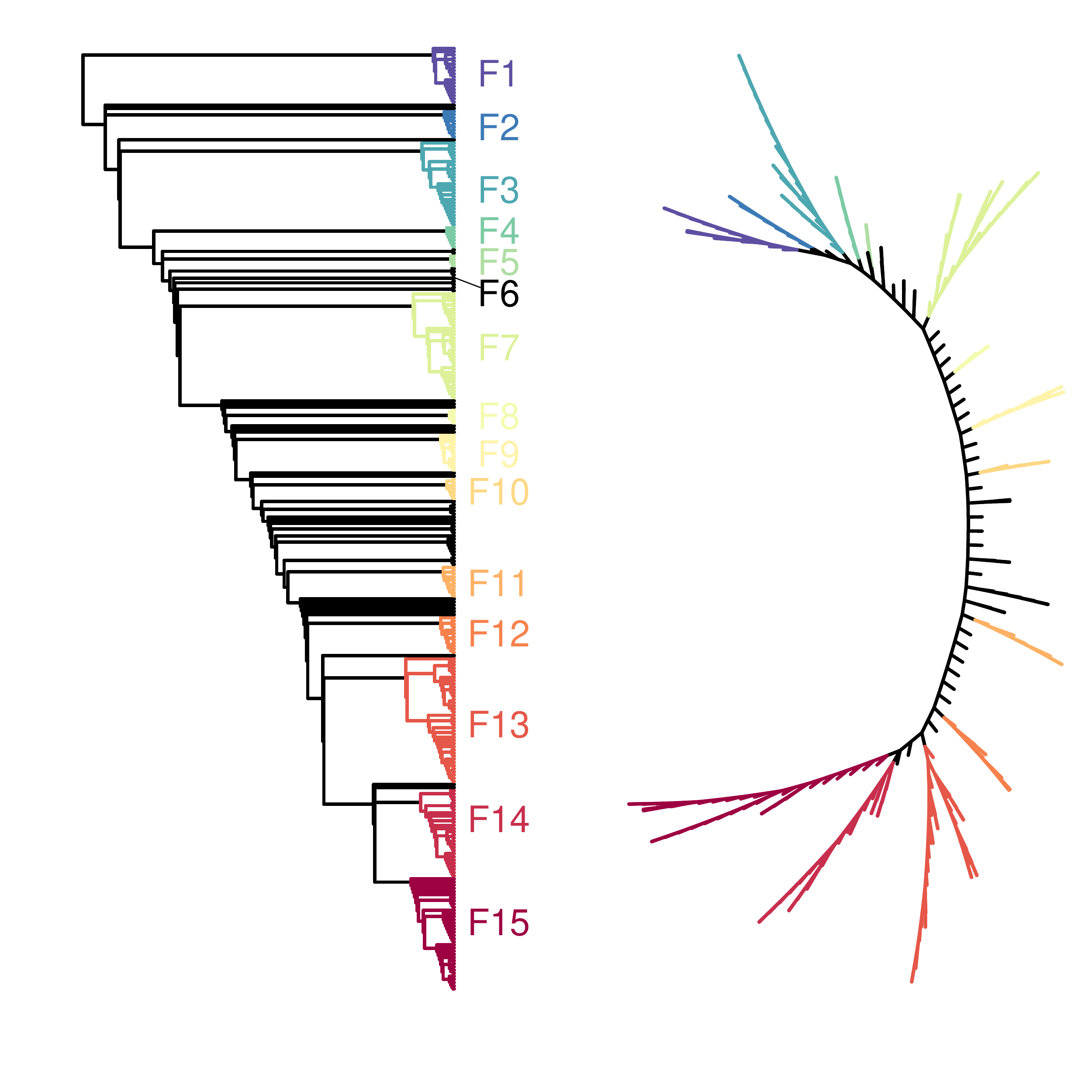}
\caption{Tree for the field sample. Only groups with more that 23 galaxies are identified.  The tree on the left has been rooted with the group (at the top) having low values of \BmV, \nV\ and \lM. Group F6 is a pre-cluster and occupies a single branch of the tree indicated by the tick mark. To the right, same tree but unrooted.}
\label{fig:TreeCont} 
\end{figure}

The tree (Fig.~\ref{fig:TreeCont}) obtained with the 1476 galaxies of the field sample is fully resolved and very regular. The rooting is the same as for the other trees in this paper, with low \BmV, \nV\ and \lM. To keep a similar number of groups as for the two other samples, we identify fifteen groups with more than 23 galaxies, gathering 1158 thus leaving 318 (22.5\%) galaxies aside (Fig.~\ref{fig:TreeCont}). We note that the group F6 with 31 galaxies is one of the 300 pre-clusters and thus occupy a single branch on the tree.
Four groups dominate (F3, F7, F13, F15, Table~\ref{tab:FDist}) with 101 to 219 galaxies, and F12 is the smallest with only 24 galaxies.

The violin plots (Fig.~\ref{fig:BoxplotCont}) show that \BmV\ is somewhat more chaotic than for the full and the cluster samples, but it is higher on average on the second part of the tree, hence suggesting an increase. \nV\ seems to increase more regularly than for the cluster sample, while the increase of \lM\ is less obvious except that the two first groups have a much lower mass. For the other variables, the notable differences with Fig.~\ref{fig:BoxplotAll} and \ref{fig:BoxplotMemb} are \muem\ that drops at F11 and then increases, and H$_{\beta}$ that is more irregular.

There are two groups with a high \DQ\ (F9 and F12) but they do not share other similar properties with A7 and C6, except may be for F12 that has a low stellar production ratio M1/M4. These two groups F9 and F12 are small but have more galaxies (respectively 33 and 24) than in A7 (19 field galaxies).

The global picture of galaxy evolution is recovered as in the two previous cases, but it seems that the evolution of the variables looks slightly less regular than for the cluster galaxies or for the full sample. This implies a difficulty to find a smooth evolutionary path between the pre-clusters very probably because the distribution of field galaxies in the parameter space is too sparse. This can be explained in two ways: i) the field sample is too small (its size is nearly half that of the cluster sample), so that the sampling performed by the hierarchical clustering in the seven parameter space is too crude so that any evolutionary path would appear less regular. However, the size of the field sample is consequent and of the same order of magnitudes as the cluster one. Or ii) the population of field galaxies is intrinsically less homogeneous so that they occupy a wider and/or sparser region of the parameter space. The first explanation is an observational bias that could possibly be corrected by more data. The second explanation has a physical consequence: there are missing links in our sample, that is galaxies that would help reconstruct the evolutionary scenario. Unfortunately, the missing links might not be observable at all, which would imply that the field galaxies evolved largely independently and lost most of the traces of their origin and their relationships to other galaxies. Nevertheless there is a significant difference between the two populations: the field sample has on average bluer, lower Sersic index and more massive galaxies (Fig.~\ref{fig:redshifts}) while we assumed that ancestral galaxies were similar but less massive. This suggest that field galaxies evolved more in mass than the cluster galaxies.
We conclude that our field sample is more heterogeneous than the cluster one, favouring an intrinsic physical origin without entirely ruling out a possible observational bias.

%\textit{to be made: clarification with statement on p6 "more diversified than field galaxies".} : made later in the discussion

\section{Discussion}
\label{Discussion}

We have proposed a phylogenetic analysis of a sample of galaxies, and obtained positive results from a methodological point of view. We must now discuss on the relevance of this approach for the study of the evolution of galaxies by first looking at the tree structures and the quality of the selected parameters for a phylogenetic reconstruction. 

We stress that the discussion below is based on the rooting of the trees according to increasing \BmV, \nV\ and \lM. This choice is based on the a priori that "primordial" galaxies could be blue (because of young stars), with a small \nV\ and not very massive.  This certainly can be disputed (see Appendix~\ref{App:hclustcompar}) especially because our sample of modest redshift very probably does not contain any galaxies resembling primordial ones, but only objects with already a long and complex diversification history. 
However, we have used the luminosity weighted age computed by \cite{Fritz2011} and notice that the low mass or low \nV\ correspond to a low such age. These types of galaxies are present with higher prevalence among the C1 and F1.
Hydrodynamical simulations seem to confirm that 
galaxies in the hierarchical scheme start with these properties \citep{Vogel2014,Genel2014,Snyder2015}.

\subsection{Influence of the parameters on the classification}
\label{paraminfluence}

In a multivariate analysis, it is always difficult to understand the origin of the clustering and the influence of the parameters. It is tempting to attribute the results to only a very few, if not one, of the parameters. But we can now check statistically whether some of them are dominant, correlated or if there exists some hidden variable explaining the clustering properties and the tree structures. 

A first approach is to look for correlations between the seven parameters used in both the hierarchical pre-clustering and the MP analyses. The Pearson correlation coefficient is always lower than 0.5 for the full sample, 0.57 (\lM\ vs \logre) for the cluster sample and 0.67 (\muem\ vs \logre) for the field sample. We conclude that there is no significant correlation that could drive our classification results, and as a consequence no hidden variable that would significantly influence several of the parameters.

This can be further checked with a Principal Component Analysis (PCA). As shown in Fig.~\ref{fig:PCA} (top left diagram), there is no really dominant eigenvector and the first four PCs explain only 70\% of the total variance, showing that the seven parameters used are non-redundant and required to cover the diversity of the galaxies of our samples. This is true for our three samples (full, cluster and field). The three other diagrams in Fig.~\ref{fig:PCA} show the projection of the data on the scatter plot drawn by the first two PCs, together with the importance (loadings) of each parameter in these two components. All parameters are approximately of equal amplitude. The main difference between the three samples is that \logre\ is close to \muem\ for the field sample, while it is in-between the latter and the five remaining parameters. The conclusion of the PCA analysis is that there is no parameter nor hidden variable that dominates the variance of the samples. In other words, the distribution of our samples in the 7D-parameter space is not very much distorted.

However, the variance, as analysed by PCA, is not necessarily related to the clustering properties of the data set \citep[e.g.][]{Chang1983}. Once the classification is established, the correlation of the parameters with respect to the class labels can be computed. This is shown in Fig.~\ref{fig:corclas} for the three samples with the pre-clusters and the MP groups. For the former classification (dashed lines), the three samples are similar with \BmV, \nV\ and \muem\ slightly dominating (with correlation coefficients between about 0.4 and 0.6) while the mass \lM\ is more important for the cluster sample (coefficient of about 0.7). Still, the correlations are not very high and are comparable for all parameters with some noticeable differences between the samples for \logre\ and \lM. Therefore, the hierarchical pre-clustering is driven more or less equally by the seven parameters.

The MP groups show more significant differences depending on the sample (solid lines in Fig.~\ref{fig:corclas}). The full and field samples are very similar except for \logre\ which has a correlation coefficient near zero for the field sample. Apart from \BmV, the cluster sample shows always significantly different correlation coefficients, being again dominated by \lM. Globally the coefficients remain mostly below 0.65 for the three samples, except for \nV\ in the field and full samples which reaches 0.8.

Clearly, the effective radius  does not play a role in the classification for both the field and cluster samples, but the Sersic Index is of great influence for the field sample. Since this difference is not seen with the pre-clusters, we must conclude that this is the evolutionary nature of the MP analysis that makes it visible.  Colour and mass are also important, the latter having a higher correlation coefficient for the cluster sample. 

Could the influence of \nV\ in the field sample only be due to a selection bias from the redshift? We do not think so because i) the correlation coefficient between the groups and the redshift is very low (0.12), ii) the correlation between redshift and \nV\ is low also (0.10), and iii) the changes of the redshift with the groups (Figs~\ref{fig:BoxplotMemb} and \ref{fig:AevolN}) do not show significant variation  along the tree except for the low redshift of group F2. This group is relatively redder for its position on the tree, but is remarkable for its mass which is the lowest of all the groups with values of the same order as group C1. In addition, since the groups have been identify in the seven-dimension space of the parameters, it is difficult to see how a Malmquist-type bias could affect only one parameter. 

There are some specific quantities that estimate which parameters explain the cladistic tree. One of these indices \citep[Rescaled Consistency Index,][]{Farris1989} synthetises the smoothness of the evolutionary behaviour of the parameters on the tree. It is a different indicator as compared to the ones described above since it is computed on the median values of the parameters within each pre-cluster and, more importantly, it is related to the evolution of the parameters along the tree rather than to the classification established from it. This index is globally lower for the field sample, confirming the apparent higher wiggling of the parameters for this sample on the violin plots (Sect.~\ref{Result4czbN}).
The "best" parameters are \BmV, \DQ, \logre\ and \nV, depending on the sample, \lM\ being always the worst, but the differences are not very important, showing that the entire tree structures are supported somewhat equally by the seven parameters. Interestingly, for the cluster tree, \logre\ is one of the two most influential parameters for the structure of the tree, but does not play a role in the classification based on the tree (see above).

The conclusion of this section is that the seven parameters used for the pre-clustering and MP analyses are not significantly correlated between each other but probably not independent. The classification is not influenced by some particular properties, but there is a clear difference between the cluster and field samples about the main drivers of the MP classification: even if colour and mass slightly dominate in both samples, the radius \logre\ and the Sersic index \nV\ do not play the same role in the origin of the groups for the two samples. However, these two structural descriptors and the stellar content seem to be the main drivers of the tree structures, hence the evolutionary scenario, the mass being not important. The existence of this difference is a clear influence of the environment which in return appears to have induced a loss of the evolutionary history of the structures of galaxies.

\subsection{Evolution along the trees}
\label{evolalongtree}

%\begin{figure}[t]
%\centering
%\includegraphics[width=\linewidth]{WINGS4czbK_evol.pdf}
%\caption{Evolution of the medians of the groups for the cluster (solid line) and field (dotted line) samples. The abcissa is the group index, going from 1 to 12 (cluster sample, bottom axis) or from 1 to 15 (field sample, top axis). Shaded regions show 1$\sigma$ dispersion around median.}
%\label{fig:Aevol} 
%\end{figure}
\begin{figure}[t]
	\centering
	\includegraphics[width=\linewidth]{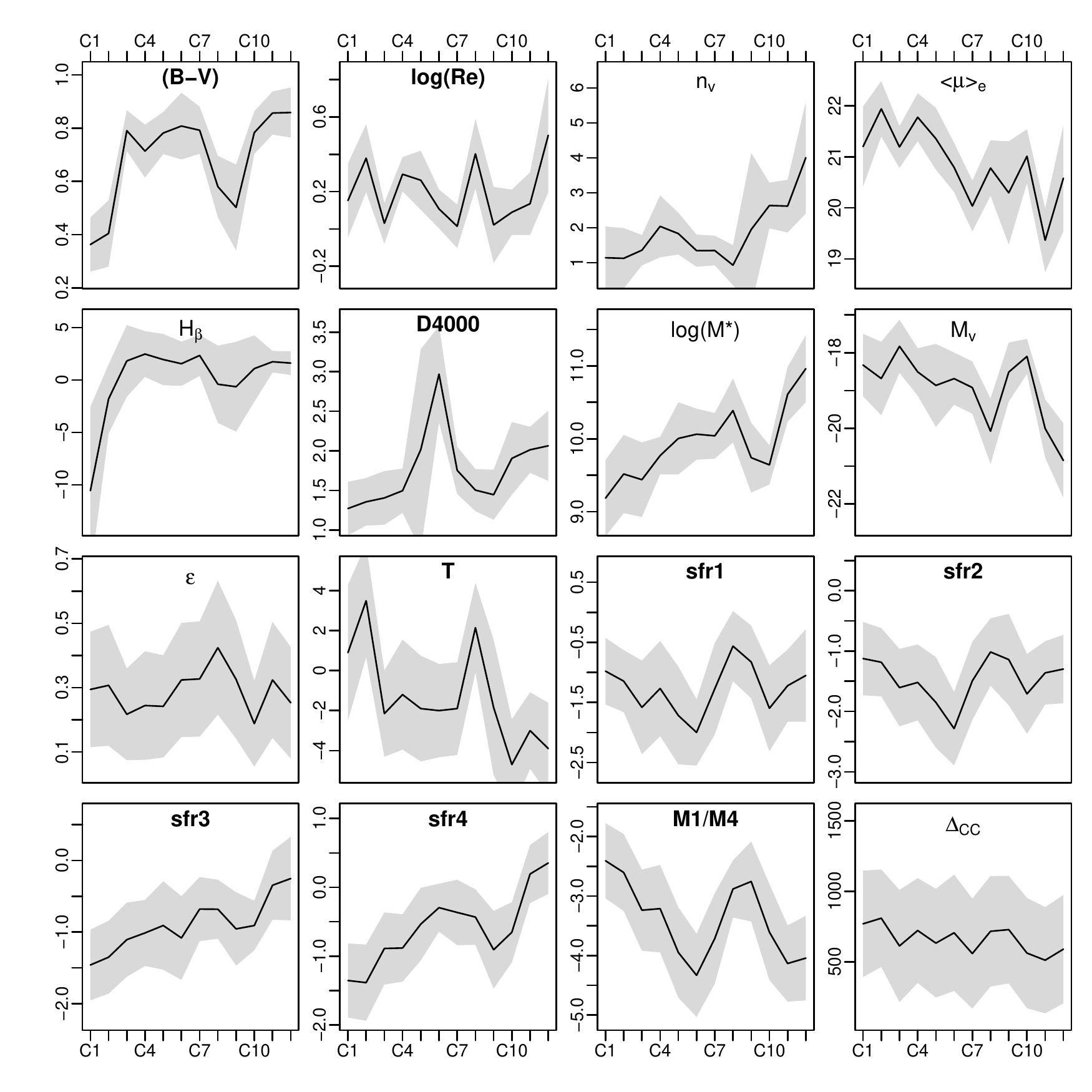}
	\caption{Changes of the medians of the groups along the tree for the cluster sample. Shaded regions show 1$\sigma$ dispersion around median.}
	\label{fig:AevolM} 
\end{figure}
\begin{figure}[t]
	\centering
	\includegraphics[width=\linewidth]{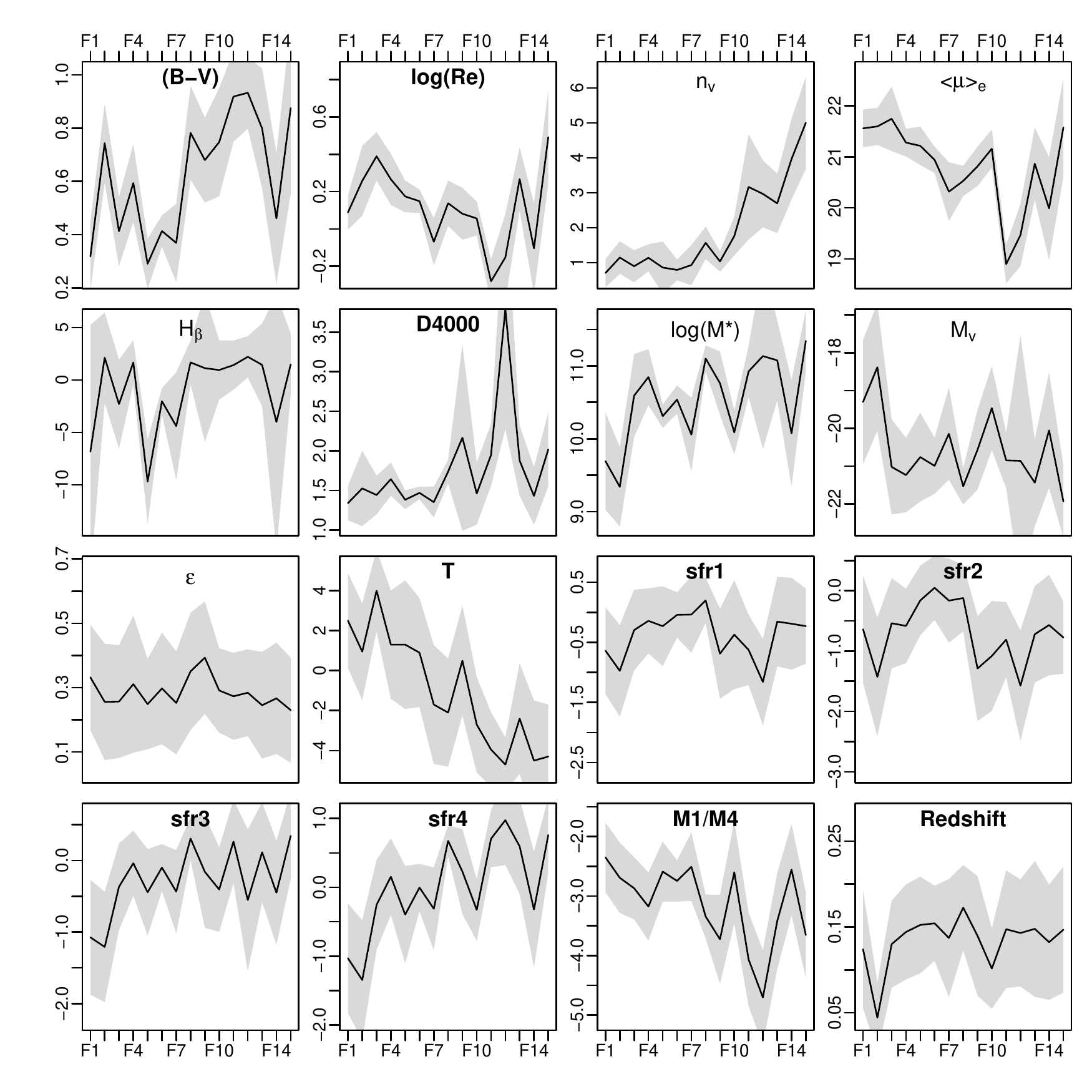}
	\caption{Idem of Fig.~\ref{fig:AevolM}\ for the field sample.}
	\label{fig:AevolN} 
\end{figure}

The evolution of the variable medians for each group along the tree for the cluster and field samples is shown in Figs~\ref{fig:AevolM} and \ref{fig:AevolN}. The evolution scale on the abscissa is simply the index of the group and thus is only indicative of the relative evolutionary stages for each tree, without any match between the two sets of groups. 

As already concluded from the violinplots, the global evolution of the variables are similar in both samples. In particular the starting and ending points of the evolution curves are relatively close, meaning that the less and the most diversified groups are similar in both samples. There are however differences in the details, the evolution for the field sample seems more wiggling (that is less monotonous), as in H$_\beta$, \lM, $sfr3$ and $sfr4$. In addition, the bumps in the curves appear sometimes opposite, as in \BmV, \mV, $sfr1$ and $sfr2$.

The two evolutions can be then interpreted as covering similar ranges of evolutionary stages, but with possibly less homogeneity (or less continuity) in the field sample (Sect.~\ref{Result4czbN}). In other words, the fact that galaxies belong to a cluster tends to smooth out somewhat their current diversity (more continuous diversity within clusters than globally) while at the same time creates an enhanced diversification from earlier more primitive galaxies (Sect.~\ref{Result4czb}). This suggests that more frequent interactions tend to accelerate galaxy "evolution" and generate more similar objects. 

This homogeneity difference is supported by the correlations between the parameters and the groups (Sect.~\ref{fig:corclas}): since the correlation coefficients are higher for the field sample (Fig.~\ref{fig:corclas}), the groups are more specific in the sense that they are better described by specific values of the seven parameters, meaning somehow less homogeneity.

The median values of \lM, $sfr1$, $sfr2$, $sfr3$ and $sfr4$ (Figs~\ref{fig:AevolM} and \ref{fig:AevolN}) are systematically higher (and lower for \mV ) for the field sample. These galaxies are thus on average more massive, more luminous and form more stars at all epochs. But at the same time they are on average bluer (lower \BmV) with a lower Sersic index \nV\ somewhat in contradiction with the general expected evolutionary trend. Hence using only one property to characterise the evolutionary stage of a galaxy is misleading. This explains why the field galaxies are less present towards the bottom of the tree (Fig.~\ref{fig:TreeAll}). This also shows that the impact of the Malmquist bias is weak since in unsupervised clustering analyses only the ranges are important, and not the exact shape of the distribution. In summary, for a similar diversification stage, field galaxies seem more massive.

There are four parameters in common with \citet{Fraix2010} who performed multivariate cluster and cladistic analyses of  56 low-redshift galaxy clusters: \muem, \logre, the mass and the distance to cluster centre even though our mass is  the original gas mass from which all stars of the galaxy formed (in other words, our mass includes the mass of stars as well as some remnant gas mass)  while theirs is the dynamical mass. The rooting is also different since we do not have the Mg index, but the two results agree on the global evolution for these parameters and that this evolution is not simply monotonically linear. 

There are also five parameters in common with \citet{Fraix2012} who performed a phylogenetic analysis of 424 early-type galaxies: magnitude (V for us vs B), \logre, \muem, $H_\beta$, and the mass (gas vs dynamical). We find only two slight differences: their mass increases while \muem\ is globally constant. However the departures from these trends are large and this may not be very conclusive.

We must be aware that the evolutions depicted by the tree depend somewhat on its rooting, and this has been chosen from a single parameter (metallicity) for the above studies, and three parameters (colour, Sersic index and mass) in our case.  If we follow the group interpretation found in Fig.~11 of \citet{Fraix2012}, we must conclude that the choices of the rooting for these three studies are satisfactory since no physical contradiction appears with our understanding of the broad picture of galaxy evolution.

\subsection{Groups and clusters}

We do not find any obvious correlation of the groups of the cluster sample with the galaxy clusters, indicating that our classification is "universal" for low redshift galaxies belonging to clusters. This is in line with our finding that cluster galaxies seem somewhat more homogeneous (Sect.~\ref{evolalongtree}).
 This is in agreement with \citet{Fraix2010} and  \citet{Valentinuzzi2011} and would seem to indicate that the formation histories of low redshift galaxies are not dramatically different from cluster to cluster. Indeed, the analysis of \citet{Fraix2012} shows that groups are mainly explained by very diverse histories.
This is supported by the barely significant dependence of the groups on the distance to the cluster center,  the intragroup variance being large (Figs.~\ref{fig:BoxplotAll}, \ref{fig:BoxplotMemb} and \ref{fig:BoxplotCont}). This is apparently not in agreement with \citet{Valentinuzzi2011} that find that some galaxy properties depend on the local environment, but this is probably because our groups are based on seven properties together and thus based on a more complete galaxy physics.

\subsection{Evolution in stellar mass}
\label{evolmass}

The separate analyses show that the evolution in stellar mass of cluster galaxies is clearer and more regular than for the field sample (Figs~\ref{fig:AevolM} and \ref{fig:AevolN}). The star formation rates ($sfr1...4$) are on average lower for cluster galaxies and the masses are lower as well. 

The stellar mass of a galaxy can in principle only increase except through harassment or strong interaction. \lM\ increases in our three analyses, the trend being much clearer in the case of the cluster sample. However, even in this case, the growth is not monotonic with the groups C9 and C10 having a much lower mass than the trend. There are three possible explanations for this drop. 

Firstly, since the tree represents the shortest path relating groups in a seven dimension space, it is possible that the projection on a one dimensional axis (here \lM) creates an apparent drop. This would be the case for instance if one of the seven parameters shows a reversal in its evolution. Fig.~\ref{fig:BoxplotMemb} does not reveal a unique culprit, none of the seven parameters being perfectly monotonic.

Secondly, our sample does not cover the entire diversity of galaxies, hence there are some missing links that could create a branch of low mass objects emerging after, say C3. The two groups C9 and C10 could then be relics of this distinct evolutionary path where mass does not increase as much as in the main trend seen from our sample. 

We note that such drops in the evolution in the dynamical mass have been found  on smaller samples by \citet{Fraix2010} and \citet{Fraix2012}. 

Lastly, the low masses of C9 and C10 can be a real loss of mass of these relatively evolved galaxies. The C9 and C10 galaxies are small and faint, but they differ in \BmV, \nV, $\epsilon$ and $T$, C9 being made of blue lenticulars and C10 of red ellipticals. C10 galaxies are more central and C9 formed more stars recently (higher $sfr1$ and $sfr2$). Hence, there might be some indication that C9 could be a stripped-off galaxies \citep{donofrioetal15}, but for C10 this is less clear.
s
So why C9 and C10 are placed nearly at the end of the tree with such low masses (and blue colour for C9)? This is because these two groups follow the main trends in three other parameters (increasing \nV\ and \DQ, decreasing \muem), while the two remaining parameters, among the seven used to build the tree, $H_{\beta}$ and \logre, do not seem to bring too much constraints (no obvious potentially strong trend). As a consequence the algorithm found the shortest path between the groups by optimizing the main trends for the maximum of parameters. 

C8 and C9 belong to the same ensemble of branches (C6 to C9) and separate at the same node.  C8 is made of late type galaxies, really more massive than C9, but as bluish as C9. This seems to be consistent with C8 having “just entered” into the cluster environment, possibly not yet stripped-off as C9. It turns out that C8 is one of the groups hosting jellyfish galaxies (Sect.~\ref{jellyfish}).

What about the field sample? The global increase in mass is grossly present but there are several departures with a wiggling shape (Fig.~\ref{fig:BoxplotCont} and Fig.~\ref{fig:AevolN}). Contrarily to the cluster sample, only one parameter (\nV), among the seven ones used for the analysis, shows a regular trend although not perfectly monotonic. This means that the algorithm was not able to find a most parsimonious solution with several regularly increasing or decreasing parameters, so it could not avoid drops in the mass evolution (known as regressions in MP analysis).

Groups C11 and C12 have high masses, are red and ellipticals, and still form some stars (average $sfr1$ and $sfr2$) even if much less that earlier on (higher $sfr3$ and $sfr4$). Clearly they are the most central galaxies. We find such kinds of galaxies in the field sample (F8, F11, F12, F13 and F15) with high masses, red with an elliptical morphology, but their $sfr3$ is not very much higher than the other groups, indicating that they formed stars only at very early stages ($sfr4$). In addition, these field sample galaxies have significantly higher $sfr$* at all epochs. So these high-mass galaxies, supposedly situated at the end of the evolutionary process of galaxies, have built up very early, and field galaxies appear to form stars at a higher rate than the cluster ones.

\subsection{Evolution in size}
\label{evolstruct}

The size of the cluster and field galaxies do not show any evolutionary trend in \logre\ (Figs.~\ref{fig:BoxplotMemb} and \ref{fig:BoxplotCont}), but the Sersic index \nV\ increases and the morphology T decreases. The effective surface brightness \muem\ becomes brighter in both samples, with a high peak at F11 and F12. So galaxies become more concentrated, more ellipticals, and much brighter, whatever their radius and their environment. 

The expected evolution in radius with redshift is likely not visible for the small range in redshift covered by our samples ($0.02<z<0.44$). Hydrodynamical simulations suggest in fact a progressive decrease of \re\ going from redshift 4 to 0 (D'Onofrio et al, in preparation).

The present data seem then to suggest that the environment has no global effect on the structure of galaxies, where for structure here we mean the effective radius and the Sersic index of galaxies. There is, on the other hand, a hint of morphological segregation in the cluster sample since the early types are at smaller \DCC. This is the well known morphology-density relation \citep{Dressleretal1987,Fasano2012}. The systematic difference however is modest, and the scatter is too large to make a definite inference.

%\subsection{Cluster environment and the gaseous component of galaxies}

\subsection{Emission line properties: gas deficit}
\label{emissionlines}

\begin{figure}[t]
\centering
\includegraphics[width=\linewidth]{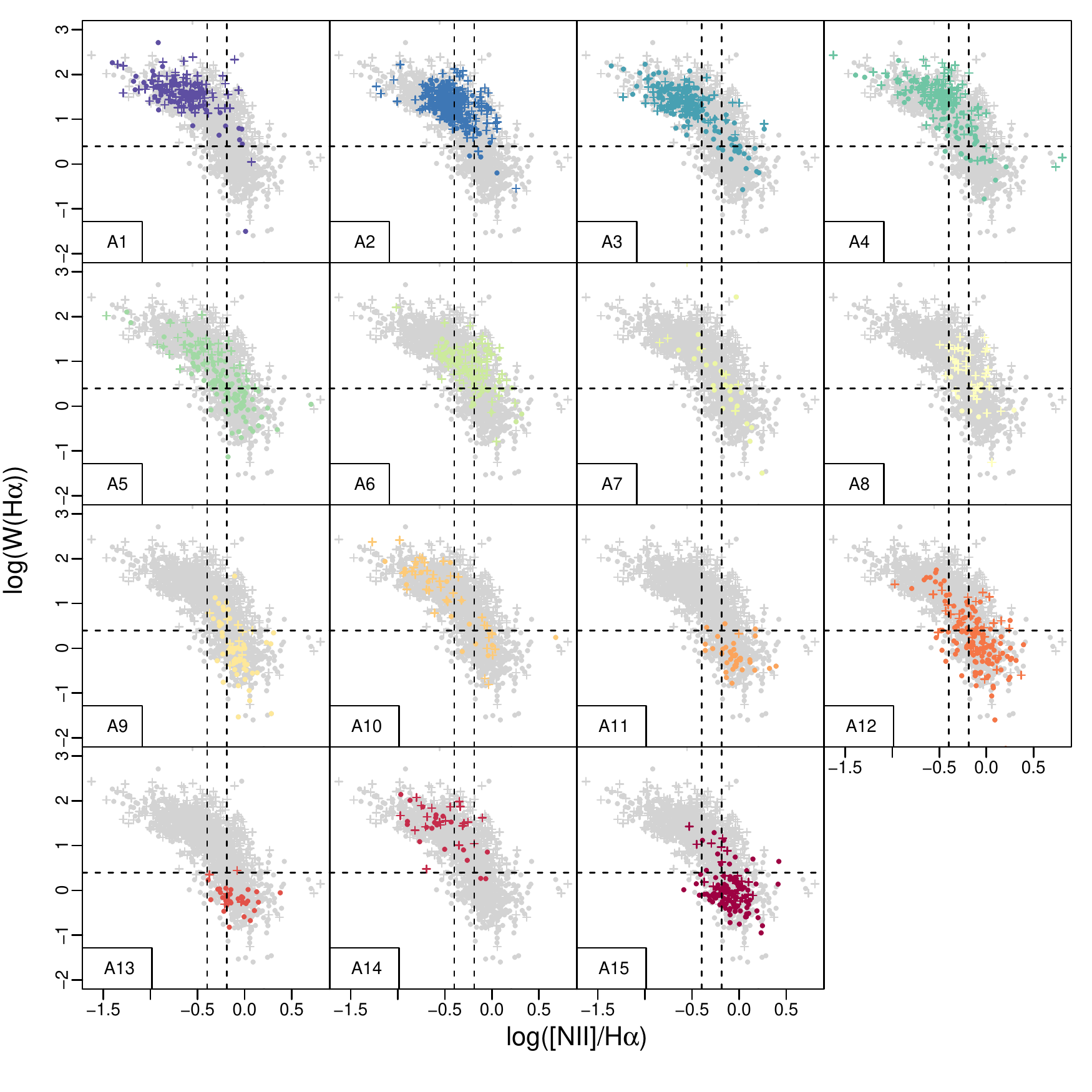}
\caption{Equivalent width of H$\alpha$ vs [NII]/H$\alpha$ . The vertical dotted lines corresponds to [NII]=/H$\alpha$=0.4 and 0.65 and the horizontal one is for W(H$\alpha$)=2.5 (see text). Dots are for cluster galaxies and crosses for field ones.}
\label{fig:HaN2Ha} 
\end{figure}

\begin{figure}[t]
\centering
\includegraphics[width=\linewidth]{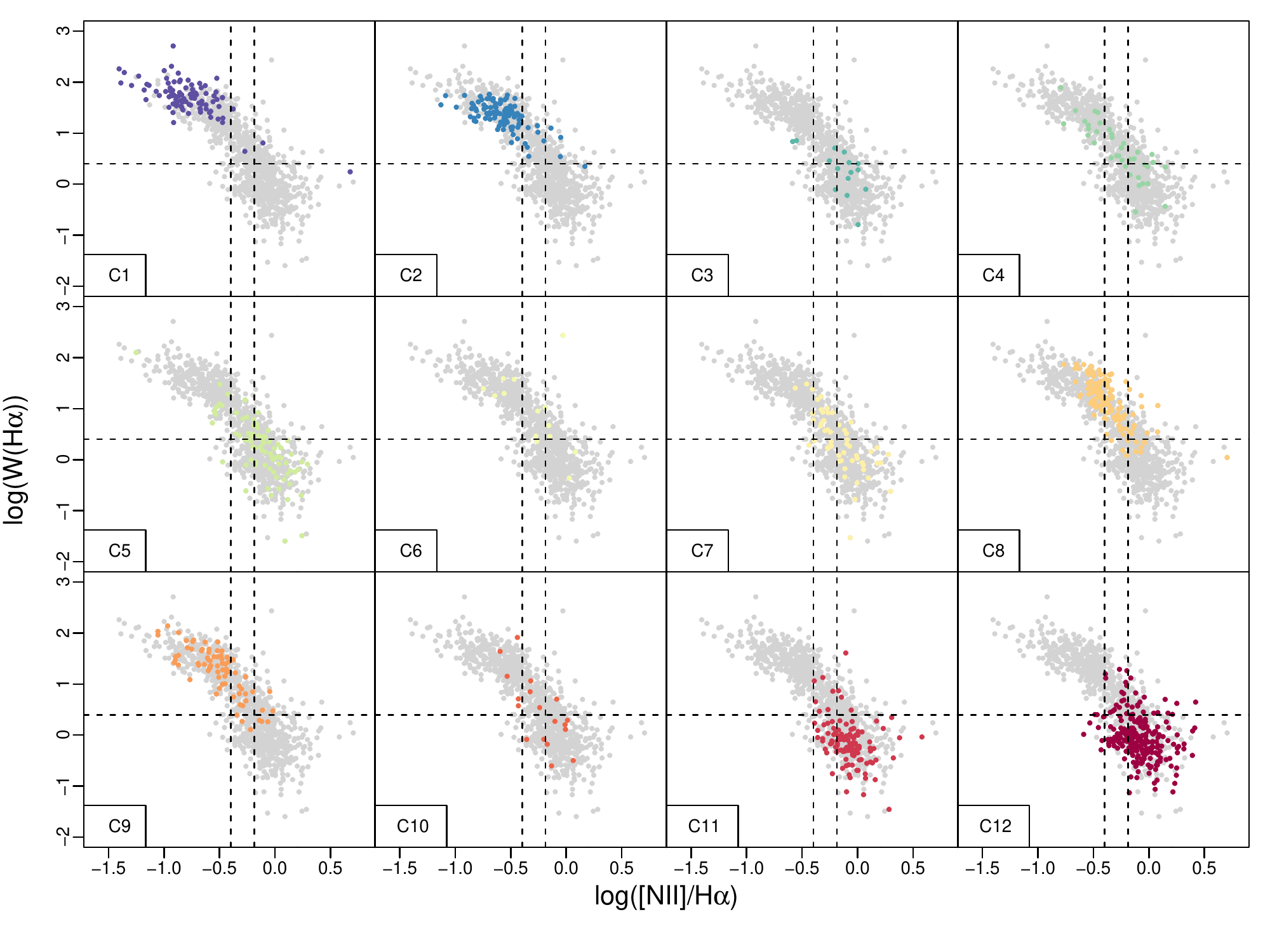}
\caption{Idem Fig.~\ref{fig:HaN2Ha} for cluster galaxies.%}
\label{fig:HaN2HaM}} 
%\end{figure}
%\begin{figure}[h!]
%\centering
\includegraphics[width=\linewidth]{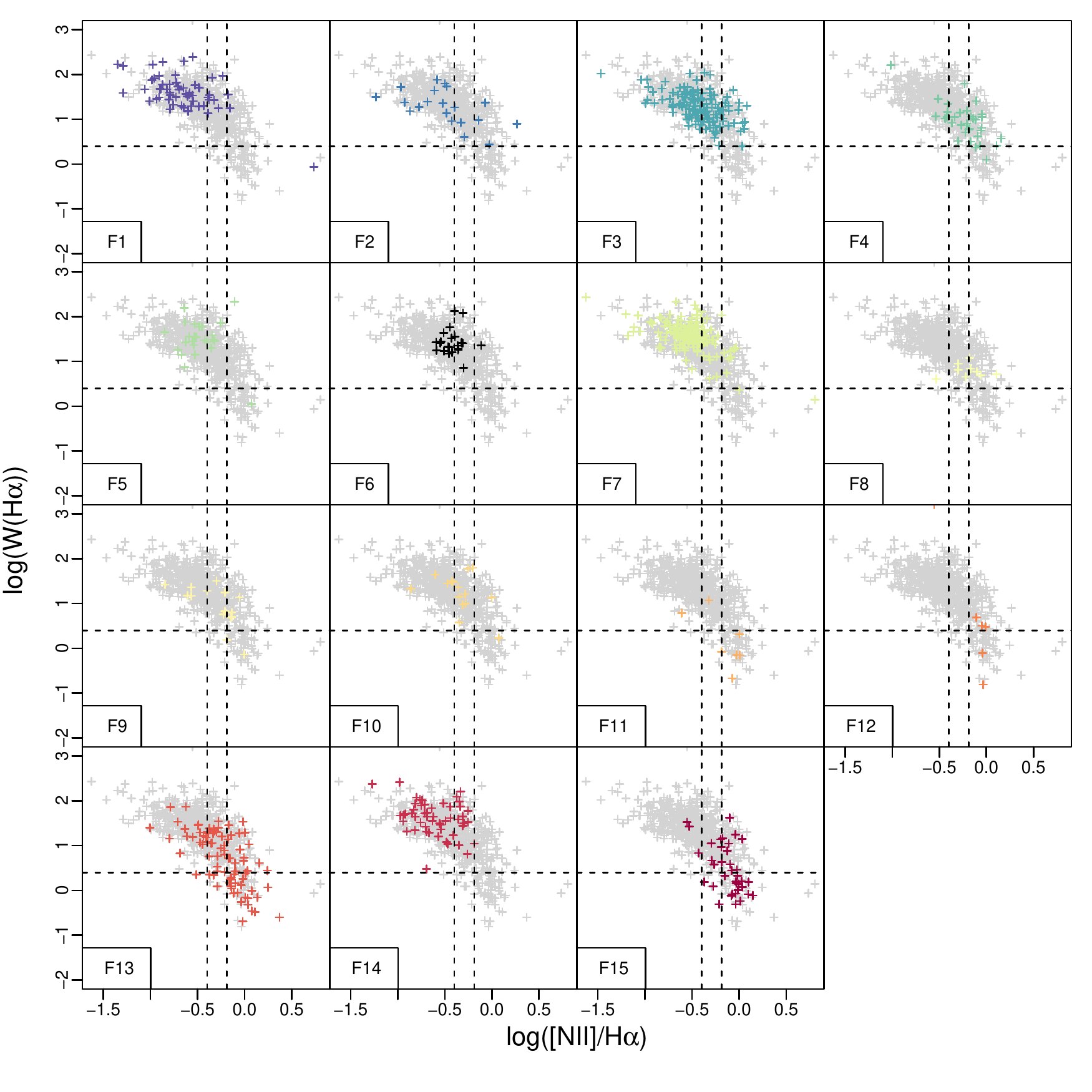}
\caption{Idem Fig.~\ref{fig:HaN2Ha} for field galaxies.%}
\label{fig:HaN2HaN}} 
\end{figure}

Figures \ref{fig:HaN2Ha}, \ref{fig:HaN2HaM}, \ref{fig:HaN2HaN} show the behavior of the equivalent width of H$\alpha$ vs. the ratio [NII]$\lambda$6584/H$\alpha$ for the full, cluster and field samples respectively. The diagrams relate the W(H$\alpha$), a measure of the amount of gas per unit stellar mass, to the ratio [NII]$\lambda$6584/H$\alpha$\ \citep{cidfernandesetal10} which is  a tracer of the ionisation mechanism. 
It    cannot be $\gtrsim 0.65$ for pure HII regions  following the  diagnostic diagrams of \citep{Kauffmann2003}.
Typical HII nuclei have [NII]$\lambda$6584/H$\alpha \lesssim 0.4$, LINERs and Seyfert-type AGN spectra have [NII]$\lambda$6584/H$\alpha \gtrsim 0.7$. 

\begin{figure}[t]
	\centering
	\includegraphics[width=\linewidth]{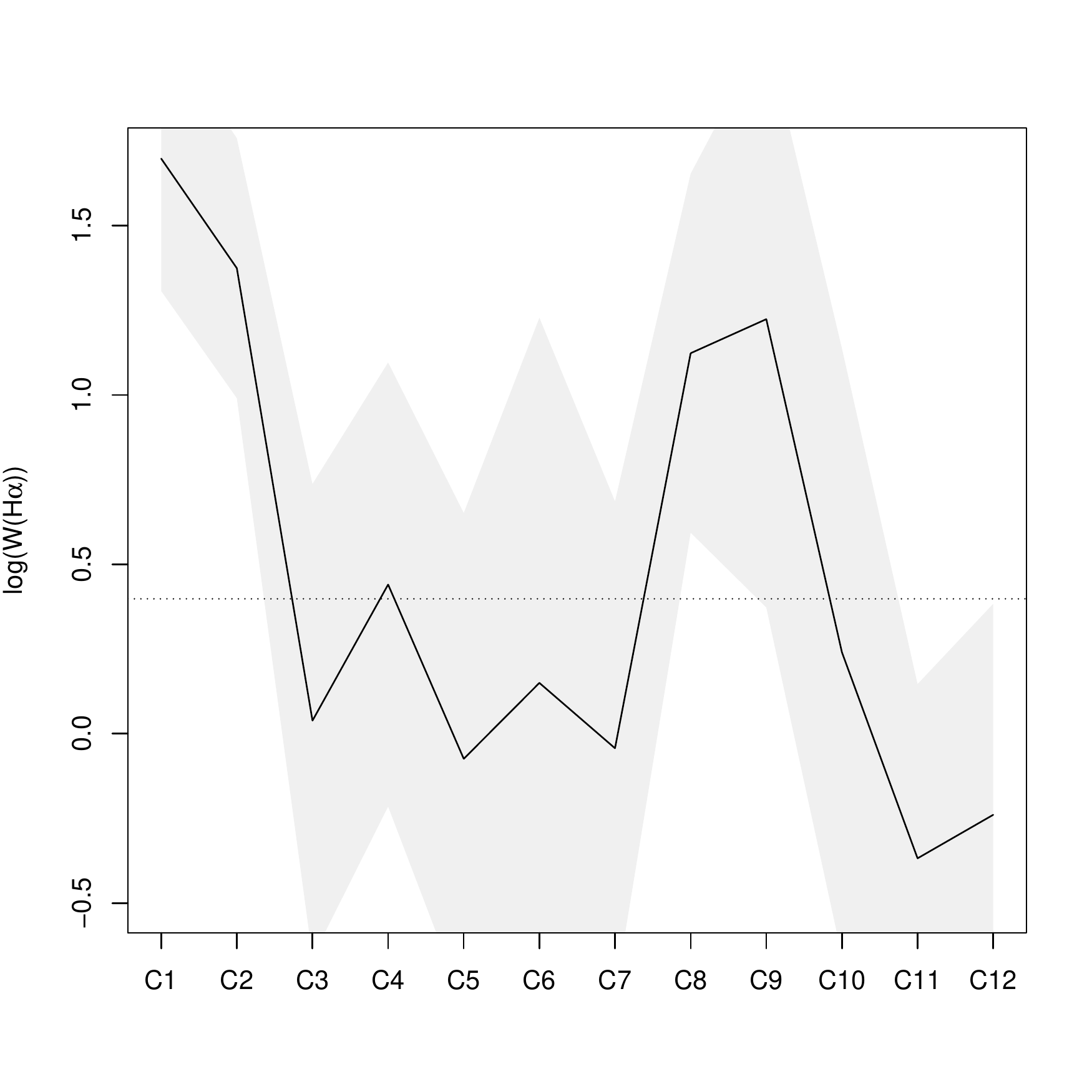}
	\caption{Evolution of the median of W(H$\alpha$) for the cluster sample. Shaded regions show 1$\sigma$ dispertion around median. The horizontal dotted line is for W(H$\alpha$)=2.5 (see text and Figs.~\ref{fig:HaN2Ha}, \ref{fig:HaN2HaM}, \ref{fig:HaN2HaN}).%}
		\label{fig:HamediansM}} 
\end{figure}

\begin{figure}[t]
	\centering
	\includegraphics[width=\linewidth]{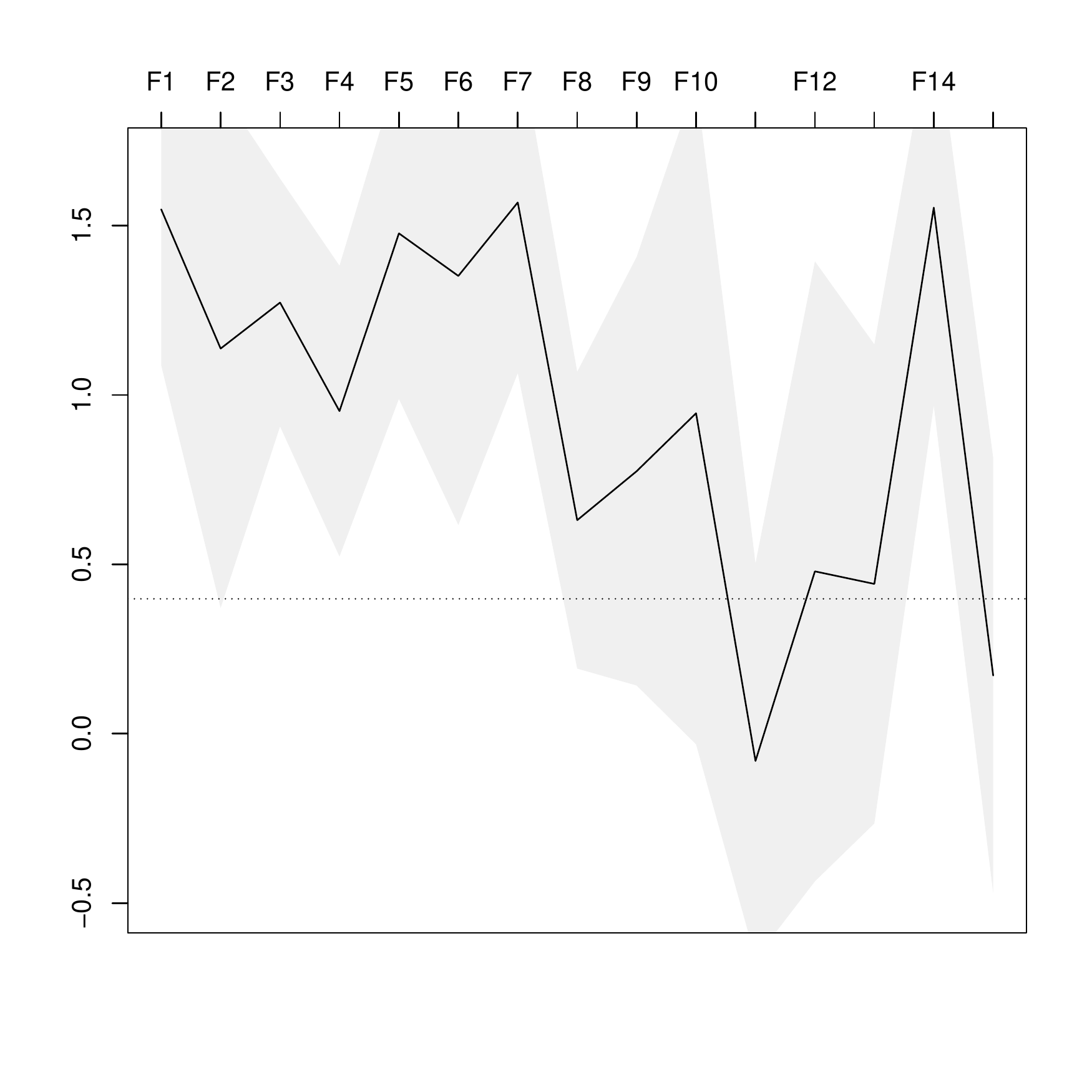}
	\caption{Idem Fig.~\ref{fig:HamediansM}for the field sample.%}
		\label{fig:HamediansN}} 
\end{figure}

There is apparently a progression in the three samples from top left to bottom right of the diagram, confirmed by the evolution of the medians of the cluster and field sample groups (Figs~\ref{fig:HamediansM} and \ref{fig:HamediansN}). This means that there are less HII regions with diversification. For the cluster sample, there are more HII regions of galaxies in the first two groups (C1-2) and in C8 and C9. The groups C1,2,8 and 9 mainly make the blue sequence of the cluster galaxies (see below).  C3,4, 6, 10 are made predominantly of transition objects between HII and LINERs. C11 and 12 are definitely dominated by LINER spectra and transition objects (e.g., HII galaxies are fully absent). However there are departures from this trend, such as C8, C9, most notably. However, C8 is predominantly made of late-type galaxies, C 9 seems to be the lower mass counterpart of C8. Both show high W(H$\alpha$) and hence large $sfr1$.  It is interesting to recall that C6 to C9 and C10 to C12 define two distinct ensembles on the tree (Fig.~\ref{fig:TreeMemb}) that we suspect to have increased their stellar mass in two different ways, the former ensemble by forming stars and the latter by accreting old stars  (see Sect.~\ref{Result4czbM}) in agreement with the line emission results.

 This trend -- that is, passing from HII dominated samples to AGN/LINER spectra through intermediate objects -- can also be seen in the F groups, with F11 and F15 the only groups strongly departing from it. F11 and 15 isolate  groups of early type galaxies with a population of HII emitters. 
 
%
%\begin{figure}[t]
%\centering
%\includegraphics[width=\linewidth]{WINGS4czbK_Ha_medians.pdf}
%\caption{Evolution of the median of W(H$\alpha$) for the cluster sample (red solid line) and field sample (dotted blue line). Shaded regions show 1$\sigma$ dispertion around median. Group indices are indicated at the bottom abcissa for the cluster sample (C1-12) and at the top one for the field sample (F1-15). The horizontal dotted line is for W(H$\alpha$)=2.5 (see text and Figs.~\ref{fig:HaN2Ha}, \ref{fig:HaN2HaM}, \ref{fig:HaN2HaN}).%}
%\label{fig:Hamedians}} 
%\end{figure}

However, there is a notable difference between field and cluster sample groups: in the latter, the W(H$\alpha$) among field galaxies is higher by a factor of a few than among cluster galaxies except for groups C8 and C9 (Figs~\ref{fig:HamediansM} and \ref{fig:HamediansN}). A large fraction of galaxies in the groups not dominated by HII nuclei falls below the limit of W(H$\alpha$)= 2.5 \AA\ that separates weak emitters from "retired" and quiescent sources according to \citet{cidfernandesetal10}.

A comparison between groups confirms the suggestion of \citet{Marzianietal2017} about a lower amount of gas in cluster galaxies, and of a population of weak, non-HII line emitters with high [NII]$\lambda$6584/H$\alpha$ associated with the red sequence of galaxies.

\subsection{Jellyfish galaxies}
\label{jellyfish}

\begin{figure}[t]
\centering
\includegraphics[width=0.49\linewidth]{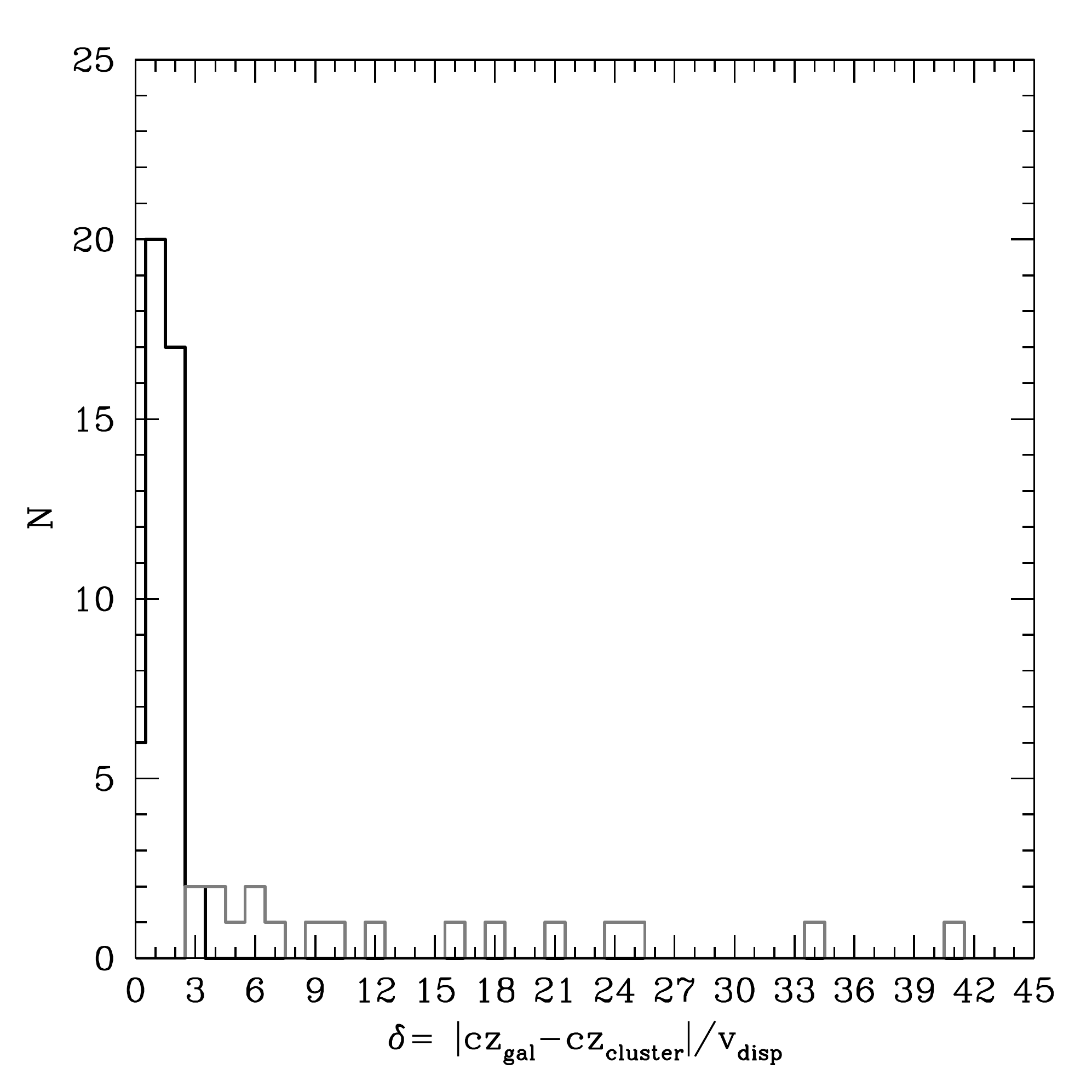}
%\caption{Jellyfish.%}
%\label{fig:jellyfish}} 
%\end{figure}
%\begin{figure}[h!]
%\centering
\includegraphics[width=0.49\linewidth]{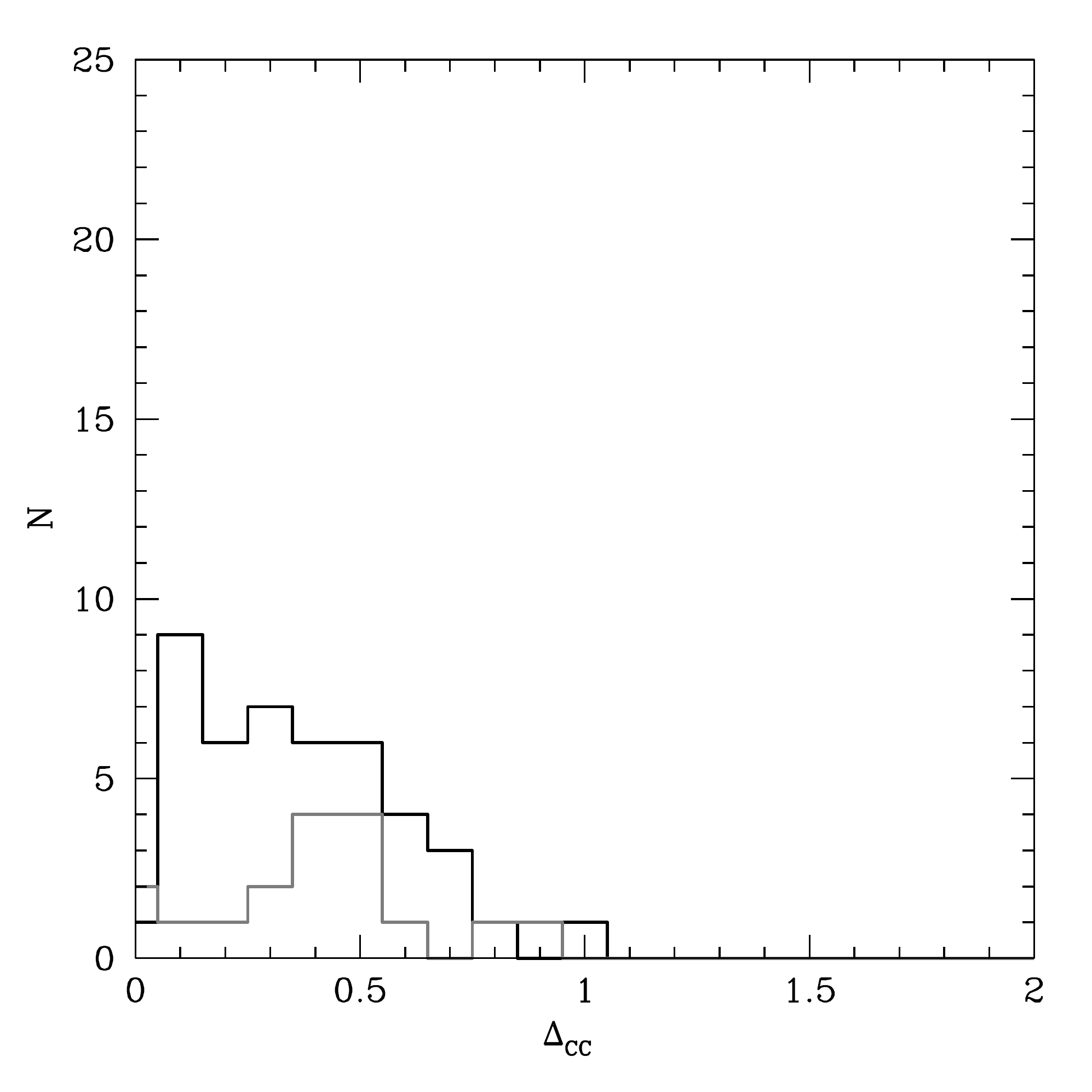}
\caption{Histograms for jellyfish galaxies that are in our cluster (black lines) or field (grey lines) samples, for the parameter $\delta$ (see text) and the distance \DCC\ to cluster center.
\label{fig:jellyfish}}
\end{figure}

A number of “jellyfish galaxies” have been identified in the WINGS database by \cite{poggiantietal16}.
These objects are outstanding examples of  galaxies which show knotted arms of material resembling the tentacles of jellyfishes. They are believed to be  galaxies that are still gas rich and have just entered the cluster environment. 

Jellyfish galaxies are found mainly in C2 and C8, with a prevalence of $\approx$ 10\%\ in each group, and in F3, F9, and F13. The groups F3 and F13 are large, and so the prevalence of jellyfish galaxies is very low, $\approx $2\%. In F9 they are $\approx$10\%\ of the sample. Among cluster group C2 is dominated by HII regions while C8 is shifted towards intermediate objects. Both groups are however predominantly above the limit of retired galaxies. The group assignment is consistent with the idea that jellyfish galaxies are characterised by ram pressure stripping of gas  by the intra-cluster medium, triggering starbursts along the tails that give jellyfish galaxies their characteristic morphology. It is also consistent with the idea that jellyfish galaxies are just entering the cluster environment.

It is remarkable that some (18) jellyfish galaxies are revealed in the field sample. This is apparently surprising since jellyfish morphology is believed to be exclusive of the cluster environment. Their clustercentric distance with respect to jellyfish galaxies classified as cluster member is only slightly skewed towards larger distances; the non-member distribution is well within the distribution of clustercentric distances for jellyfish galaxies that are cluster members. The field sample galaxies do not meet a radial velocity criterion that implies that their radial velocity $cz$ should be different from  the $cz_{\mathrm{cluster}}$ by less than  3 times the velocity dispersion of the cluster. Looking at the distribution of the variable $\delta = \left| cz-cz_{\mathrm{cluster}}\right| /\sigma $, we note that there are about 6 galaxies that are within $\delta \lesssim 6 \sigma$ (Fig~\ref{fig:jellyfish}). Considering that jellyfish galaxies are probably high velocity intruders \citep{jaffeetal18}, these galaxies could be physically associated with the cluster. The remaining   jellifishes have $\delta$\ variables that are too large for a physical association. Even if their nature remains to be clarified on a source-by-source basis, they may be true jellyfish galaxies occurring in groups or low-mass halos \citep{poggiantietal16}, or  be due to gravitational interactions among field galaxies that are known to lead to tidal tails with star forming knots \citep[e.g., ][]{ducrenaud13} loosely resembling the tails of cluster jellyfishes.

\subsection{The \lM, \nV, and M/L relation}

\begin{figure}[t]
\centering
\includegraphics[width=\linewidth]{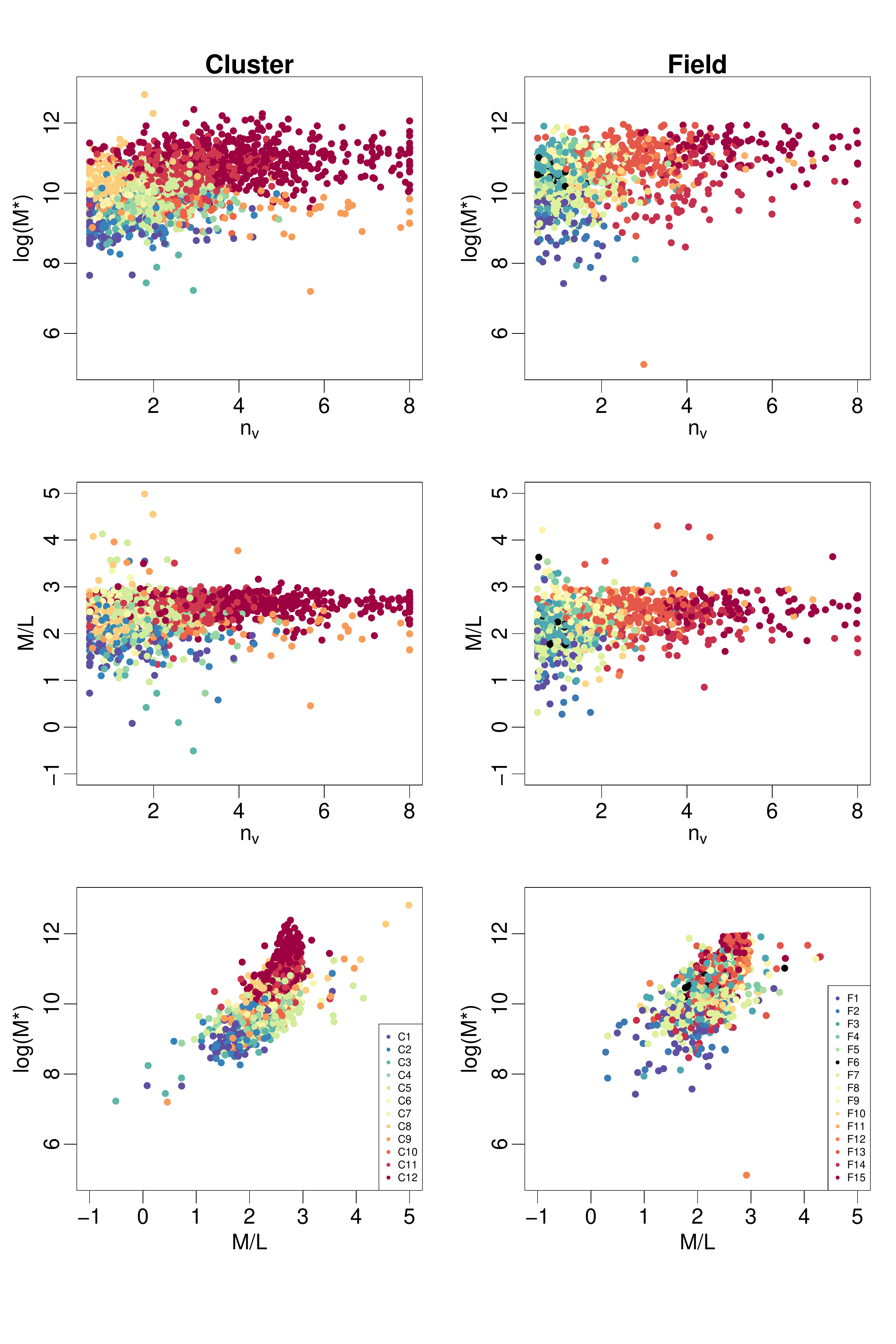}
\caption{Scatter plots between \nV,  \lM\ and M/L for cluster (left column) and field (right column) galaxies.}
\label{fig:nMMoL}
\end{figure}

A relation between the stellar mass M, \nV, and M/L has been found by \citet{DOnofrio2011}. We confirm this correlation  (Fig.~\ref{fig:nMMoL}) for the cluster and field samples with a clear sequence along the tree. Not much difference is present between the two samples. However, the correlation does not hold within most of the groups, suggesting that this correlation is probably due to the co-evolution of these variables that is much influenced by the evolutionary history of the galaxies.

The shape of galaxies and their stellar population are therefore linked each other statistically. Nature does not permit the existence of blue ellipticals and red spirals at the present epoch. This relation is therefore another example of co-evolution among galactic components, the most famous being that between the black-hole mass and the bulge mass \citep{Magorrian1998}.

The existence of co-evolving components is particularly interesting in a framework where galaxies form progressively through merging events, that are necessarily random in their nature. This implies that behind causality there is a well defined evolving flux of galaxy structures. The memory of disturbing events, due for instance to minor mergers, should be rapidly lost, while in case of catastrophic events such as major mergers, galaxies enter in a new configuration that rapidly forget the previous galaxy properties.

Co-evolution should be at the origin of many of the observed scaling relations.

\subsection{The \lM\ vs \logre\ relation}
\label{massradius}
 
 \begin{figure}[t]
 \centering
 \includegraphics[width=\linewidth]{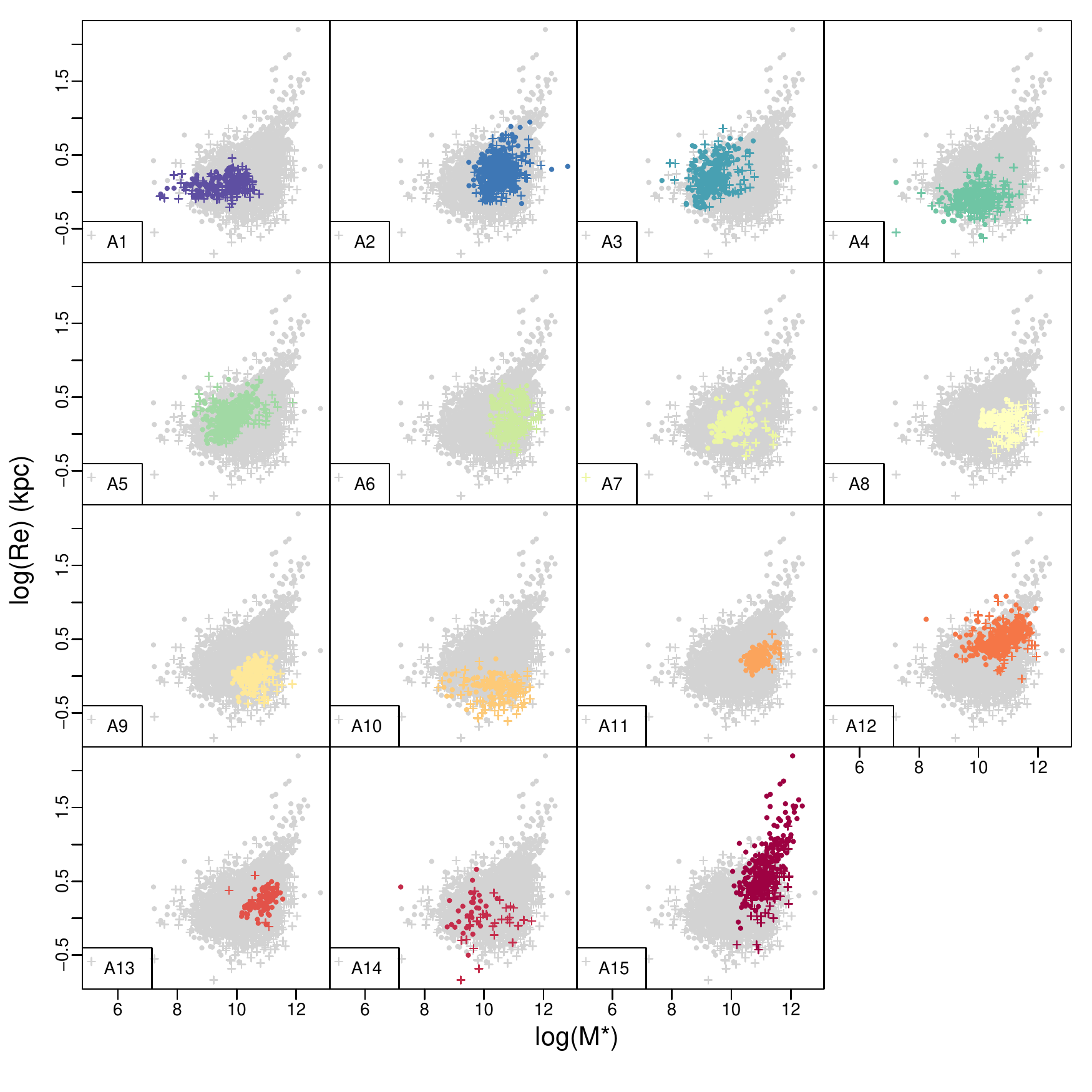}
\caption{Effective radius vs mass for the groups of the full sample. Gray points are the galaxies of the full sample.}
 \label{fig:allRmass} 
 \end{figure}
 
 \begin{figure}[t]
 \centering
 \includegraphics[width=\linewidth]{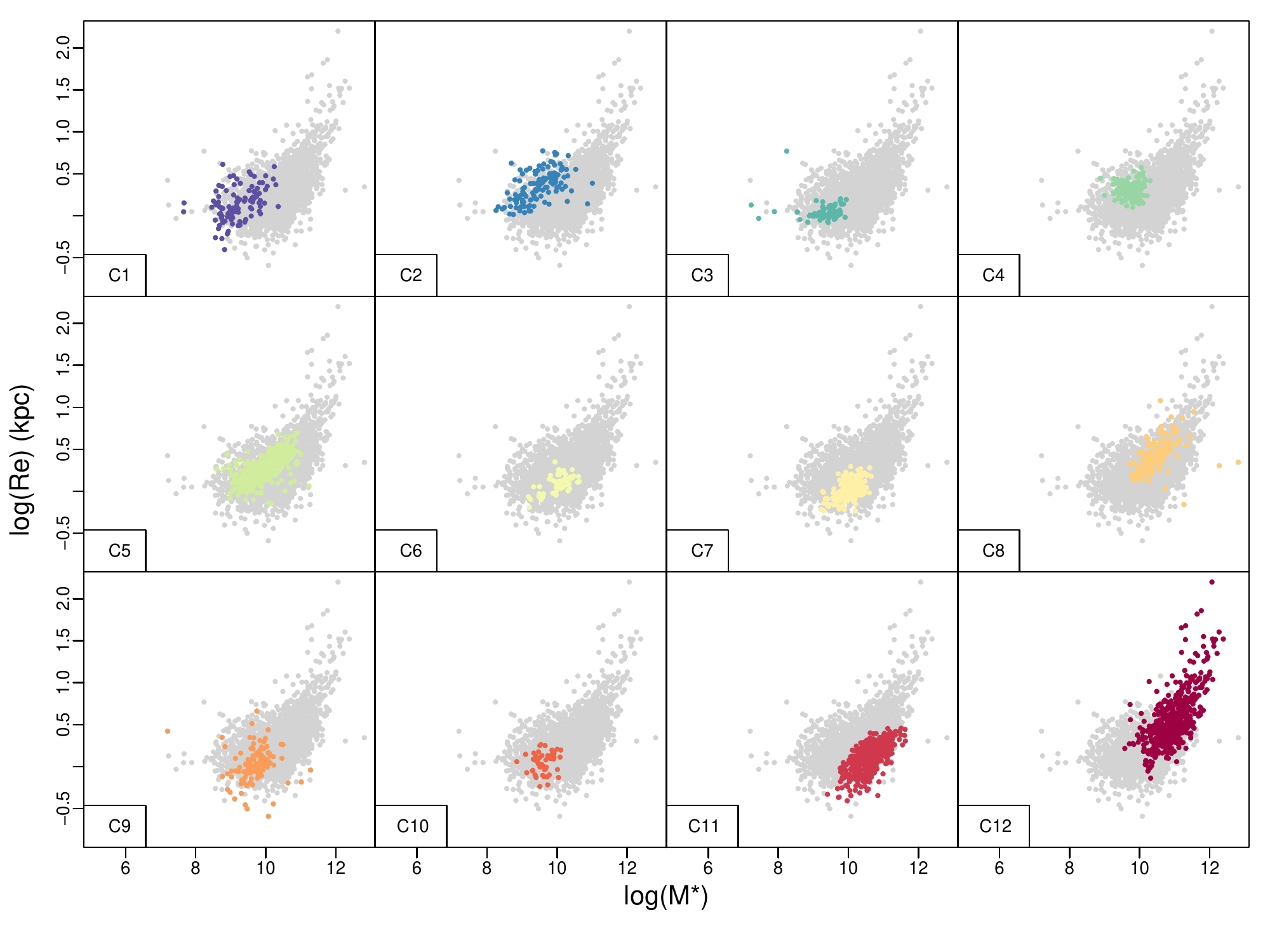}
 \caption{Effective radius vs mass for the groups of the cluster sample. Gray points are the galaxies of the cluster sample only.%}
 \label{fig:MembRmass} }
% \end{figure}
%
% \begin{figure}[t]
 %\centering
 \includegraphics[width=\linewidth]{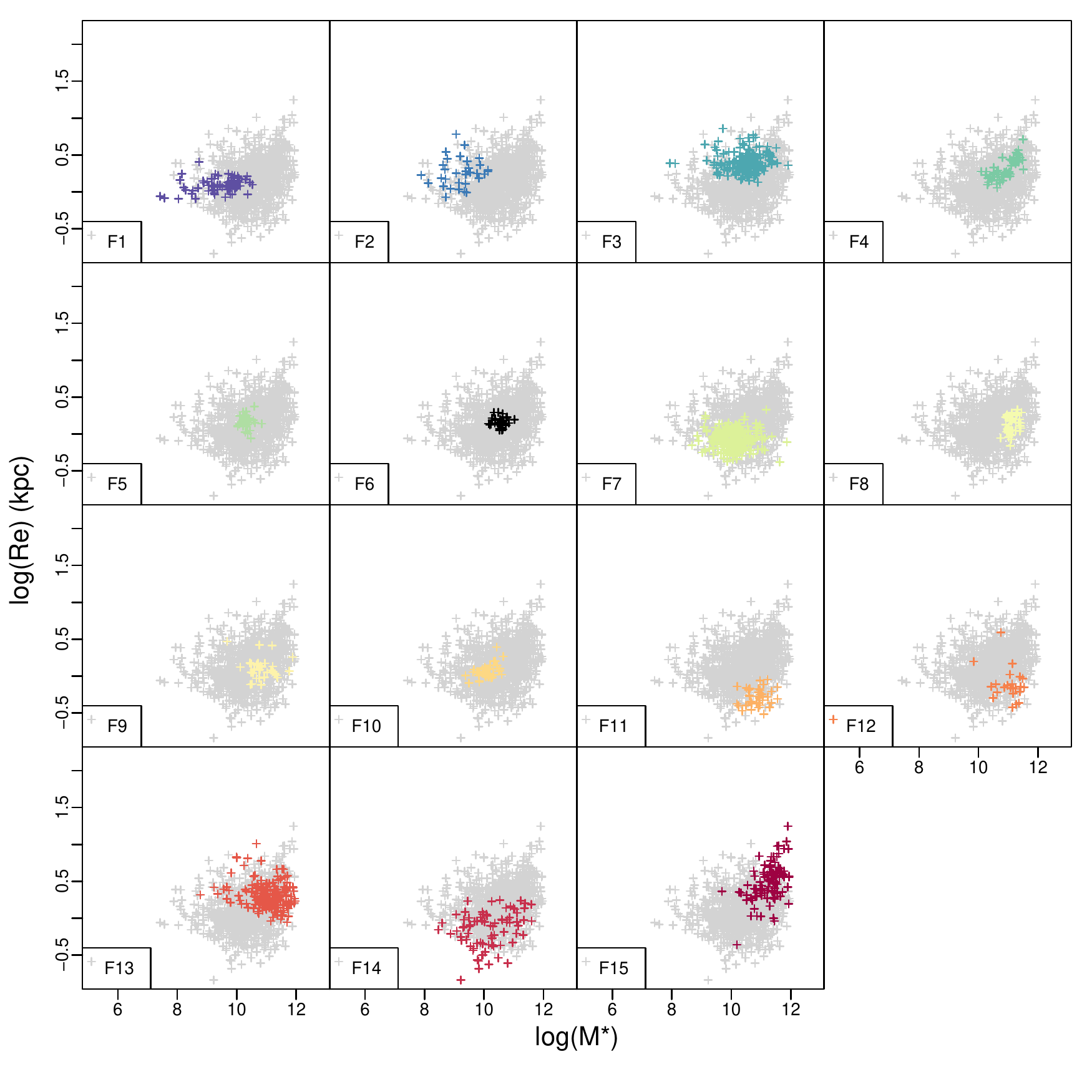}
 \caption{Effective radius vs mass for the groups of the field sample. Gray points are the galaxies of the field sample only.%}
 \label{fig:ComparisonRmass}  }
 \end{figure}

%\section{Mass, radius, surface brightness}

The \logre\ vs \lM\ scatter plot shows that the galaxies in A15 and C12 are the only ones that define a clear Mass-Radius relation ((Fig.~\ref{fig:allRmass} and \ref{fig:MembRmass}). These are the most diversified groups with the most massive objects, in good agreement with numerical simulations \citep{Taylor2016}. However, this is not true for the field sample where no such relation is present in any of the groups (Fig.~\ref{fig:ComparisonRmass}). The group distribution is highly discriminated for the three samples, and is identical to the one found in \citet{Fraix2010} and \citet{Fraix2012}. This can be explained by the different assembly histories of the galaxies, through transformation events such as mergers \citep[e.g.][]{Robertson2006} or AGN feedback \citep{Taylor2016}.  We note that the sample studied by \citet{Fraix2010} is made of cluster galaxies only \citep[699 early-type galaxies in 56 clusters, ][]{Hudson2001}, and the one studied by \citet{Fraix2012} is a priori a mixture of field, group and cluster galaxies \citep[424 galaxies from ][]{Ogando2008} unfortunately to unknown proportion to us.

The almost complete absence of a linear relation between \logre\ and \lM\ for field galaxies is very important. It is a clear suggestion that the relation has its origin in the peculiar cluster environment. The large number of dry merging events should be the key physical mechanism producing large radii for big mass objects. Such events should be almost absent in the field. 

The  \logre\ vs \lM\  relation appears almost flat for galaxies with M$^*<10^{10}$ M$_{\sun}$ both in ETGs and LTGs.
This is also seen in numerical simulations, such as Illustris \citep{Vogel2014}. The linear relation appears only for the
most massive galaxies in clusters, where the radius likely increases for the high number of minor merging events while the
stellar mass remains almost the same. The linear relation \logre-\lM\  is expected for the total mass M, that is Dark Matter plus baryonic mass.
Many effects can determine the absence of such a relation. The most likely explanation, according to models,
is that winds from supernovae and feedback effects from the central AGN have expelled most of the gas and quenched the star
formation, disconnecting R$_{\rm{e}}$, which depends on the dimension of the potential well of the total mass, from the stellar mass.

The F12 group is quite peculiar, being made by small size high mass ($\simeq 10^{11}$ M$_{\sun}$) elliptical galaxies,
with an anomalous high \DQ\ and the highest $sfr4$. The number of these objects are rare in other
surveys at different redshifts, such as that of Fornax \citep{Venhola2018}, 
that of isolated galaxies \citep{FernandezLorenzo2013}, the SDSS \citep{Shen2003,Nair2010}, 
GAMA \citep{Lange2015} and SAMI \citep{Scott2017}. However, super-dense galaxies are known to exist either in
clusters \citep{Valentinuzzi2010} and in the field \citep{Poggianti2013}. In the field the frequency of
these objects with masses larger than $10^{10}$ M$_{\sun}$ is estimated around 4.4\%.
In our sample we have 24 members of the F12 groups over a total number of field galaxies
of 1158 (~2\% frequency). If we add the F11 group, that is also made of quite massive and small
objects, we reach a frequency of ~5\%.
These value are in line with the frequency estimated by \citet{Poggianti2013},
but in any case we prefer to warn the reader about a possible discrepancy with the above cited surveys.

\subsection{The Kormendy \muem\ vs \logre\ relation}

 \begin{figure}[t]
 \centering
 \includegraphics[width=\linewidth]{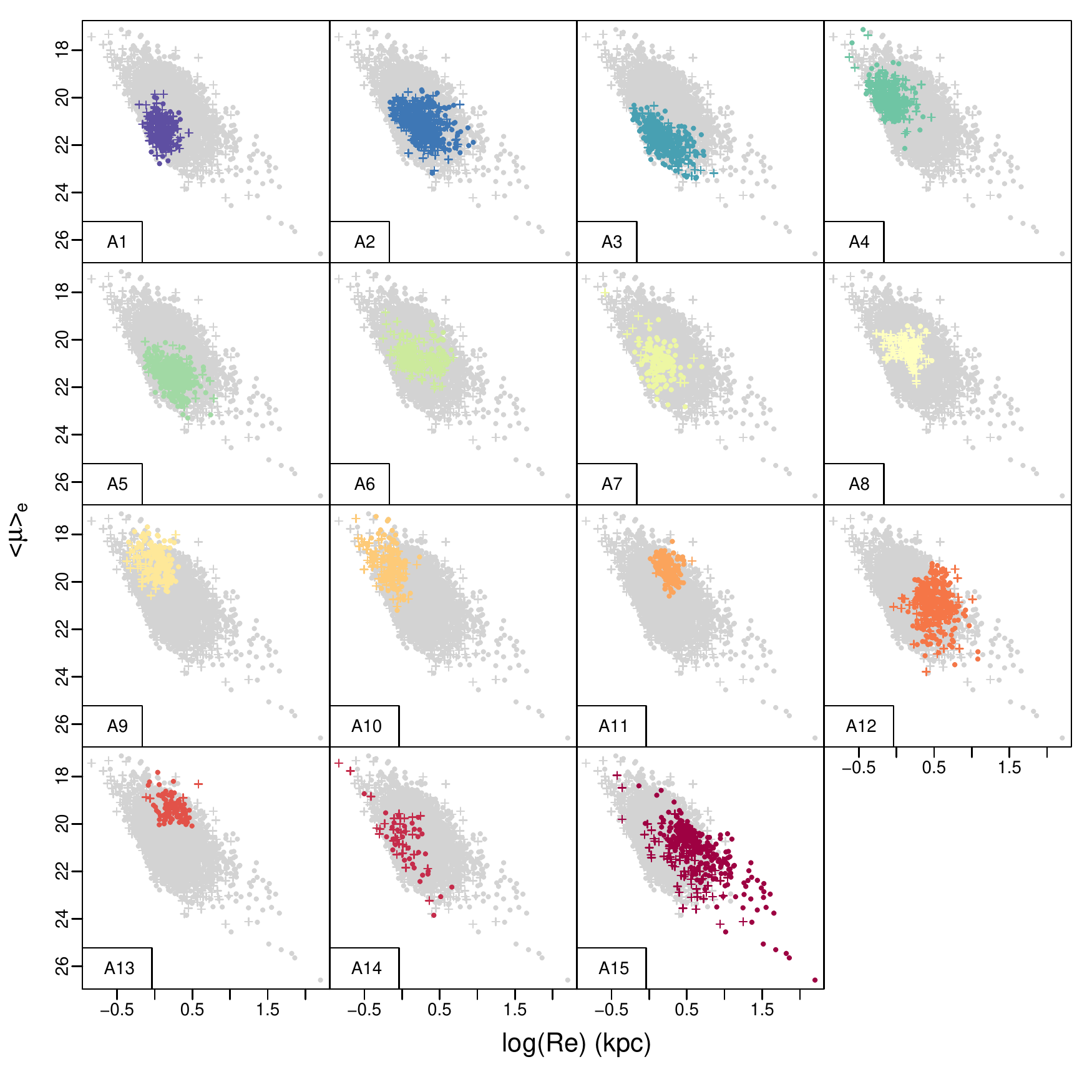}
\caption{Effective radius vs effective surface brightness for the groups of the full sample. Gray points are the galaxies of the full sample.}
 \label{fig:allRmue} 
 \end{figure}
 
 \begin{figure}[t]
 \centering
 \includegraphics[width=\linewidth]{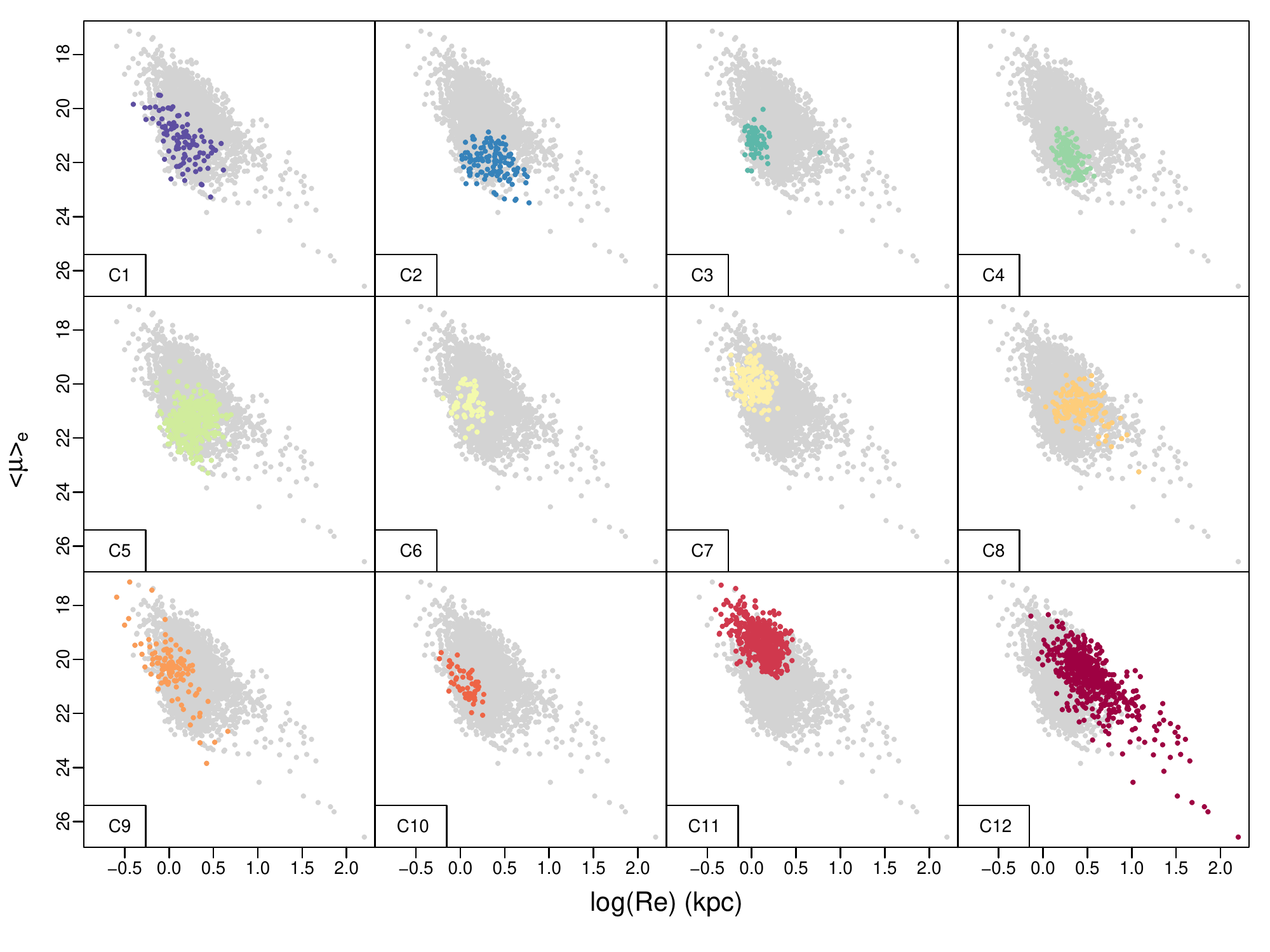}
 \caption{Effective radius vs effective surface brightness for the groups of the cluster sample. Gray points are the galaxies of the cluster sample only.%}
 \label{fig:MembRmue} }
% \end{figure}
%
% \begin{figure}[t]
 %\centering
 \includegraphics[width=\linewidth]{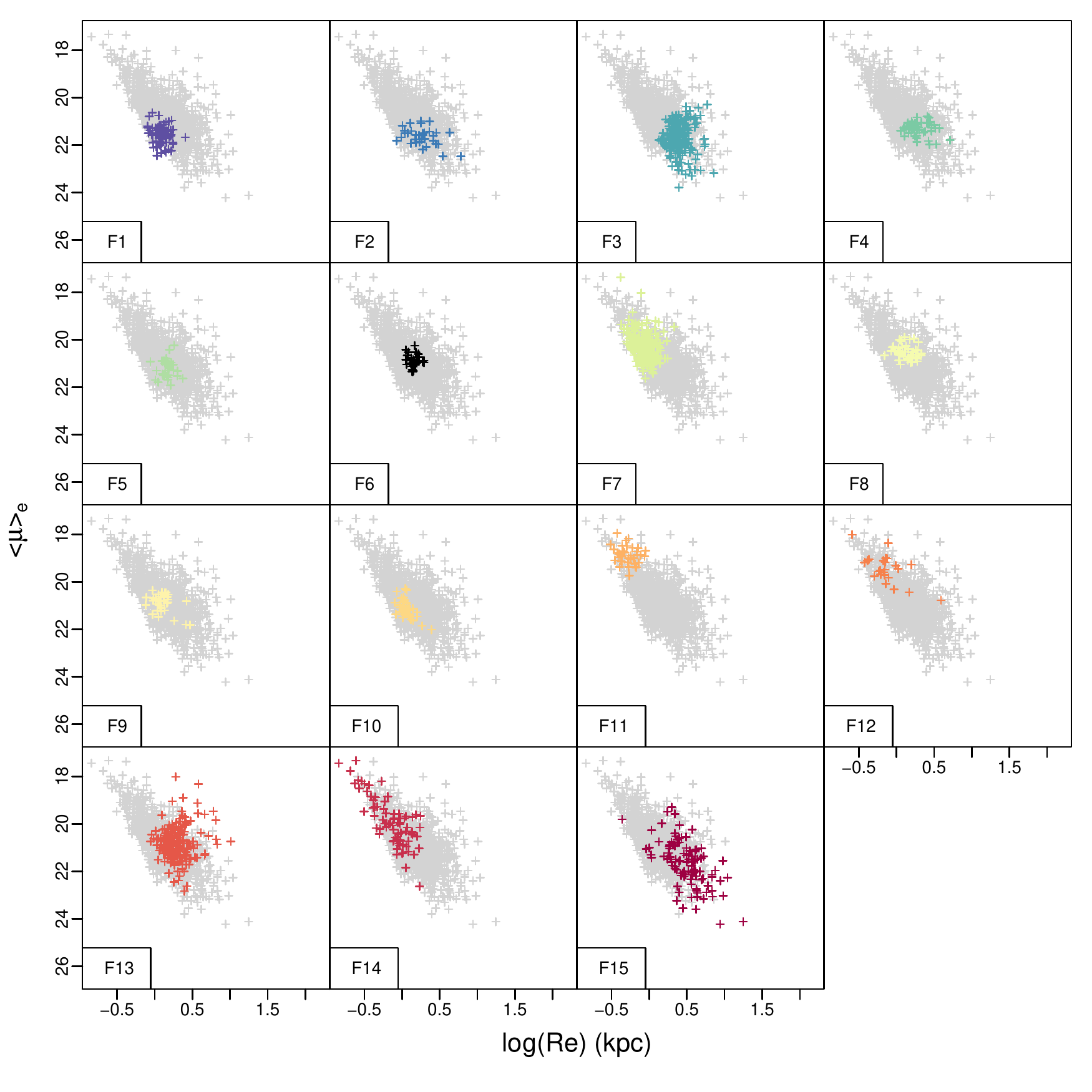}
 \caption{Effective radius vs effective surface brightness for the groups of the field sample. Gray points are the galaxies of the field sample only.%}
 \label{fig:ComparisonRmue}  }
 \end{figure}

The Kormendy relation \muem\ vs \logre\ \citep{Kormendy1977} is visible on the three samples (Figs.~\ref{fig:allRmue}, \ref{fig:MembRmue}, \ref{fig:ComparisonRmue}) but with a large dispersion and it is not present in most of the groups. Even if the most diversified groups (A15, C12, F15) show a clear relation, some other groups of massive galaxies do not (A8, A11, A13, C11, F12, F13) while some groups of less massive galaxies do (A3, A10, A14, C1, F14). Mass is thus not the only ingredient necessary for this relation to hold. 

In addition, there is a lot of dispersion in this plot when the entire sample is considered, and it is very significantly reduced when only individual groups are considered. This is also striking on the \logre\ vs \lM\ plots (Sect.~\ref{massradius}). The reason of this dispersion is the mixture of different populations, at different stages of evolution, quite alike the emission line diagnostic plots (Sect.~\ref{emissionlines}) where the different sources of ionisation can only be separated by a multivariate analysis \citep[or by combining several such diagrams, see e.g.][]{Souza2017}. 

It must be noticed that our groups pave the full sample distribution without much overlap. For instance, considering the Kormendy diagram for the cluster sample (Figs.~\ref{fig:MembRmue}), the regions to the top right are occupied by the three more massive groups (C8, 11, 12), so the dispersion perpendicular to the global relation is mainly due to mass. This is also true for the full and field samples. This means that by dividing the samples in two mass ranges, we find two parallel relations. Hence the Kormendy relation is not due to the mass nor to the environment, but to some co-evolution that depends on mass.

\subsection{Color-magnitude diagram}
\label{CMD}

\begin{figure}[t]
\centering
\includegraphics[width=\linewidth]{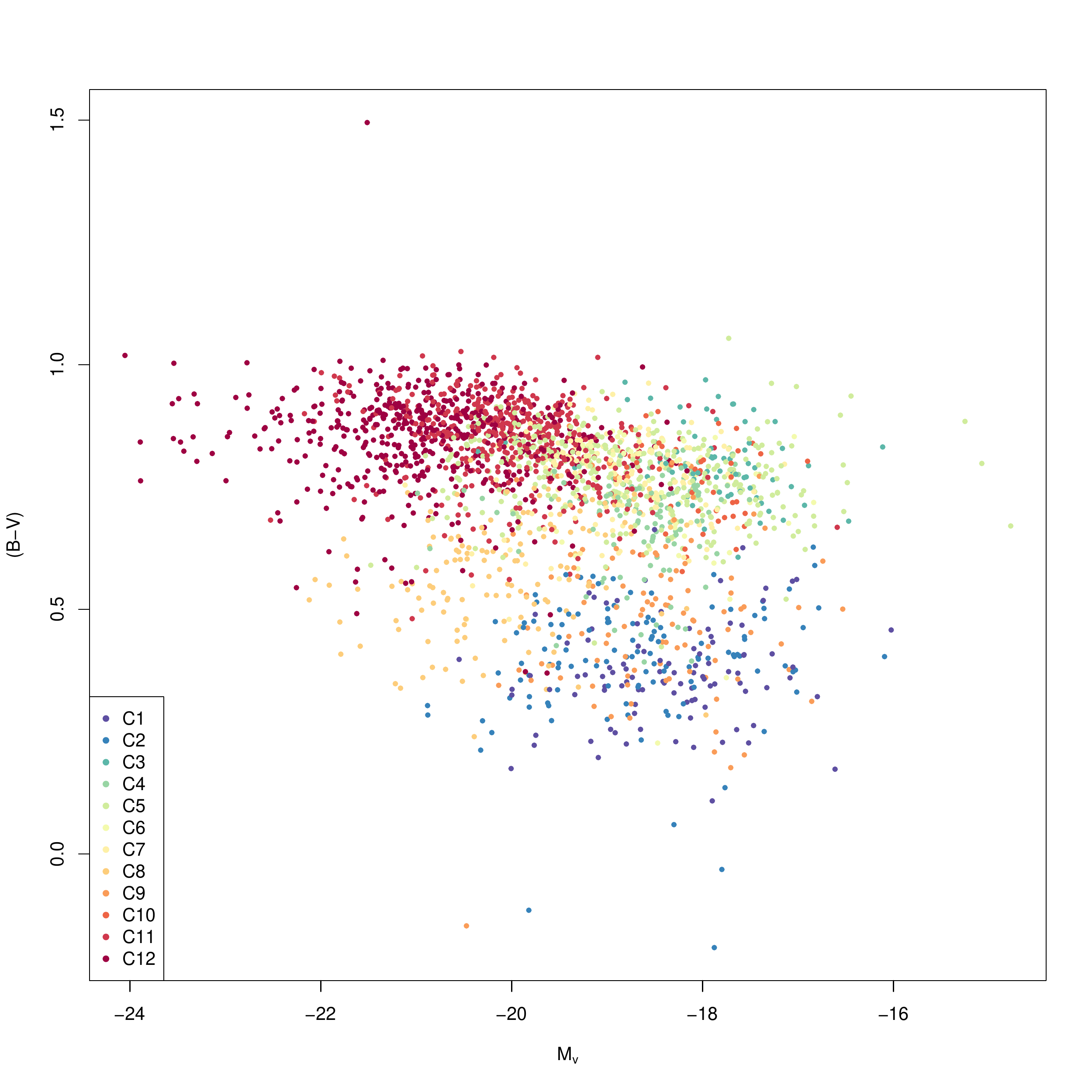}
\caption{Colour-magnitude diagram for the cluster sample.%}
\label{fig:CMDM} }
\includegraphics[width=\linewidth]{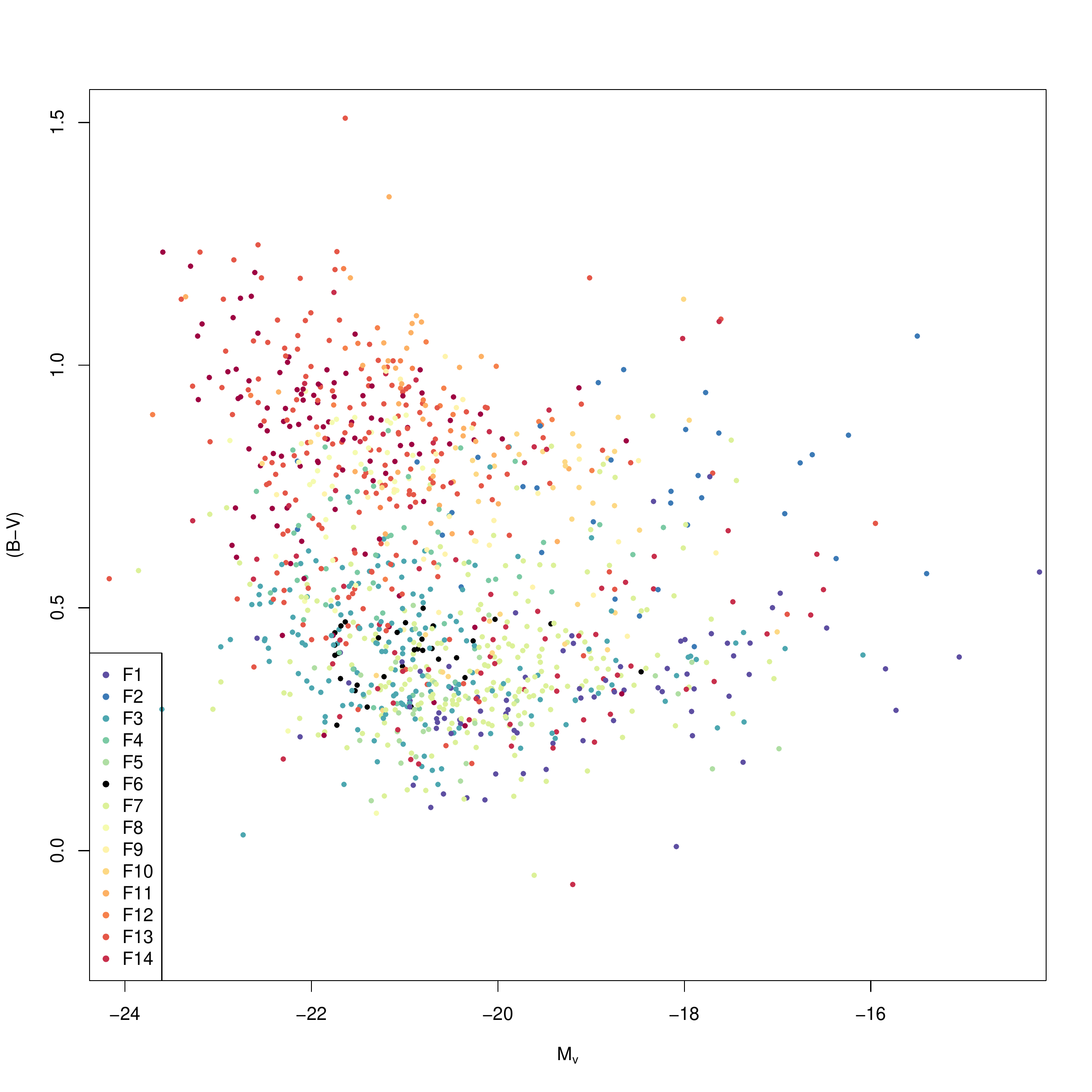}
\caption{Colour-magnitude diagram for the field sample.%}
\label{fig:CMDN} }
\end{figure}

The colour-magnitude diagram for the cluster galaxy sample (Fig.~\ref{fig:CMDM}), shows two main regions that are occupied by groups: C1, 2 8 and 9, that corresponds to the blue sequence, and a longer and more populated region corresponding to the red sequence. The projection of the tree onto this diagram (Figs.~\ref{fig:treeMembCMD}) depicts the evolutionary path: the evolution goes first vertically between the two regions. After C5, there is a bifurcation corresponding to the two ensembles seen in Fig.~\ref{fig:TreeMemb}): one along the red sequence (C10 to C12), and the other one (C6 to C9) going downwards with C8 and C9 clearly in what is often called the green valley. Hence, we have somewhat expected evolution from the blue to the red sequence, and subsequently along the red sequence towards brighter galaxies, but the green valley is not a transitional region, rather it is the destination of another branch of the evolutionary path of galaxies.

The field sample (Fig.~\ref{fig:CMDN}) is remarkably different, the redder and less luminous part being poorly populated, while the blue and more luminous part is quite dense. There seems to be a shift of a large part of galaxies between the two diagrams. Even if this might partially be due to a Malmquist bias, it appears that the beginning of the evolution from F1 to F2 is very similar to the case of the cluster sample (Figs.~\ref{fig:treeCompCMD}). However, the subsequent trajectory passes through the green valley and then to the upper left part of the red sequence.

This diagram is another example of the influence of the environment on the galaxy properties and their evolution. This is also another good proof that the Maximum Parsimony approach is able to reveal the characteristics of the evolution of different galaxy populations. In clusters the red and blue sequence are better defined and we can clearly identify the class of objects in the green valley, that are likely rejuvenated by episode of star formation or wet merger events. Note in particular the extended sequence of red and faint objects present in the cluster sample. They are almost absent in the field. These objects, once entered in the cluster environment rapidly loose their gas and stop their star formation becoming red. In contrast, luminous and blue galaxies are numerous only in the field sample, probably because in clusters stellar formation has been hindered or suppressed.

\section{Conclusions}
\label{Conclusion}

This study aims at investigating the effect of cluster environment on the evolutionary scenario of galaxies. We have used a consequent sample of homogeneous data, from the WINGS catalogue, with about 2600 galaxies which are members of clusters, and about 1500 ones which are field galaxies. 

We used Maximum Parsimony, a phylogenetic technique also known as cladistics. It is designed to find evolutionary paths between objects or classes and has now been often used in astrophysics. However, the sample is too large for this technique to be applied directly. We reduced the amount of objects by performing a pre-clustering using a hierarchical clustering method in order to pave the data set according to its extent and not to its density distribution. We have found that 300 pre-clusters is a good compromise between representativity and computational constraints of the MP analysis. We used seven parameters for these two machine learning computations: \BmV, \logre,  \nV, \muem, $H_{\beta}$, \DQ, \lM. They are all fully documented, and we have checked that there is no redundancy nor strong correlation in this set.

We have obtained an evolutionary tree for the two samples combined and separately. The result with all galaxies does not show distinct evolutionary paths for cluster and field galaxies. Rather, it shows that the cluster environment may somehow accelerate the evolution of galaxies by making them more diversified (different) from primitive galaxies. Also, field galaxies appear more massive at similar evolutionary stages and form more stars. 

On the tree for the cluster cluster sample, there is a distinct sub-structure, a distinct evolutionary path, that seems to gather rejunevated or stripped-off galaxies, and goes into the green valley (or even blue sequence) from the red sequence on the CMD diagram.  

The two separate analyses reveal subtle evolutionary differences. Cluster galaxies appear more homogeneous since the groups of the field sample are characterised by slightly more specific and varied properties. The classifications we obtained from the cluster and field trees are only partly driven by mass and colour. The Sersic index is highly correlated with the groups for the field galaxies, and it appears to be one of the main drivers of the evolutionary tree structure together with the the effective radius and the stellar content, the mass being here not important at all.  

The classification obtained from the tree of the cluster sample does not appear to depend on the clusters, implying some "universality" of the diversity of cluster galaxies. However, our sample is at low redshift, and the number of galaxies in each cluster is sometimes limited.

An important outcome of a multivariate analysis is the investigation of scaling relations, and in particular the understanding of their dispersion. We find that many, if not all, such relations are explained by co-evolution, and that a proper classification makes these relations appearing as a mixture of galaxy populations at different stages of evolution. For instance, we show that the dispersion in the Kormendy relation is in fact due mainly to at least two parallel relations with two different masses. Also the Kormendy relation is not due to mass since it appears in some low-mass groups as well. Also, the mass vs radius relation holds only for highly diversified and massive groups, but not for the field sample.

Finally, we find a striking difference in the colour-magnitude diagrams for cluster and field galaxies, and the evolutionary trees projected on this diagram shows that evolution does not go simply from the blue sequence to the red one. Even if it is true for some cluster galaxies, the others tend to go back to the blue sequence, through the green valley. This latter process is the one chosen by field galaxies.

This paper is another demonstration that unsupervised machine learning clustering  is able identify known classes of objects by precising the properties, and to stick out more peculiar types of objects. Importantly, the phylogenetic approach is able to provide evolutionary scenarios to explain the physical origins of all these classes. Of course, investigating the diversification of galaxies at a redshift of nearly zero is limited and frustrating. We have however shown that the evolutionary paths are somewhat complicated: even though galaxies globally become redder, more massive, less disky, there are many other possible trajectories, such as field galaxies which are bluer, more concentrated but more massive than cluster ones, or cluster galaxies that are stripped-off, etc. Also, there are many more properties of galaxies other than colour, mass and morphology. What we need now is to go to higher redshifts, trying to find more primitive galaxies, and better map the evolutionary paths of present day objects. It is important to recall that even if cluster galaxies undergo more frequent interactions than field ones, the latter certainly were subject to such transforming events long time ago.

\begin{acknowledgements}
We thank the referee for thorough and constructive comments that improved significantly this paper.
\end{acknowledgements}	
	
\bibliographystyle{aa}
\bibliography{WINGS}

\begin{appendix}
\section{Number of hierarchical clusters and tree rooting}
\label{App:hclustcompar}

\begin{figure}[ht]
\centering
\includegraphics[width=\linewidth]{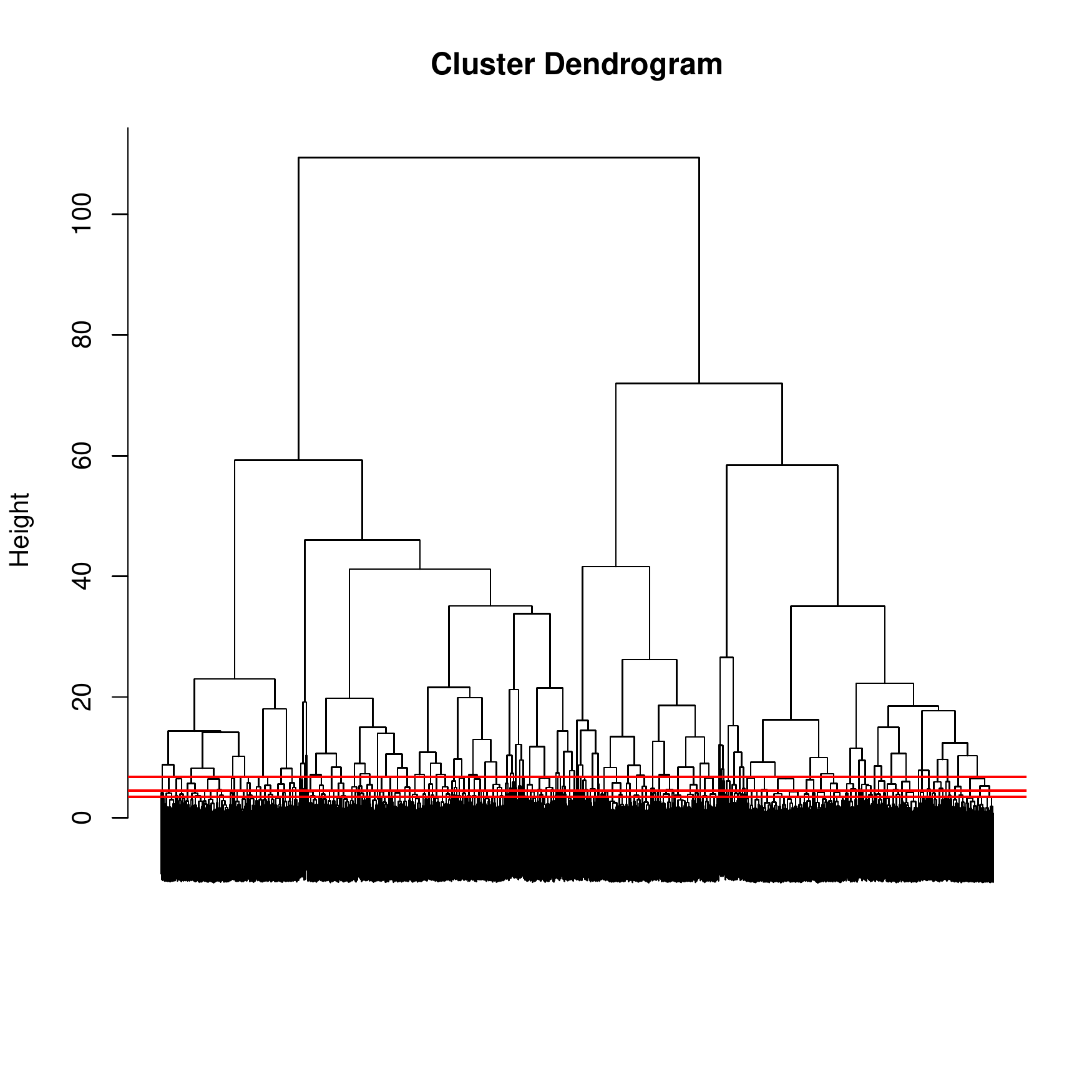}
\caption{Dendrogram of the hierarchical clustering algorithm for the full sample. The horizontal red lines indicate from top to bottom the cuts defining the 100, 200 and 300 pre-clusters respectively.}
\label{fig:dendrogram}
\end{figure}

\begin{figure*}[ht]
\centering
\includegraphics[width=\linewidth]{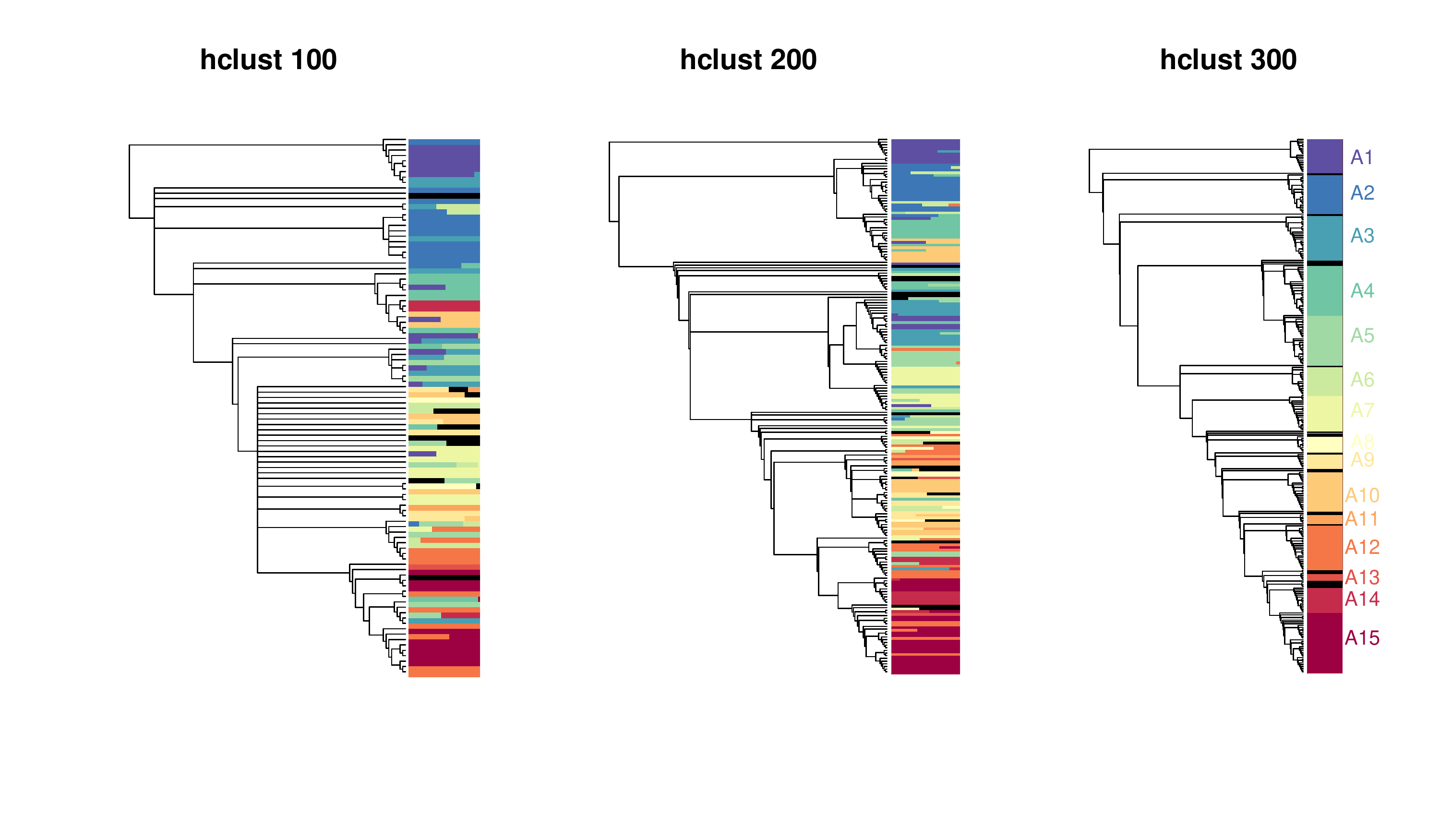}
\caption{Trees obtained for the full sample with 100 (left), 200 (middle) and 300 (right) pre-clusters (see Sect.~\ref{MPlarge}). The tree on the right is the one shown in Fig.~\ref{fig:TreeAll}. The coloured bars at each tip of the trees give the distribution of the galaxies of the corresponding pre-cluster  in the groups defined in this paper (the classes A). }
\label{fig:hclustcompar}
\end{figure*}

\begin{figure*}[ht]
\centering
\includegraphics[width=0.8\linewidth]{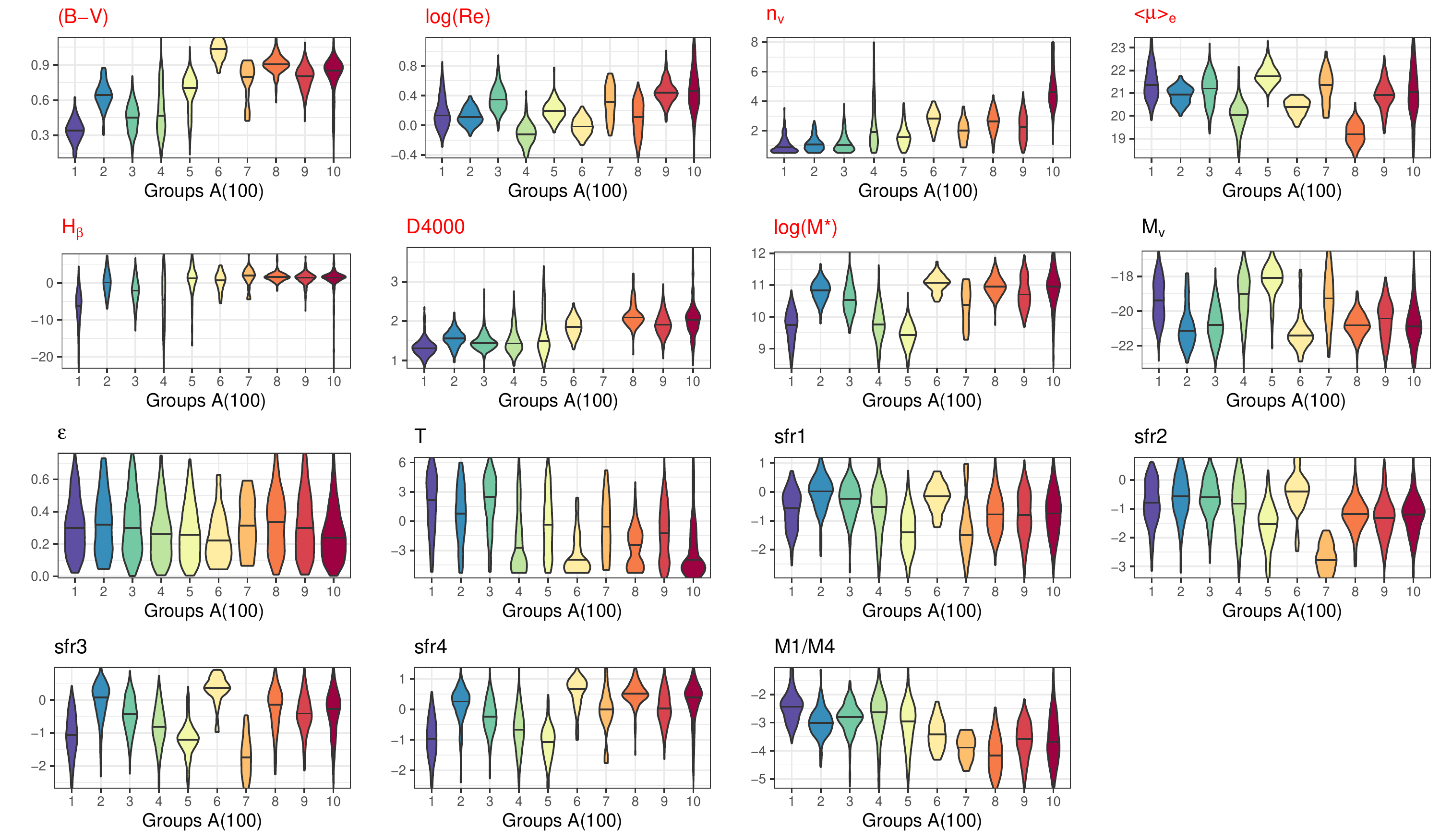}
\caption{Violinplots for groups defined on the tree with 100 pre-clusters (left in Fig.~\ref{fig:hclustcompar}) in the same way as described in Sect.~\ref{groupdef}. This figure should be compared with Fig.~\ref{fig:BoxplotAll} and Fig.~\ref{fig:hclustcomparviolins200}. The violinplot of group 7 for parameter \DQ\ spreads between 5 and 9 and is thus well above the diagram.}
\label{fig:hclustcomparviolins100}
\end{figure*}

\begin{figure*}[ht]
\centering
\includegraphics[width=0.8\linewidth]{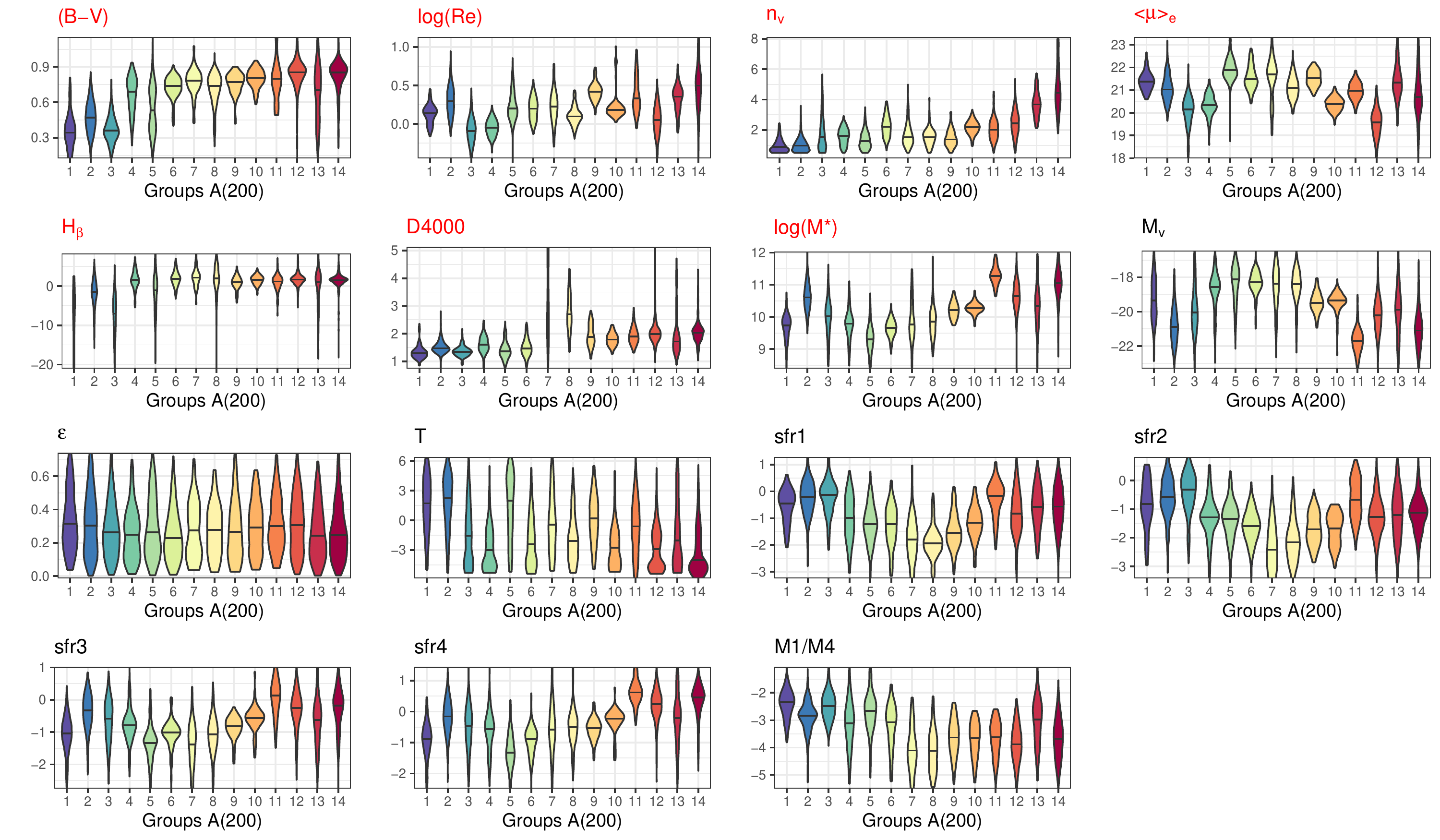}
\caption{Violinplots for groups defined on the tree with 200 pre-clusters (middle in Fig.~\ref{fig:hclustcompar}) in the same way as described in Sect.~\ref{groupdef}. This figure should be compared with Fig.~\ref{fig:BoxplotAll} and Fig.~\ref{fig:hclustcomparviolins100}.}
\label{fig:hclustcomparviolins200}
\end{figure*}

\begin{figure*}[ht]
	\centering
	\includegraphics[width=0.6\linewidth]{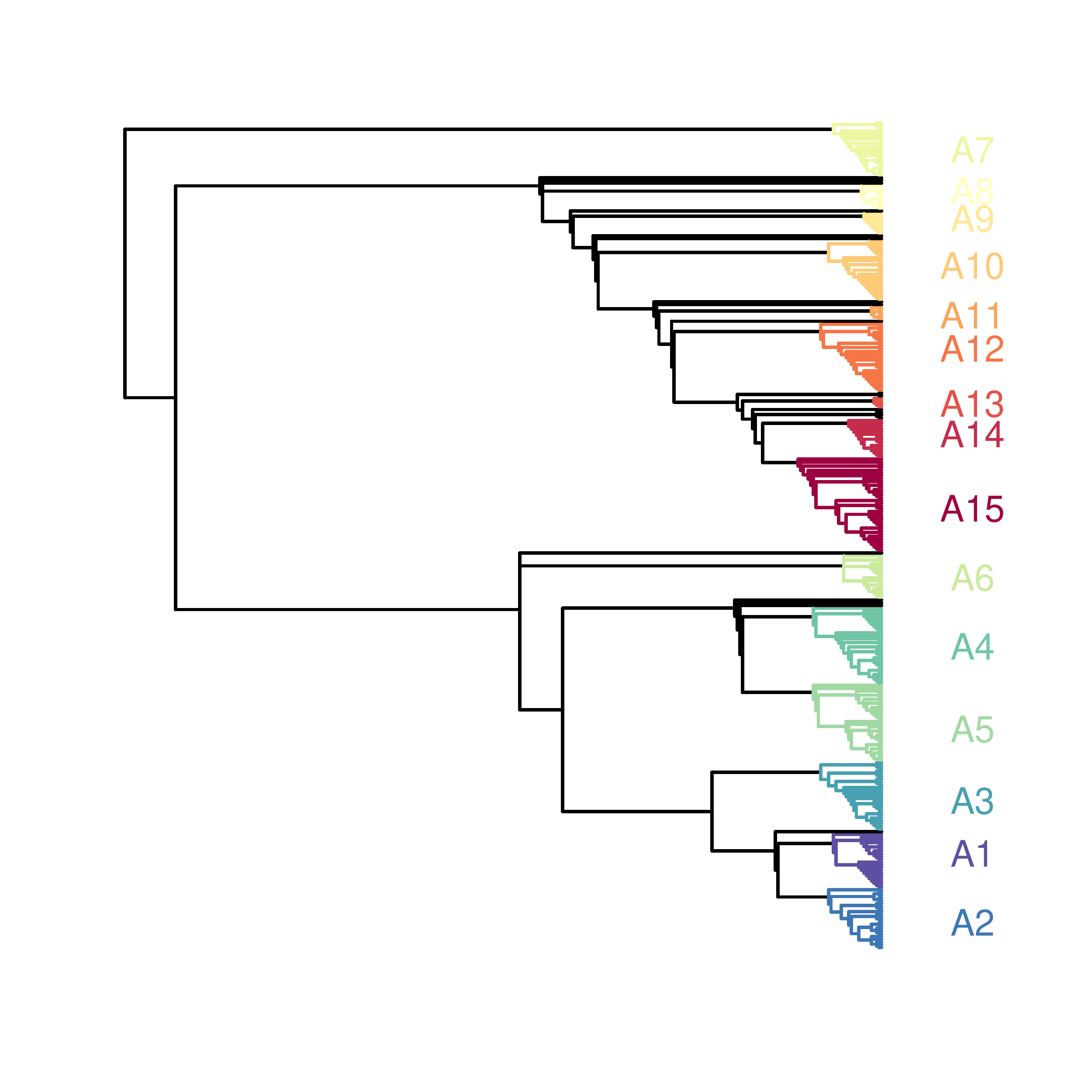}
	\caption{Same tree as in Fig.~\ref{fig:TreeAll} for the full sample but rooted with group A7. Colours and names of groups are identical.}
	\label{fig:treeotherroot}
\end{figure*}

\begin{figure*}[ht]
	\centering
	\includegraphics[width=0.8\linewidth]{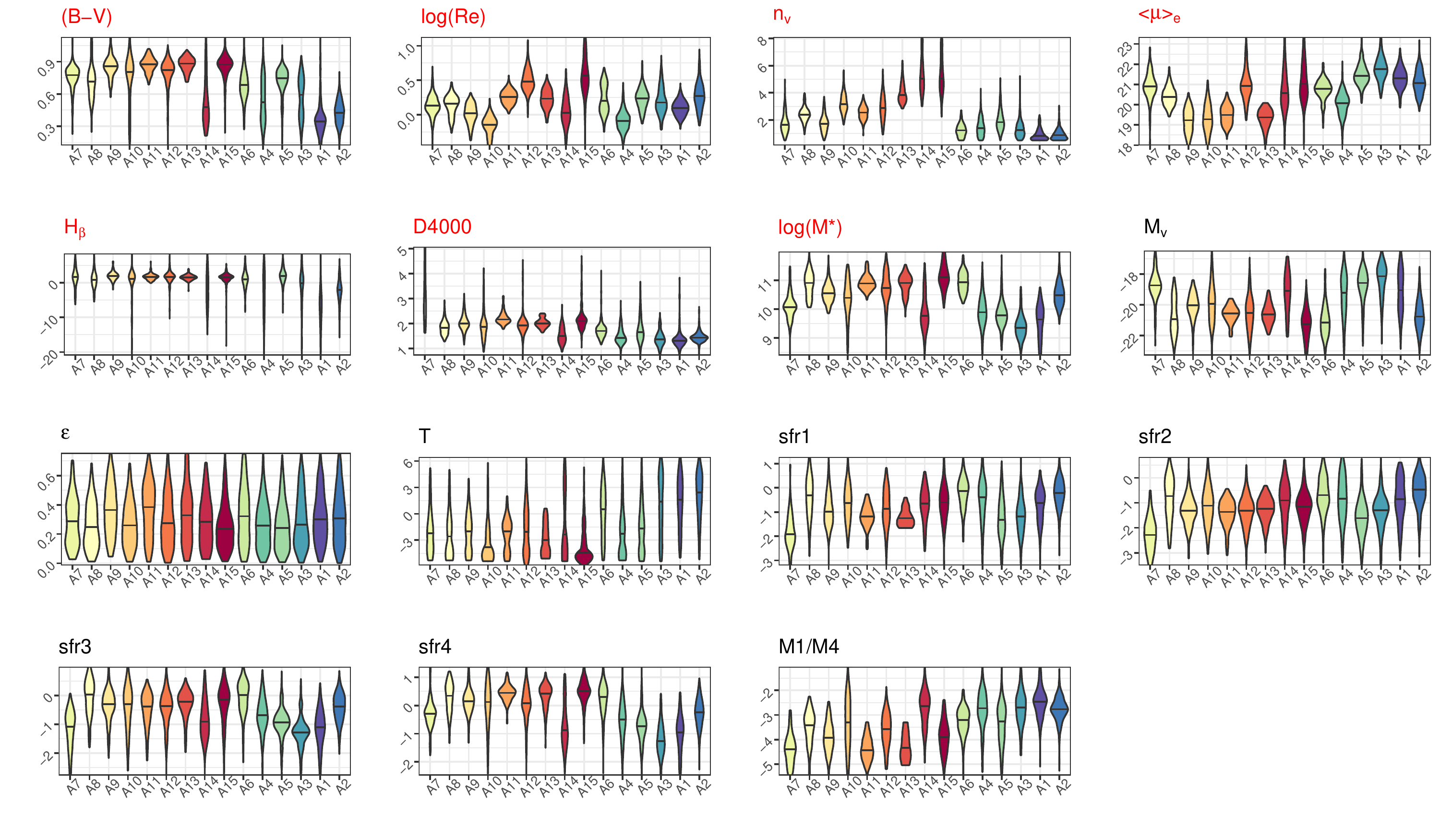}
	\caption{Same plot as in Fig.~\ref{fig:BoxplotAll} for the full sample but rooted with group A7 according to tree in Fig.~\ref{fig:treeotherroot}. Colours and names of groups are identical..}
	\label{fig:violinotherroot}
\end{figure*}

In this Appendix, we investigate the influence of the choice of the number of pre-clusters and the tree rooting on the result of the MP analysis.

We have explained in Sect.~\ref{MPlarge} and Sect.~\ref{groupdef} how the number of pre-clusters and the contours of the phylogenetic groups have been chosen. We want to show here the robustness of our results with respect to these choices. For this purpose we have performed the same cladistic analysis for the full sample with 100 and 200 pre-clusters. 

Due the hierarchical nature of their definition, the three sets of pre-clusters match: a pre-cluster of the 100-set contains one to several pre-clusters of the 200-set and so forth (see Fig.~\ref{fig:dendrogram}). As a consequence, the medians of the parameters used for the MP analysis similarly span the entire variance of the full sample in the three cases. We can thus expect that the evolutionary paths (the relationships between classes) should not be very different.

This is confirmed on the MP trees obtained using the three pre-cluster sets (Fig.~\ref{fig:hclustcompar}): for each tree, and for each pre-cluster (at tree tips) we show the distribution of galaxies in the groups defined in this paper from the tree in Fig.~\ref{fig:TreeAll}, reproduced here on the right, using the same colour code. The tree shapes in the three cases are not identical, but most of the groups can be well identified on the trees of the 100- (left) and 200 (right) pre-cluster sets, indicating that the majority of the galaxies belonging to one of the A groups remain together. The agreement is better for the 200 pre-cluster tree although part of groups A5 and A10 are displaced.

The colours on the three trees clearly show that the diversification scheme is very similar as well. By defining groups from the 100- and 200-pre-cluster trees, the violinplots (Fig.~\ref{fig:hclustcomparviolins100} and \ref{fig:hclustcomparviolins200}) confirms this similarity when compared with Fig.~\ref{fig:BoxplotAll}. This implies that the main conclusions of our study are not modified,

The MP analysis yields an unrooted tree since no direction of changes of the parameters has been provided. To interpret the tree as an evolutionary scenario, an arrow of diversification must be defined by choosing a root for the tree. This root must be as similar as possible as the common ancestor to the objects under study. In this paper, we have used three parameters the evolution of which is thought to be well understood: colour, driven by the age of the stellar population as well and metallicity, Sersic index which represent the dynamics, and the mass which tend to increase. The choice of the root can always been disputed, and only the consistency of the final tree can in principle support it or not.

To provide astronomers an idea of the consequences of a different root, we show here a tree rooted with group A7 (Fig.~\ref{fig:treeotherroot}) with the corresponding sequence of violinplots (Fig.~\ref{fig:violinotherroot}). The tree shows two main branches splitting at group A7, each representing a kind of lineage. Interestingly, the proportion of cluster galaxies is on average significantly higher in the branch A8 to A15 than in the branch A6 to A1 (Table~\ref{tab:ADist}) although this is not a clearcut separation.

The sequence of the changes of the variables along the tree (Fig.~\ref{fig:violinotherroot}) should be read with caution, the two branch must be seen as two parallel separate evolutionary paths. The striking behaviour is the decreasing colour, that is galaxies become bluer, especially in the branch A6-A2, which is not compatible with the ageing of stars and the increasing metallicity. This fact by itself rules out the choice of group A7 as root. The mass is higher in the branch A8 to A15 than in A7, but is slightly lower in the branch A6 to A2. The morphology also in the second branch clearly evolves towards spiral types. We think that the changes within each branch is not in good agreement with our knowledge of the physics of the galaxy evolution.

One could have chosen for instance only mass to root the tree, using group A3, but this would have not changed significantly our result, and it is doubtful that the evolution (and the diversification) of galaxies would depend on only one single property.

\clearpage

\section{Statistical correlations among the variables}
\label{statvariables}

These figures are described in Sect.~\ref{paraminfluence}.

\begin{figure*}[ht]
\centering
\includegraphics[width=\linewidth]{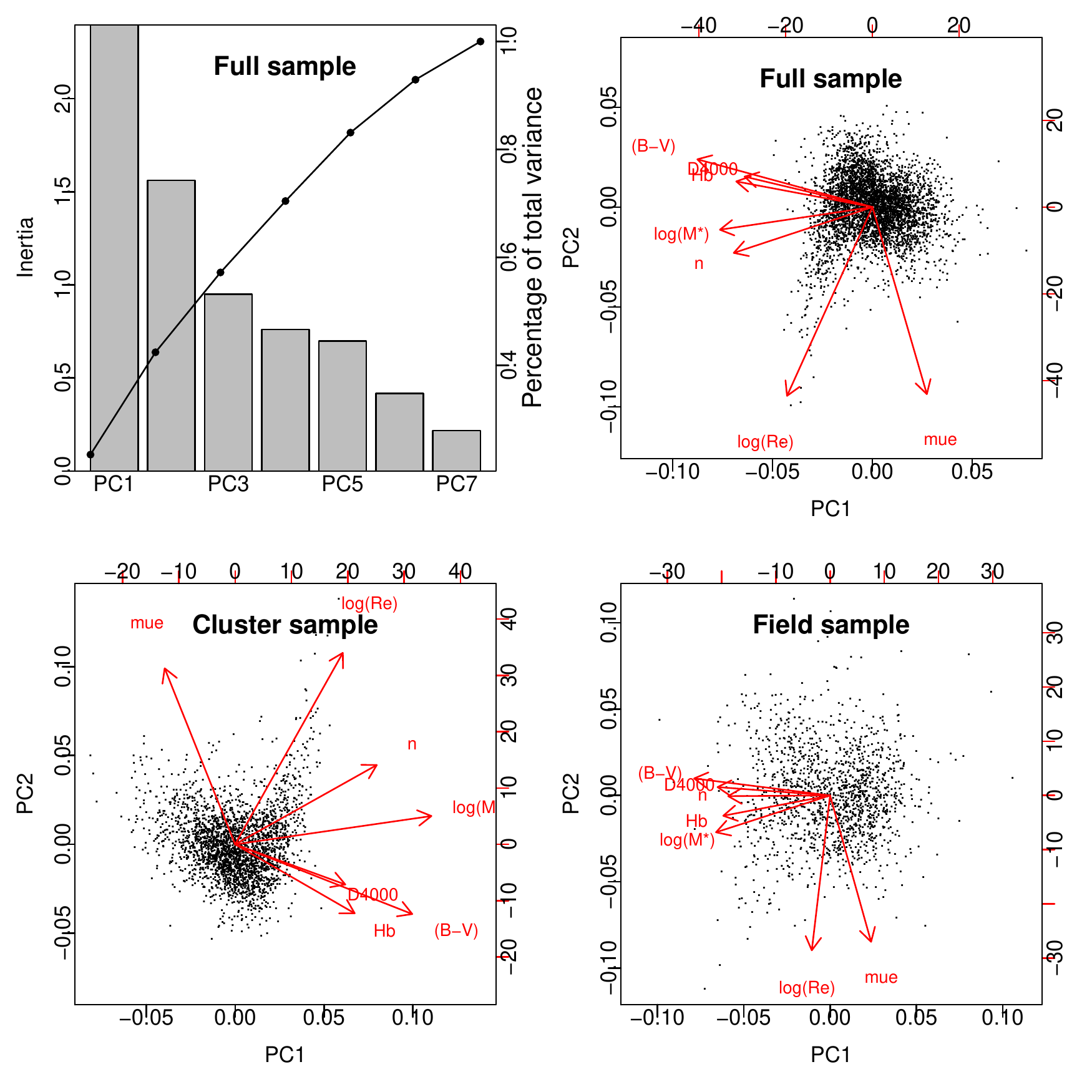}
\caption{Principal Component Analysis using the seven parameters used for the clustering and MP analyses (Sect.~\ref{paraminfluence}). \textit{(Top left)} Screeplot for the full sample showing the variance (eigenvalues) of the seven Principal Components (PCs). The curve represents the cumulative percentage of the total variance. The screeplot for the cluster and field sample are very similar. \textit{(Top right)} Projection of the data on the plane of the first two PCs with arrow showing the loadings and direction of the parameters. The same diagram are plotted for the cluster (\textit{bottom left})  and field (\textit{bottom right}) samples.}
\label{fig:PCA}
\end{figure*}

\begin{figure}[ht]
\centering
\includegraphics[width=\linewidth]{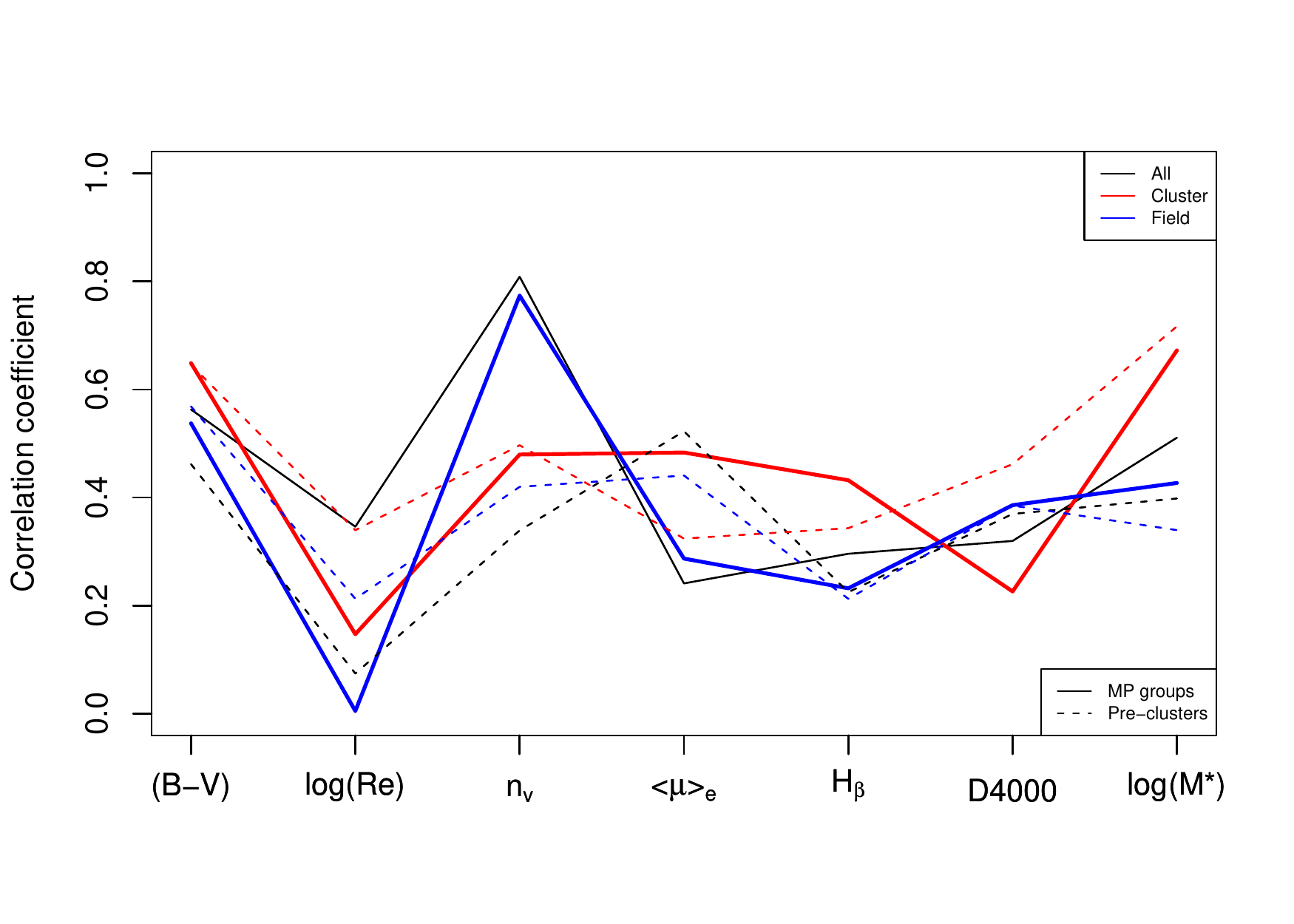}
\caption{Correlation (absolute value) of the seven variables with the groups (solid lines) and the pre-clusters (dashed lines; Sect.~\ref{paraminfluence}).}
\label{fig:corclas}
\end{figure}

\clearpage

\section{Projection of the tree on the CMD diagram}
\label{treeonCMD}

These figures are described in Sect.~\ref{CMD}. The trees in Fig.~\ref{fig:CMDM} and \ref{fig:CMDN} are projected on the bivariate plot \BmV\ vs \mV. The tips of the trees are the medians of these parameters for each group. The internal nodes are given values by minimizing the squared changes of the parameters along the branches  \citep[squared-change parsimony ancestral state reconstruction,][]{Maddison1991}.

\begin{figure*}[ht]
	\centering
\includegraphics[width=\linewidth]{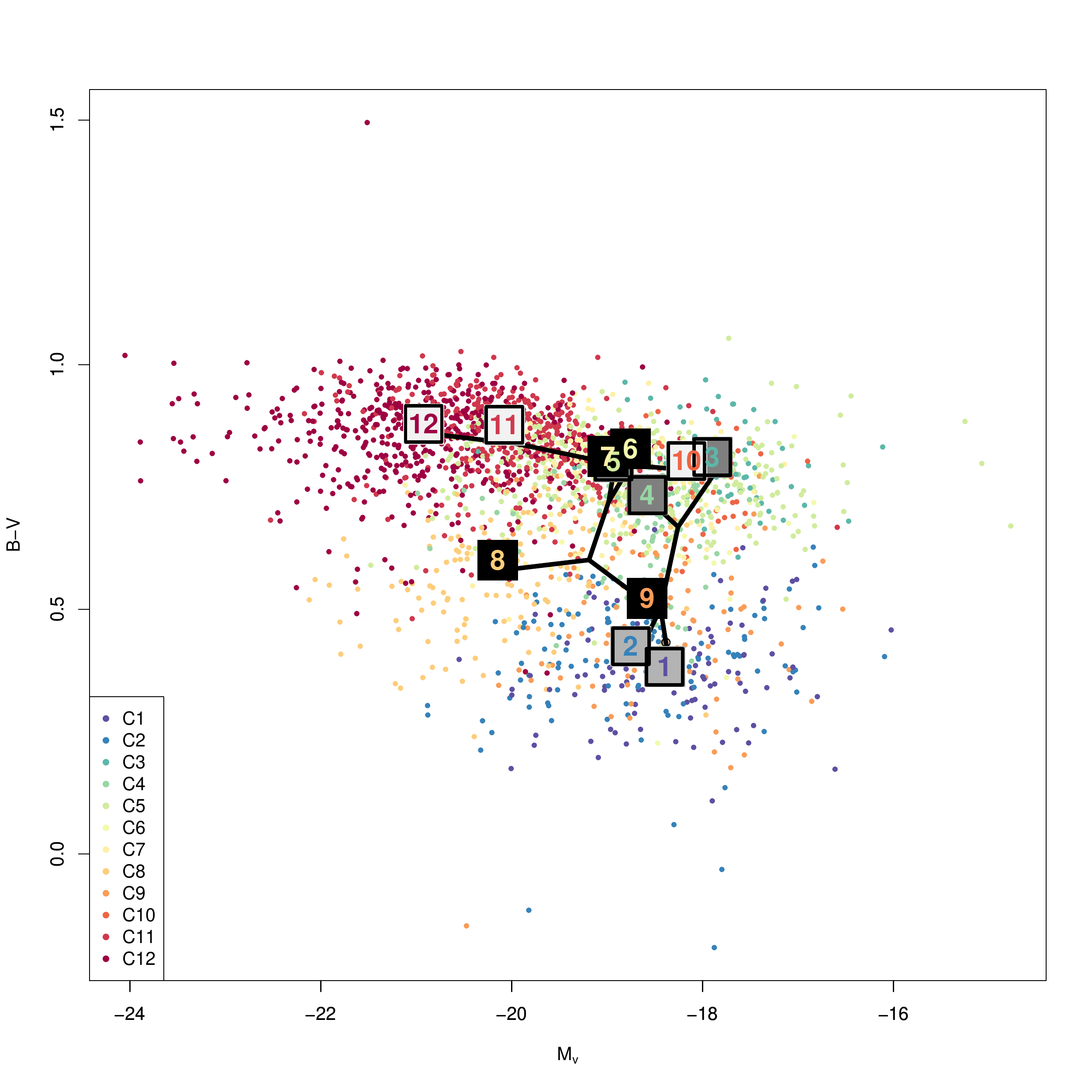}
\caption{Tree projected onto the CMD diagram for the cluster sample.}
\label{fig:treeMembCMD}
\end{figure*}

\begin{figure*}[ht]
\centering
\includegraphics[width=\linewidth]{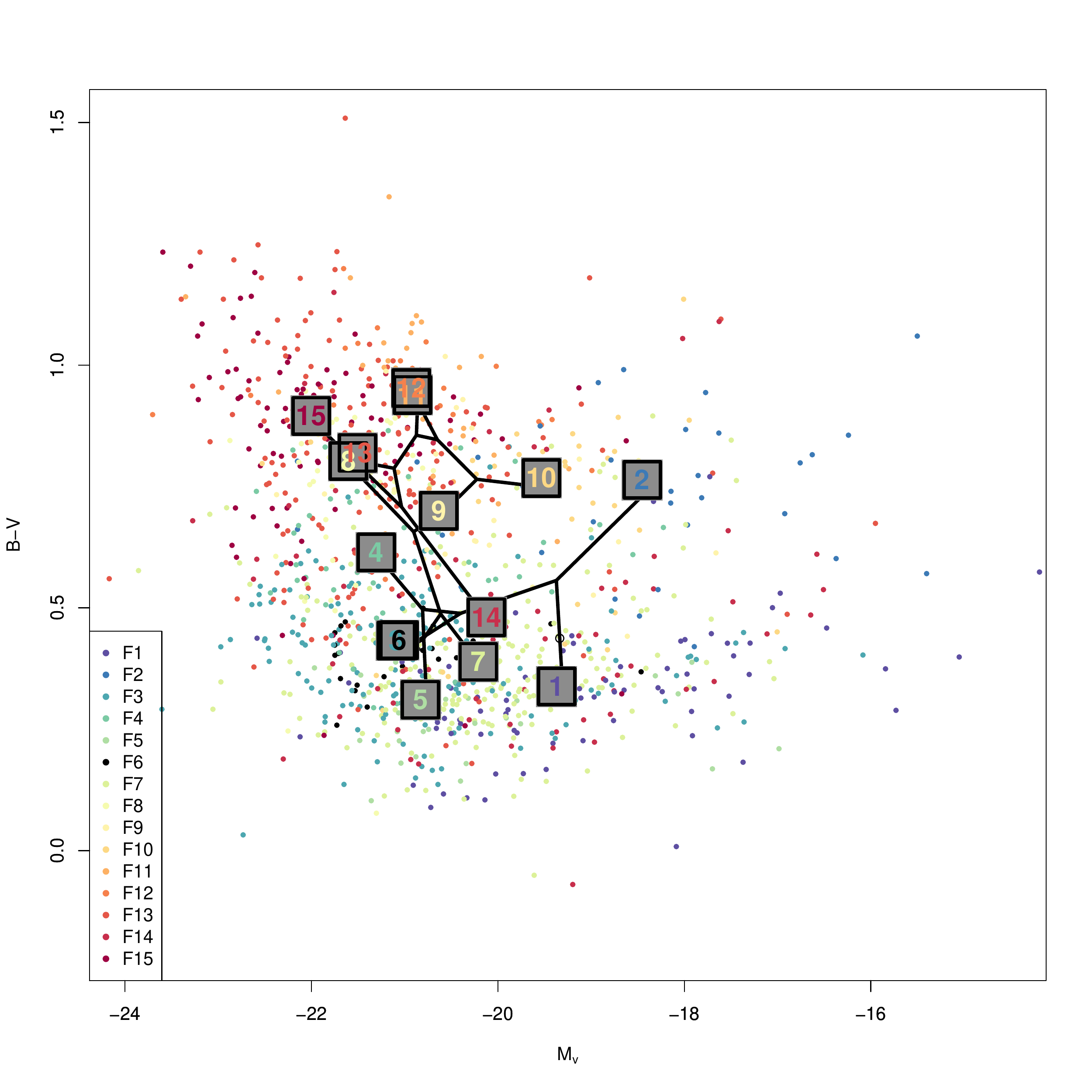}
\caption{Tree projected onto the CMD diagram for the field sample. The median values of the groups F3 and F6 are superimposed on this plot.}
\label{fig:treeCompCMD}
\end{figure*}

\end{appendix}

\end{document}